\documentclass{emulateapj}

\slugcomment{Not to appear in Nonlearned J., 45.}

\shorttitle{Spitzer spectra of Seyfert galaxies}
\shortauthors{Tommasin et al.}

\begin{document}


\title{Spitzer-IRS high resolution spectroscopy of the 12$\mu$m Seyfert galaxies: \\
  I. First results}


\author{Silvia Tommasin\altaffilmark{1}, Luigi Spinoglio}
\affil{Istituto di Fisica dello Spazio Interplanetario, INAF, Via Fosso del Cavaliere 100, I-00133 Roma, Italy}

\author{Matthew A. Malkan}
\affil{Astronomy Division, University of California, Los Angeles, CA 90095-1547, USA}

\author{Howard Smith}
\affil{Harvard-Smithsonian Center for Astrophysics, 60 Garden Street, Cambridge, MA 02138}

\author{Eduardo Gonz\'alez-Alfonso}
\affil{Universidad de Alcal\'a de Henares, Departamento de Fisica, Campus Universitario, E-28871 Alcal\'a de Henares, Madrid, Spain}

\and

\author{Vassilis Charmandaris\altaffilmark{2}}
\affil{University of Crete, Department of Physics, GR-71003 Heraklion, Greece}


\altaffiltext{1}{also at the Physics Department of Universit\'a di Roma, La Sapienza, Roma, Italy}
\altaffiltext{2}{IESL/Foundation for Research and Technology - Hellas,
  GR-71110, Heraklion, Greece and Chercheur Associ\'e, Observatoire de
  Paris, F-75014, Paris, France}


\clearpage

\begin{abstract}

The first high resolution \textit{Spitzer} IRS 9-37$\mu$m spectra of 29 Seyfert galaxies (about one quarter) 
of the 12$\mu$m Active Galaxy Sample are presented and discussed. 
The high resolution spectroscopy was obtained with corresponding off-source observations. 
This allows excellent background subtraction, so that the continuum levels and strengths of weak 
emission lines are accurately measured. The result is several new combinations of emission line ratios, 
line/continuum and continuum/continuum ratios that turn out to be effective diagnostics of the strength 
of the AGN component in the IR emission of these galaxies. The line ratios [NeV]/[NeII], [OIV]/[NeII], 
already known, but also  [NeIII]/[NeII] and [NeV]/[SiII] can all be effectively used to measure the 
dominance  of the AGN. We extend the analysis, already done using the 6.2$\mu$m PAH emission 
feature, to the equivalent width of the
11.25$\mu$m PAH feature, which also anti-correlates with the dominance of the AGN. 
We measure that the 11.25$\mu$m PAH feature has a constant ratio with the H$_2$ S(1) irrespective 
of Seyfert type, approximately 10 to 1. Using the ratio of 
accurate flux measurements at about 19$\mu$m with the two spectrometer channels, having 
aperture areas differing by a factor 4, we measured the source extendness and correlated it with 
the emission line and PAH feature equivalent widths. The extendness of the source gives another 
measure of the AGN dominance and correlates both with the EWs of [NeII] and PAH emission. 
Using the rotational transitions of H$_2$ we were able to estimate temperatures (200-300K) and 
masses (1-10 $\times$ 10$^{6}$ M$_{\sun}$), or significant limits on them, for the warm molecular 
component in the galaxies observed.
Finally we used the ratios of the doublets of [NeV] and of [SIII] to estimate the gas electron density, 
which appears to be of the order of n$_e$ $\sim$ 10$^{3-4}$ cm$^{-3}$ for the highly ionized 
component and a factor 10 lower for the intermediate ionization gas. 

\end{abstract}



\section{Introduction}
Mid-infrared (mid-IR) spectroscopy provides a powerful tool to
investigate the nature and physical processes in active galactic
nuclei (AGNs) and in the starburst dominated regions frequently
associated to them. 
Because of the large variety of fine structure lines present in the mid-IR, 
covering a wide range in ionization/excitation conditions and gas density \citep[e.g.][]{sm92} 
mid-IR spectroscopy of the Narrow Line Regions (NLR) in AGNs can  
add information not available from classical optical spectroscopy, 
especially when dust extinction is high. Furthermore, the
electronic transitions responsible for the infrared emission lines of
various elements are less sensitive to uncertainties in temperature
than the corresponding optical lines. 
Moreover the brightest H$_2$ rotational lines, that can be used to quantify the presence and excitation of warm  
molecular gas, as well as a prominent Polycyclic Aromatic Hydrocarbons \citep{pule89}, 
hereafter PAH, feature at 11.25 $\mu$m are in the mid-IR.
The first detailed mid-IR spectroscopic studies of AGNs and Starburst
galaxies using multiple ionic transitions of various elements were
performed by various authors \citep[][for a
review]{stu02, spi05, ver03, ver05} with the \textit{Short Wavelength
Spectrometer} (SWS) \citep{deg96} onboard of the
\textit{Infrared Space Observatory} (ISO) \citep{kes96}. However, the improved sensitivity of
the Infrared Spectrometer (IRS) \citep{hou04} on the 
\textit{Spitzer Space Telescope} now enables a detailed
investigation of the nature and physical processes in large samples of galactic nuclei from
nonstellar (AGN) and stellar (starburst) power sources. 

Even at low resolution, the IRS spectra of the first few classical AGNs 
already showed the diversity of the mid-IR spectral features: silicate absorption and emission, 
PAH emission and strong fine structure lines \citep{wed05}.

\citet{buc06} examined 51 low resolution IRS spectra of 12$\mu$m selected Seyfert galaxies \citep{rms93},
exactly the sample we are considering in our study. They report a few major findings: 
(1) the sample contains a very wide range of continuum types, with no more than about 3 galaxies being closely similar to one another, 
however principal component analysis applied to their data suggests that the relative contribution of starburst emission may be the 
dominant cause of variance in the observed spectra; (2) the starburst component in the sample objects does not contribute 
more than 40\% of the total IR flux density; 
(3) Seyfert 1's have higher ratios of infrared to radio emission \citep[see also][]{rme96}; (4)  the Seyfert 2 galaxies typically show stronger starburst contributions than Seyfert 1's.

\citet{Gor07} found a strong correlation between the [NeV]14.3$\mu$m and the [NeIII]15.5$\mu$m lines in 
Narrow Line Regions (NLR) of AGNs, spanning 4 orders of magnitude in luminosity. This would imply a 
very narrow range in ionization parameter (-2.8$<$logU$<$-2.5) for simple constant density photoionization 
models. \citet{dud07} discussed the ratio of the doublets of [NeV] and [SIII] in a heterogeneous sample of 
active galaxies. Finally ULIRG galaxies, more than Seyfert galaxies, have been so far the object of  
systematic spectroscopic studies with Spitzer IRS \citep{hig06,arm07,des07,far07}.

In this article we present the first  high resolution IRS spectra of 29 galaxies  from the original list 
of Seyfert galaxies of the "12$\mu$m Galaxy Sample" (12MGS) \citep{rms93}, an IRAS-selected 
all-sky survey flux-limited to 0.22 Jy at 12$\mu$m.

The sample selection is briefly described in \S 2, the observations are described in \S 3, the 
results are presented in \S 4, the diagnostics diagrams using line ratios, equivalent widths and 
other observed quantities are presented and discussed in \S 5 and a comparison between Seyfert 1's and 
Seyfert 2's is given in \S 6. The conclusions are given in \S  7.

\section{The Seyfert galaxies of the 12$\mu$m Galaxy Sample}

This sample is essentially a bolometric flux-limited survey outside the galactic plane, because of the 
empirical fact that all {\it galaxies} emit a
constant fraction of their total bolometric luminosity at 12$\mu$m. This fraction is $\sim$ 15\% for 
AGNs \citep{sm89} and $\sim$ 7\% for normal 
and starburst galaxies, independent of star formation activity \citep{spi95}. Moreover, the 12MGS is 
less subject to contamination by high star-formation rate objects than other infrared samples defined at longer wavelengths \citep{hm99}.
The 12MGS contains: 53 Seyfert 1s and 63 Seyfert 2s \citep{rms93}. A unique advantage of our 
sample is that it already has the best and most complete 
set of observations at virtually every other wavelength: full IRAS and near-IR coverage \citep{rms93, spi95}, 
X rays \citep{rus96}, optical spectroscopy, 
radio \citep{rme96}, optical/IR imaging \citep{hun99, hm99},  ISOPHOT data on half of the Seyfert galaxies \citep{sam02}. 
In the recent years 10$\mu$m imaging \citep{gor04}, K band and 2.8-4.1$\mu$m slit spectroscopy \citep{ima03, iah04} 
optical spectropolarimetry \citep{tra01, tra03} and radio observations \citep{the00, the01} have been collected for most 
of the Seyfert galaxies in our sample. Low resolution IRS Spitzer spectra have been collected on 87 Seyfert galaxies of 
the 12$\mu$m sample and  presented for a large fraction of them (51 objects) \citep{buc06}.

\section{Observations and data reduction}

The Spitzer IRS observations presented here have been collected with the Spitzer Cycle 3 GTO 
program "IRS spectroscopy of a complete sample of Seyfert galaxies in the Local Universe" 
(P30291, PI. G. Fazio). In following papers we will present the remaining data of this program 
as well as the results of a complete reduction of the other high resolution spectra of the 
12$\mu$m Seyfert galaxies of the Spitzer archive. We refer to another future paper for 
the comparison of the data with photoionization models.

A complete description of the IRS onboard of Spitzer Space Telescope can be found 
in \citet{hou04}. For our observations we used the two high resolution modules of IRS, 
that allow observations at the wavelength range from 10  to 37$\mu$m with resolutions 
R=$\lambda$/$\delta\lambda$=600. The short wavelengths in the range 9.7-19.6 $\mu$m 
are covered by the Short-High (SH) module and the 18.7-37.2 $\mu$m range 
by Long-High (LH).

We acquired the targets using the red (22$\mu$m) IRS peak-up camera in high accuracy 
mode to locate the mid-IR centroid of the source and then offset to the appropriate slit 
using the standard IRS staring observing mode. For each observation of a Seyfert nucleus, 
we also collected an observation of an off-source position at a distance of 2\arcmin~in declination, 
to be able to subtract the background emission from each on-source observation. To allow 
for redundancy in the data, each source was observed multiple times (cycles) at two positions 
(the nods) along each slit at 1/3 and 2/3 of its length, both with the requested integration time.

In Table 1 the journal of the observations is presented, which gives for each source its 
equatorial coordinates, the Seyfert type, the redshift, the date of the observation and 
the total on-source integration times in the SH and LH slits.  As it can be seen from the table, 
for every observation, we have obtained multiple cycles so that each of the two nod 
positions of each spectrograph (SH - LH), both on and off-source position, are sampled 
multiple times. This not only increases the signal to noise ratio but is also essential to 
correct for glitches in the pixel response, mainly due to cosmic rays hits.

The data were processed by the {\em Spitzer} Science Center (SSC) pipeline version 15. 
The three-dimensional data cubes were converted to two-dimensional slope images after 
linearization correction, subtraction of darks, and cosmic-ray removal. To reduce the 
data, we used the software package Spectroscopy Modeling Analysis and Reduction 
Tool (SMART) \citep{hig04}, specifically designed for the reduction of IRS spectra. 
The starting files are the basic calibrated data (BCD), which are provided by the SSC. 
The first step in the data reduction consisted in removing the bad-responsing pixels by 
cleaning the BCD images with the rogue pixel masks of the IRS campaign relative to 
each observation.

Off-source images were simply averaged to obtain a reliable background image at each 
nod position of each spectrograph. The extraction of the spectra from the two dimensional 
cleaned BCD images was done as "full aperture extraction" (see the SMART manual), 
using the off-source averaged spectra as "sky images" . The resulting spectra of each nod
position of each spectrograph were compared to remove the border effects of each order 
and the glitches.  As many cleaned spectra as cycles were obtained for each nod position. 
Finally, the cleaned spectra from the same nod were averaged, using weighted means with 
2.5$\sigma$ clipping, to get one final single cleaned spectrum.

We note that three galaxies in Table 1, which were classified as Seyfert galaxies 
at the time of the 12$\mu$m sample selection \citep{rms93}, have been reclassified 
\citep{tra03} as LINERs or startburst galaxies: these are: MCG-03-34-63, NGC7496, 
NGC7590. The final Seyfert list of this study contains therefore 13 type 1's (including 
both galaxies of the pair of MRK1034) and 14 Type 2's. We therefore
use a different symbol for the 3 non-Seyfert galaxies in the 
plots and exclude them from any analysis comparing type 1's with type 2's. 

\section{Results}

The only emission feature which is spectroscopically resolved is the 11.25$\mu$m PAH blend, 
while the fine structure and H$_2$ emission lines are not resolved with IRS.
The line fluxes have been measured using Gaussian fits, done within the SMART 
software package. In Table 2 we report the line fluxes, together with their 1$\sigma$ 
statistical errors, of the common ionic fine structure lines, while we list in Table A1, 
in the Appendix, the line fluxes of weak lines that have been detected only on a few 
galaxies. In Table 3 we list the fluxes and 1$\sigma$ errors of the pure rotational 
transitions of H$_2$ and the PAH integrated fluxes, measured with the moment fit, 
as well as the PAH equivalent widths. When the signal to noise ratio 
was less than 3, we have quoted in the tables 3$\sigma$ upper limits. 

The PAH emission feature is detected in all sources except for  
IRAS00521-7054, where the red part of the silicate 
absorption feature could hide the PAH emission. Figures 1 show the measured 
high resolution IRS spectra. 

\subsection{Source extendness}

Although the continua have flux densities generally rising with wavelength, the larger aperture for 
the LH spectra results in some
cases in a substantially higher continuum than measured at the red end of the SH spectrum,
especially in IRAS00198-7926, NGC424 (with major and minor diameters: 1.8'$\times$0.8'), 
NGC513 (0.7'$\times$0.3'), 
NGC931 (3.9'$\times$0.8'), NGC1125 (1.8'$\times$0.9'), NGC4501 (6.9'$\times$3.7'), 
NGC4748, MKN897 (0.6'$\times$0.6') and NGC7590 (2.7'$\times$1.0'). This is due to the difference 
in the two entrance slits of the spectrometers 
(4".7$\times$11".3 and 11".1$\times$22".3 for SH and LH, respectively). 
This increase in area covered by a factor of 4 has only 
minor effects when the source is more compact.  In particular, the nonstellar
AGN continuum is probably unresolved by even the smaller aperture \citep{em86, emr87}. 

Because our spectra have been carefully background-subtracted, and there is no 
contamination from zodiacal emission in our slits,  by measuring the flux values at 
the longest wavelength of the SH spectrograph and at the shortest one of the LH, 
we can use the ratio of these two fluxes as a measure of how much each of our 
galaxies deviates from an unresolved point source.  We call the ratio between 
the larger (LH) to the smaller (SH) aperture fluxes \textit{extendness factor},
exactly the inverse of the "compactness" ratio defined by \citet{em86, emr87}, 
which is the ratio of small beam to large-beam 12$\mu$m photometry. 
We have reported in table 4 the photometric fluxes measured at 19$\mu$m in SH 
(using a bandwidth of about 1 $\mu$m) and the fluxes 
measured at 19.5$\mu$m in LH (in a similar bandwidth), excluding the [SIII]18.71$\mu$m 
line and the additional noise frequently 
present at the border of the orders. The error given is the standard deviation of the linear fit 
we performed. 
We divided our sources in three classes: I) the compact sources with R$<$ 1.15 (10 galaxies); 
II) moderately extended sources with  
1.15$<$R$<$1.5 (14 galaxies); and III) very extended sources with R$>$ 1.5 (6 sources). 
We allow for compact sources an increase of 10\% in flux from the SH and the LH waveband, 
because all the spectra of our targets are moderately rising with wavelength. 

\subsection{H$_2$ excitation diagrams: H$_2$ temperatures and masses}

From the measured H$_2$ rotational line fluxes, we derived in Figures 2 the excitation 
diagrams, using the Boltzmann equation, which relates the level population to the temperature,
following the method described by \citet{rig02}. In Table 5 we report the 
H$_2$ temperatures, as derived from the excitation diagrams of Fig. 2. We estimated 
the temperatures, or their upper or lower limits, from the slopes obtained connecting two 
adjacent measurements in the excitation diagrams, namely the T(0-1) from the 
S(0)-S(1) fit, the  T(1-2) from the S(1)-S(2) and the T(2-3) from the S(2)-S(3). 

From the H$_2$ line column densities, following \citet{rig02}, we then estimated 
the warm H$_2$ masses, considering the excitation temperature as derived from 
the S(0)-S(1) fit and the S(1) line flux. The estimated mass is related to the IRS 
aperture and it assumes that the S(1) emission fills the beam. If the galaxy is more 
extended the masses derived have to be considered as lower limits, on the contrary, 
if the H$_2$ emission is more compact than the IRS SH beam, the derived numbers 
have to be considered upper limits. 

For eight galaxies we can derive an average mean temperature from the temperature 
upper and lower limits given in Table 5: MRK1034 NED02 (T $\sim$ 188K), MRK6 
(T $\sim$ 246K), NGC3516 (T $\sim$ 263K), NGC4501 (T $\sim$ 300K), NGC4748 
and NGC4968 (T $\sim$ 250K), MCG-03-34-63 ( T $\sim$ 217K), NGC7496 (T $\sim$ 240K). 
For NGC513, the pair of detected lines infer a temperature of 191K. Five galaxies show 
a much flatter slope between the S(2) to S(3) lines or S(1) to S(2) than between the S(0) to 
S(1) lines: these are ESO545-G013, MRK9, NGC4602, IRASF22017+0319 and NGC7590: 
this reflects distinct temperature components in the gas: a warmer component of about or 
more than 400K and a cooler one of about or less than 300K. For the remaining 15 galaxies 
only a broad range in temperature, as wide as 100-150K or more, can be given.

There is no obvious correlation of the excitation temperatures of H$_2$ with Seyfert type
or any other properties in Table 2.

The estimated H$_2$ masses are significantly smaller than those found by the same procedure
in ultra-luminous infrared galaxies \citep{hig06}, but are comparable to those observed
in more normal spiral galaxies \citep{rou07}.  Thus there is no indication of an
unusually large reservoir of molecular gas in the centers of our Seyfert galaxies.

\section{AGN Diagnostic diagrams}

\subsection{Line ratios versus PAH equivalent widths}

Ratios of emission lines of different ionization potential for the same element have been used 
to determine the hardness of the underlying radiation field in galaxies \citep[for example see][]{sm92, ver03}. 
This allows us to quantify how much IR emission arises from O and B stars, and how
much from an AGN, which has much stronger EUV and X-ray emission.
The mid-IR lines of neon [NeII]12.81$\mu$m, [NeIII]15.56$\mu$m, [NeV]14.32$\mu$m, with 
ionization potentials of 21.6eV, 41.0eV and 97.1eV, are among the prime candidates for tracing 
emissing from AGN \citep[e.g.][]{stu02}. Their proximity in wavelength suggests that effects 
of differential extinction are minimized. In addition the high ionization potential of [NeV] 
ensures that it has been excited by photons originating from the accretion disk of an AGN. 
Since the [NeV] line is rather faint, in particular for the ISO SWS spectrometers the 
[OIV]25.91$\mu$m line with an ionization potential of 54.9eV has also been 
used \citep[e.g.][]{ge98, arm07,far07}. We note though that [OIV]25.91$\mu$m 
is not as clear a diagnostic since it could also be excited by shocks, or extremely metal-poor
star clusters \citep{lut02}. 
 
With ISO spectra of ULIRGS, \citet{ge98} used the inverse correlation of [NeV]/[NeII] and [OIV]/[NeII] with equivalent width
of the 6.2$\mu$m PAH emission to infer the proportion of IR emission produced by the central AGN.
Recently, \citet{des07} have shown that the EW of the 6.2$\mu$m PAH emission is systematically weaker in the most luminous ULIRGs, particularly those with warm IRAS colors.  The PAH EW can be small even if their optical spectra show no hint of an AGN. They suggest that these galaxies may host strong star formation as well as obscured AGN activity.
Instead of the PAH 6.2$\mu$m emission feature, commonly used in low resolution shorter 
wavelength spectra, our data allowed an accurate measurement of the 11.25$\mu$m PAH 
emission feature, which is $\rm$0.14 \% of the L$_{\rm IR}$ \citep{soi02}. 
We then use the [NeV]14.3$\mu$m/[NeII]12.8$\mu$m line ratio, the [OIV]14.3$\mu$m/[NeII]12.8$\mu$m line ratio, the  [NeV]14.3$\mu$m/[SiII]34.8$\mu$m line ratio and the [NeIII]15.5$\mu$m/[NeII]12.8$\mu$m 
line ratio as a function of the PAH 11.25$\mu$m equivalent width in the diagrams of 
Fig. 3 (\textit{a}, \textit{b}) and Fig. 4 (\textit{a}, \textit{b}), respectively. 

Throughout this paper, we have looked at the correlations using the following method: first we applied to the data
a weighted least squares (hereafter w.l.s.) fit, to derive a slope for each data set, considering both the Seyfert 1's and 2's separately and all the Seyfert's together. To estimate the significance of the correlation found we have then applied a bootstrap method (with a 1000 times resampling). From each bootstrapped sample we have computed the slope with a w.l.s. fit and then used the mean value of these slopes and the standard deviation. We considered valid a correlation if the resulting slopes from the two methods, i.e. simple w.l.s. fit and bootstrapped w.l.s. fit,  were consistent within a threshold of $\pm$ 0.25. For each fit where a correlation is found, we used the mean value of the slope, as derived from the bootstrap method, and its standard deviation to give a confidence interval.

For all the Seyfert galaxies, but especially for Seyfert 2's, it appears from Fig. 3 and Fig. 4 
that there is a clear trend  between  the [NeV]14.3$\mu$m/[NeII]12.8$\mu$m, [NeV]14.3$\mu$m/[SiII]34.8$\mu$m 
and [NeIII]15.5$\mu$m/[NeII]12.8$\mu$m line ratios and the PAH equivalent width. The ratio of the high-ionization 
lines, excited by the AGN, to the low-ionization line excited in HII regions ([NeII]12.8$\mu$m) and also to  
[SiII]34.8$\mu$m, increases by about two orders of magnitude as the PAH equivalent width decreases. 
However we do not find any significant correlation. 

Because of the poor statistics, we cannot exclude that such correlations could appear with a larger sample of objects, 
and we postpone to the next articles, which will present new spectra, any firm conclusion.

\subsection{Line Equivalent Width Diagnostics}

As with the ratios of high- to low-ionization lines, the equivalent width can also function in the same way 
to measure AGN dominance. The ratio of an H II-region dominated line to the underlying continuum 
dominated by the AGN should decrease systematically as the AGN produces an increasing proportion of
the IR luminosity.
In Fig. 5
the [NeII]12.8$\mu$m  equivalent width as a fuction of the  PAH equivalent width is presented. 
The equivalent width of the PAH shows an increasing trend with the [NeII] equivalent width, however no
correlation is present.

\subsubsection{Line equivalent widths versus extendness factor}

\citet{em86} and \citet{emr87} first considered the extendness of the CfA 
Seyfert galaxies as a useful AGN measure, using the ratio of 10$\mu$m ground-based small-aperture flux 
to the IRAS 12$\mu$m flux.
We have correlated the extendness factor at 19$\mu$m defined in \S 4.1 and listed in Table 4 with the 
equivalent widths of the key lines measured in our spectra (listed in Table A2 in the Appendix) 
and the PAH feature. 
We present the results in figures 6-8. 
In particular we show in Fig 6 (\textit{a} 
and \textit{b}, respectively) the EW of the 11.25$\mu$m emission feature and that 
of the [NeII]12.8$\mu$m line against the extendness factor. We can see that the
trend in these two diagrams is the same: increasing the size of the 19$\mu$m 
emitting region, both the PAH and the [NeII] equivalent widths increase with a very 
similar relationship. This confirms that the amount of star formation, given by the 
integrated number of HII regions due to young stars, which is measured by the [NeII] 
line, or by the strength of the PAH emission over the continuum nicely correlates 
with the source extendness. For a compact source both emissions are minimized, 
because of the dominance of the point-like AGN.

In Fig. 7 the EW of the [NeIII]15.5$\mu$m line against the extendness factor is presented. 
This diagram does not show a strong correlation as the previous ones, because 
the [NeIII] emission lines is not simply related to the amount of star formation/young stars, 
but it includes contributions from the AGN: being  
photoionized by the AGN radiation field in the NLR. 

In Fig. 8 (\textit{a} and \textit{b}, respectively) the [NeV]14.3$\mu$m  and [OIV]25.9$\mu$m 
line equivalent widths as a function of the extendness parameter are presented. There is no 
correlation left because, as we expect, the emission in these lines is mostly due to the AGN, 
just like most of the underlying continuum. With a significant scatter, the average equivalent widths
of the [NeV] and [OIV] lines of 0.02 and 0.07$\mu$m represent ``pure" AGN emission.
 
\subsubsection{Line ratios as a function of luminosity and spectral index}

One of the reasons why the 12$\mu$m band has been used to select a complete sample 
of Seyfert galaxes in the local universe was because this band is carrying a constant fraction 
of the bolometric luminosity of Seyfert galaxies \citep{sm89}. We consider therefore that 
for our sample the 12$\mu$m luminosity is a good approximation of the total luminosity 
and use this to see correlations with the power of the AGN, as indicated by the line ratio 
of [NeV]/[NeII] and to a lesser extent by the ratio [NeIII]/[NeII]. We show the 
[NeV]14.3$\mu$m/[NeII]12.8$\mu$m  line ratio versus 12$\mu$m luminosity in Fig. 9.  
Applying our statistical method, no correlation is found.

Another method that has been used with the large-beam wideband photometry (e.g. IRAS), 
followed the discovery that local Seyfert galaxies, and AGNs in general at low 
redshift, have a "warm" mid-to-far-IR spectral index \citep[see, e.g.][for the 12$\mu$m 
selected sample]{rms93}. 
We show in Fig. 10 (\textit{a} and \textit{b}, respectively) how the 60-25$\mu$m spectral 
index flattens as the two ionization-sensitive line ratios  [NeV]/[NeII] and [NeIII]/[NeII]
increase. 
The higher these line ratios, the flatter --or warmer-- is the spectral index. Because a warmer spectral 
index is produced by an AGN continuum which is getting 
brighter than the cold dust component of a spiral galaxy, we can interpret these 
plots saying again that the dominance of the AGN over the total power of a galaxy is 
well measured by its ionization indicators.

Using our statistical tests, we have found that the [NeV]/[NeII] ratio correlates with the spectral 
index for all Seyfert galaxies (see the caption of Fig. 10 for the numerical results), while the 
[NeIII]/[NeII] ratio shows an increasing trend with the spectral index but not a statistical correlation.

One advantage of the photometric indicators of AGN strength is that since they
do not require spectroscopy, they will be much easier  
to measure for faint galaxies at high redshifts.

\subsubsection{Neon line ratios versus the [SIII] to [SiII] ratio}

In their study of the SINGS Spitzer legacy program, \citet{dal06} found 
that the [SIII]33.5$\mu$m/[SiII]34.8$\mu$m line ratio can be used to distinguish AGNs 
from starbursts. We therefore compared this ratio against the [NeV]14.3$\mu$m/[NeII]12.8$\mu$m 
line ratio and against the  [NeIII]15.5$\mu$m/[NeII]12.8$\mu$m line ratio, to look for 
differences between the two Seyfert populations. We plotted in Fig. 11 (\textit{a}, \textit{b}) 
the two line ratio diagrams.
Although of low statistical significance, it appears that the [SIII] to [SiII] ratio 
is somehow separating Seyfert 1's from Seyfert 2's, in the sense that most of the former (8/12) 
have [SIII]33.5$\mu$m/[SiII]34.8$\mu$m $<$ 0.6, while most of the Seyfert 2's (8/10) have 
this ratio $>$ 0.6. This threshold is also the one shown in the \citet{dal06} published diagram. 
However, the lack of correlation in Fig. 6 shows that the [SIII]/[SiII] ratio is
less effective at picking out the strength of the AGN emission in detail. 
This is because it only spans a range of
less than one order of magnitude, while the Ne line ratios span nearly two orders
of magnitude.
 
\section{Comparison of Seyfert 1's and 2's}

The unification hypothesis states that Seyfert 1 and Seyfert 2 galaxies are intrinsically the same, 
but appear different at some wavelengths because of the much larger extinction along our 
lines-of-sight to the active nuclei in Seyfert 2's. At some infrared wavelength, the AGN 
emission becomes more isotropic, so that Seyfert 1 and Seyfert 2 of the same intrinsic 
luminosity should appear identical. Thus a fair test of the unification hypothesis requires matching 
Seyfert 1's and 2's of comparable bolometric luminosity, and this is one reason we are analyzing the 
12$\mu$m sample, which is relatively unbiased, i.e. equally sensitive to Seyfert 1s or Seyfert 2s having the 
same bolometric fluxes.

The equivalent widths of the PAH 11.25$\mu$m or [NeII]12.8$\mu$m  emission features --- 
which are unrelated to the AGN --- are normalized by the underlying nuclear continuum at 12$\mu$m, 
which is predominantly coming from the active nucleus.  This is why these EWs are anticorrelated 
with the relative dominance of the AGN component. As expected, the EW of high-ionization lines 
such as  [NeV] or [NeIII] is not correlated with AGN dominance because both those emission lines, 
and the underlying nuclear continuum, is dominated by the AGN.  
The AGN itself produces [NeV] and [NeIII] that are about 0.03\% of its bolometric 
power; since the 12$\mu$m  continuum is 10\% of the bolometric power for Seyfert galaxies 
this results in the "pure AGN" EWs of these lines of 0.3 $\mu$m.

By normalizing the 12$\mu$m continuum to other observables which we believe are emitted 
roughly isotropically from AGN, we can compare directly the mid-IR continuum emitted from 
Seyfert 1's and Seyfert 2's.
Although it has long been known that Seyfert 2 nuclei are relatively weaker in their 12$\mu$m continuum
\citep[e.g.][]{em86, emr87} the difference with Seyfert 1's is not large in our sample. 

We observe that the average equivalent width (EQW) of the 11.25$\mu$m PAH for 
Seyfert 1's is 0.38$\pm$0.39 $\mu$m while for Seyfert 2's it is 0.44$\pm$0.39 $\mu$m. 
These two values differ by less than 1$\sigma$. 
This can be understood if some obscuration is still present in the mid-IR in Seyfert 2's.
One popular possibility is that it is higher since the plane of the AGN torus is along our 
line of sight. 
But the effect must be quite weak. 
The EW's of the PAH at 11.25$\mu$m and that of the [NeII]12.8$\mu$m are about  20\% larger 
in the Seyfert 2's, while the AGN-dominated [Ne III] and [NeV]
EWs are also about 20\% larger in Seyfert 1's. Assuming the basic tenet of unification--that 
Seyfert 1's
and Seyfert 2's are intrinsically the same, this implies that only 20\% of the 12$\mu$m 
continuum from Seyfert 2's has
been absorbed or blocked by geometric effects. This relatively modest implied optical 
depth to the 12$\mu$m  continuum of Seyfert 2's is far lower than the values required 
for "standard" thick dusty torus models \citep{pikro92, pikro93}.
However, recent models have allowed for the more likely possibility that the 
dust distribution is clumpy, so that some sight-lines to Seyfert 2's could
intercept relatively little absorption \citep{eli06}.
This, combined with the likelihood of mid-IR continuum emission from 
optically thin dust above the torus in both Sy 1's and 2's, means that 
our data may not rule out these more sophisticated unification models.

\subsection{Host Galaxy ISM}

Photometric mid-IR studies of \citet{emr87} \citet{maio98} indicated that Seyfert 2 
galaxies more often have enhanced star formation than Seyfert 1's. This finding has 
also been supported by the low resolution IRS spectroscopy of \citet{buc06}, who 
conclude that Seyfert 2 galaxies typically show stronger starburst contributions in 
their IR spectra than the Seyfert 1 galaxies. This result, if confirmed in a sample
which is not compromised by selection effects, 
would be in disagreement with the predictions of the unified scheme \citep{ant93}.  

The PAH and H$_2$ emission fluxes and luminosities, 
according to \citet{rig02}, can discriminate between starburst and AGN.
In Fig.12, we show the diagram with fluxes and luminosities (\textit{a} and \textit{b}, respectively). 
The emission from small dust grains and warm molecules appears to be only correlated 
when the luminosities are taken, and not using the fluxes. Considering separately the Seyfert 1's and 2's, 
we find similar correlations, therefore there is no contradiction with the unification idea.

We notice, however, that the correlation between luminosities, that does not appear between the fluxes, 
could be induced by the distance factor and therefore it may not be real. 

\subsection{Line ratios as density indicators}

\citet{dud07} performed the first systematic study of the [NeV](14.3$\mu$m/24.3$\mu$m) 
and [SIII](18.7$\mu$m/33.5$\mu$m) mid-infrared fine structure doublets of a heterogeneous 
sample of Seyferts (type 1's and 2's), LINERs and quasars, from the Spitzer archive. They 
found that the [NeV] line ratio is often below the theoretical low-density limit, using a model 
that ignored absorption and stimulated emission. This appeared particularly true for the type 2 AGNs.  
They claim that the ratios below the low-density limit are due to differential infrared extinction to 
the [NeV] emitting region caused by the obscuring torus, and that this ratio can trace the inclination 
angle of the torus to our line of sight. Because the amount of extinction, the temperature of the 
emitting gas and the  effects of absorption and stimulated emission are all unknown, they conclude 
that the [NeV] ratio cannot be used to infer the gas density. Similarly, they caution that the 
[SIII] line ratio is also useless because mainly originated from the star forming regions of the host 
galaxies and the different slit aperture sizes used to observe the two lines prevent meaningful 
conclusions about the gas density in nearby galaxies. 

Even taking into account the \textit{caveats} reported by \citet{dud07}, we nevertheless plotted in 
Fig. 13  these two line ratios, which are in principle sensitive to the gas electron densities. 
The dashed lines show the low density limits for both ratios and the solid lines indicate the values 
of the density, as derived from the models of \citet{dud07} assuming a temperature of 10,000K. 
Most of the sample galaxies lie close to the low density limits.  This supports 
a thin spatially extended "Coronal Line Region", rather than a very small "Intermediate Line Region" 
\citep{sm92}. The densities derived from the [NeV] ratio are of the order of 10$^3$ cm$^{-3}$. 
It appears that type 2 Seyfert's have {\it higher} densities of the highly ionized gas compared to type 1's. 
Half of the Seyfert 2's have n$_{e}$ $\geq$ 10$^{3.5}$ cm$^{-3}$, while all Seyfert 1's have  
densities  lower than this value. The densities derived from the [SIII] ratio are slightly lower, 
clustering around n$_{e}$ $\sim$ 10$^{2.6}$ cm$^{-3}$.

We do not find the same results as reported from \citet{dud07}: only two Seyfert 2 galaxies 
(NGC1125 and NGC3660) are below the low density limit derived for the [NeV] line ratio, 
and the same number of Seyfert 1's  (ESO012-G021 and NGC4748, that becomes four 
including the objects having upper limits, namely MRK817 and NGC4602). We notice, however, 
a greater number of galaxies falling below the [SIII] line ratio density limit. While the [NeV] 
line ratio cannot be affected by aperture effects, because the [NeV] emission surely originates 
from the AGN, we cannot rule out the presence of these effects for the [SIII] line ratio, which is 
excited also in HII regions. About this latter, however, we can exclude such effects in 11 galaxies 
(about 1/3 of the objects), for which we have the [SIII]18.71$\mu$m line measured in the LH slit 
(see Table 2). For the other objects with a low [SIII](18.7$\mu$m/33.5$\mu$m) ratio, we have 
used the extendness factors (Table 4) to correct the [SIII]17.81 $\mu$m line flux, assuming 
that increasing the aperture will increase by the same proportion continuum and line fluxes. 
Because the extendness factors were defined by measuring the continuum around the wavelength 
of 19 $\mu$m, we basically assume that the equivalent width of the [SIII] line does not change 
with the aperture. We show in the diagram of Fig.13, both the measured 
and the aperture-corrected [SIII] line ratios directly above them.
By applying this simple correction, we find that only 4 galaxies (IRAS00198-7926 
and NGC513, which have been measured with the LH slit, and ESO141-G055 and NGC6890) 
have a [SIII] ratio below low density limit. 

 \section{Conclusions}

We have identified a family of IRS observables which quantify the proportion of the 
total IR emission coming from a Seyfert nucleus, all of which are intercorrelated with 
each other. The ratios of ionic fine structure lines [NeV]/[NeII] and [OIV]/[NeII] 
were already proposed to measure the importance of the AGN component. We also 
see that [NeIII]/[NeII] and OIV or [NeV]/[SiII]35$\mu$m can be used
to quantify the AGN dominance. 

It was also known from ISO that the equivalent width of the 6.2$\mu$m (and 7.7$\mu$m) 
PAH features is inversely related to the AGN dominance; we find that the same holds for the 
equivalent width of the 11.25$\mu$m PAH feature. We also discovered two additional IRS 
observables: the equivalent width of [NeII]12.8$\mu$m and the 
extendness of the 19$\mu$m continuum, which also quantify the dominance of the AGN 
component compared with the emission from the underlying spiral galaxy. All of these 
observables are correlated with each other, since they are measuring this same astrophysical 
quantity, and they are all correlated with the hardness of the far-IR continuum, since a more 
dominant AGN component is already known to be correlated with hotter dust.

There is no clear indication that recent star formation is much more important on average 
in the Seyfert 2's in our sample, compared with the Seyfert 1's.
Although the Seyfert 1's generally tend to have more dominant AGN than the Seyfert 2's, there 
is a strong overlap between these two classes.  The relatively small difference between the averages 
of these observables for Seyfert 1's and Seyfert 2's indicates that those two AGN categories are not 
extremely different in the mid-IR range. Thus for example, the AGN in Seyfert 2's are clearly 
observable in their 10$\mu$m continuum emission, and in [NeIII], [OIV] and [NeV], to almost 
the same degree as in Seyfert 1's, which is not consistent with some proposed torus models for 
Seyfert 1/2 unification.

Finally, we do not fully confirm the observational claims of \citet{dud07} : 75 - 80\%  of our 
Seyfert 1's and Seyfert 2's show [SIII] or [NeV] densities larger than the low-density limit. In fact, 
after applying plausible aperture corrections to the [SIII] line ratio, only three Seyfert 2's and one 
Seyfert 1 have a [SIII] line ratio below the density limit. A similar result is also found for the 
[NeV] ratio, for which we have two Seyfert 1's (that became four including the upper limits) and 
2 Seyfert 2's below the low density limit. These few cases do not in our view require enormous 
dust extinction values.

\acknowledgments

This work is based on observations made with the Spitzer Space Telescope 
which is operated by the Jet Propulsion Laboratory and Caltech under a 
contract with NASA. HAS acknowledges support by NASA grant NAG5-10659. 
ST acknowledges support by ASI.
We thank Giovanni Fazio and members of the IRAC team for contributing 
Guaranteed Time to obtaining these data. We thank Mrs. Erina Pizzi of IFSI-INAF 
for the preparation of the postscript figures of the article. Finally we wish to thank 
the anonymous referee for the very useful comments and suggestions.


\appendix

\section{Additional Observational Measurements}

\subsection{Weak ionic fine structure lines}

In the IRS high resolution spectra presented, besides the common lines reported in Table 2 and 3, 
we have also measured a few  fine structure lines in the spectra of a few galaxies. These include 
the [ArII]13.10$\mu$m and the [MgV]13.52$\mu$m lines, as well as five [FeII] lines and the 
[FeIII]22.93$\mu$m line. Among the [FeII] lines, we notice that the line at 25.99$\mu$m can 
be blended with the  much brighter 25.89$\mu$m [OIV] line, because the spectral resolution is 
only about half of the lines separation. Except in the few cases where the lines were well separated, 
we did not attempt to deblend them, but we only gave a measurement of the [OIV] line. We refer to 
a future paper, where we will compare the data of our galaxies with photoionization models, for more 
accurate measurements of the [FeII] 25.99$\mu$m line. We report in table A1 the fluxes, derived 
from gaussian fits, of the weak lines detected.

\subsection{Equivalent widths of the key fine structure lines}

We report in Table A2 the equivalent widths (in $\mu$m) of the brightest lines, which have been 
used in the diagrams and correlations presented in the main text.

\clearpage


\begin{turnpage}
\begin{deluxetable}{lcccccclcc}
\tabletypesize{\scriptsize} 
\tablecaption{Journal of
Spitzer IRS observations} \label{tbl-1} \tablewidth{0pt}
\tablehead{ & \colhead{R.A.} & \colhead{Dec.}  &  & &  &  &  & {SH } & {LH}  \\
\colhead{NAME}& (J2000.0) & (J2000.0)    & \colhead{type}  & \colhead{z}  & \colhead{F$_{12\mu m}$}  &  \colhead{F$_{25\mu m}$} & \colhead{Obs. Date}  &  Int.time  & Int.time \\
& h~m~s    &  $\deg$ ~'~" &  &   &   (Jy)    &   (Jy)      &    &  (sec.)  & (sec.)   \\
}
\startdata
IRAS00198-7926  &  00:21:57.0   & -79:10:14 &  Sy2  & 0.072800  &  0.28  &   1.15  & 2006/06/28  &   4$\times$30 & 2$\times$60\\
ESO012-G021     &  00:40:47.8   & -79:14:27 &  Sy1  & 0.032840  &  0.17  &   0.25  & 2006/06/28  &   4$\times$30 & 2$\times$60\\
IRAS00521-7054  &  00:53:56.2   & -70:38:03 &  Sy2  & 0.068900  &  0.28  &   0.80  & 2006/06/29  &   4$\times$30 & 2$\times$60\\
ESO541-IG012    &  01:02:17.5   & -19:40:09 &  Sy2  & 0.056552  &  0.20  &   0.36  & 2006/08/03  &   4$\times$30 & 2$\times$60\\
NGC0424         &  01:11:27.5   & -38:05:01 &  Sy2  & 0.011764  &  1.10  &   1.73  & 2006/06/28  &   2$\times$30 & 4$\times$14\\
NGC0513         &  01:24:26.8   & +33:47:58 &  Sy2  & 0.019544  &  0.17  &   0.28  & 2006/09/17  &   2$\times$30 & 4$\times$60\\
MRK1034 NED01   &  02:23:18.9   & +32:11:18 &  Sy1  & 0.033633  &  0.25  &   0.69  & 2006/09/19  &   2$\times$30 & 4$\times$14\\
MRK1034 NED02   &  02:23:22.0   & +32:11:49 &  Sy1  & 0.033710  &\nodata & \nodata & 2006/09/19  &   2$\times$30 & 4$\times$14\\
ESO545-G013     &  02:24:40.2   & -19:08:27 &  Sy1  & 0.033730  &  0.17  &   0.35  & 2006/09/08  &   4$\times$30 & 2$\times$60\\
NGC0931         &  02:28:14.5   & +31:18:42 &  Sy1  & 0.016652  &  0.61  &   1.31  & 2006/09/19  &   2$\times$30 & 4$\times$14\\
NGC1125         &  02:51:40.4   & -16:39:02 &  Sy2  & 0.010931  &  0.17  &   0.83  & 2006/09/08  &   4$\times$30 & 2$\times$60\\
ESO033-G002     &  04:55:59.6   & -75:32:27 &  Sy2  & 0.018100  &  0.20  &   0.45  & 2006/07/27  &   4$\times$30 & 2$\times$60\\
MRK0006         &  06:52:12.2   & +74:25:37 &  Sy1  & 0.018813  &  0.22  &   0.69  & 2006/11/16  &   4$\times$30 & 2$\times$60\\
MRK0009         &  07:36:57.0   & +58:46:13 &  Sy1  & 0.039874  &  0.21  &   0.44  & 2006/11/16  &   4$\times$30 & 2$\times$60\\
NGC3516         &  11:06:47.5   & +72:34:07 &  Sy1  & 0.008836  &  0.43  &   0.89  & 2006/11/16  &   2$\times$30 & 4$\times$14\\
NGC3660         &  11:23:32.2   & -08:39:30 &  Sy2  & 0.012285  &  0.19  &   0.22  & 2006/06/27  &   2$\times$30 & 4$\times$14\\
NGC4501         &  12:31:59.0   & +14:25:10 &  Sy2  & 0.007609  &  1.02  &   1.28  & 2006/06/27  &   2$\times$30 & 4$\times$14\\
NGC4602         &  12:40:36.5   & -05:07:55 &  Sy1  & 0.008469  &  0.54  &   0.57  & 2006/06/27  &   2$\times$30 & 4$\times$14\\
TOLOLO1238-364  &  12:40:52.9   & -36:45:22 &  Sy2  & 0.010924  &  0.64  &   2.26  & 2006/07/03  &   2$\times$30 & 4$\times$14\\
NGC4748         &  12:52:12.4   & -13:24:54 &  Sy1  & 0.014630  &  0.17  &   0.37  & 2006/07/01  &   4$\times$30 & 2$\times$60\\
NGC4968         &  13:07:06.0   & -23:40:43 &  Sy2  & 0.009863  &  0.39  &   1.05  & 2006/06/29  &   2$\times$30 & 4$\times$14\\
MCG-03-34-063   &  13:22:19.0   & -16:42:30 &	    & 0.021328  &  0.95  &   2.88  & 2006/06/29  &   2$\times$30 & 4$\times$14\\
MRK0817         &  14:36:22.1   &  58:47:39 &  Sy1  & 0.031455  &  0.33 &   1.18  & 2007/03/20  &   4$\times$30 & 2$\times$60\\
IRASF15091-2107 &  15:11:59.8   & -21:19:02 &  Sy1  & 0.044607  &  0.23  &   0.50  & 2007/03/08  &   4$\times$30 & 2$\times$60\\
ESO141-G055     &  19:21:14.1   & -58:40:13 &  Sy1  & 0.036000  &  0.24  &   0.35  & 2007/04/30  &   4$\times$30 & 2$\times$60\\
NGC6890         &  20:18:18.1   & -44:48:23 &  Sy2  & 0.008069  &  0.34  &   0.65  & 2006/10/25  &   4$\times$30 & 2$\times$60\\
MRK0897         &  21:07:45.8   & +03:52:40 &  Sy2  & 0.026340  &  0.37  &   0.86  & 2006/11/16  &   4$\times$30 & 2$\times$60\\
IRASF22017+0319 &  22:04:19.2   & +03:33:50 &  Sy2  & 0.061100  &  0.29  &   0.72  & 2006/11/19  &   4$\times$30 & 2$\times$60\\
NGC7496         &  23:09:47.2   & -43:25:40 &	    & 0.005500  &  0.35  &   1.60  & 2006/06/28  &   2$\times$30 & 4$\times$14\\
NGC7590         &  23:18:55.0   & -42:14:17 &	    & 0.005255  &  0.56  &   0.83  & 2006/06/28  &   2$\times$30 & 4$\times$14\\
\enddata
\end{deluxetable}
\end{turnpage}



\clearpage

\begin{turnpage}
\begin{deluxetable}{lccccccccccc}
\tabletypesize{\scriptsize} 
\tablecaption{Common ionic fine structure lines}\tablewidth{0pt} 
\tablehead{& \multicolumn{5}{c}{Line fluxes (10$^{-14}~erg~s^{-1} cm^{-2}$) in SH} && \multicolumn{5}{c}{Line fluxes (10$^{-14}~erg~s^{-1} cm^{-2}$) in LH}\\
 \cline{2-6}\cline{8-12}\\
\colhead{NAME} &
\colhead{[SIV]} & \colhead{[NeII]} & \colhead{[NeV] } & \colhead{[NeIII]}&
\colhead{[SIII]} && \colhead{[SIII]} & \colhead{[NeV]} & \colhead{[OIV]} & \colhead{[SIII]} &
\colhead{[SiII]}\\
 &(10.51$\mu m$) &(12.81$\mu m$) &(14.32$\mu m$)&(15.56$\mu m$)&(18.71$\mu m$)&&(18.71$\mu m$)&(24.32$\mu m$)&(25.89$\mu m$)& (33.48$\mu m$)&(34.82$\mu m$)}
\startdata
IRAS00198-7926  &  8.10$\pm$0.40   &   6.19$\pm$0.32 &  12.27$\pm$0.28  & 14.03$\pm$0.31 & \nodata	         &&  5.41$\pm$0.33  & 11.38$\pm$0.39 &  33.03$\pm$0.49 & 17.13$\pm$1.59  & \nodata     \\ 
ESO012-G021	   &  2.5$\pm$0.31    &  11.95$\pm$0.22 &   3.19$\pm$0.26  &  6.42$\pm$0.30 & 5.63$\pm$0.45 &&  8.47$\pm$0.16  &  4.61$\pm$0.11 &  15.98$\pm$0.18  & 11.17$\pm$0.55  &  26.8$\pm$0.96 \\ 
IRAS00521-7054  &  1.38$\pm$0.15 &   5.80$\pm$0.25  &   5.78$\pm$0.29  &  8.13$\pm$0.42 & \nodata	         &&   \nodata	     &  2.42$\pm$0.26 &   8.63$\pm$0.37  &  3.75$\pm$0.98    & \nodata     \\ 
ESO541-IG012	   &  2.03$\pm$0.39 &   1.87$\pm$0.24  &   2.21$\pm$0.24  &  2.02$\pm$0.34 & \nodata	         &&  1.65$\pm$0.20  &  1.16$\pm$0.33 &   4.98$\pm$0.3   & \nodata 	            &  4.66$\pm$0.94 \\ 
NGC0424 	           &  8.98$\pm$0.44 &   8.70$\pm$0.38  &  16.10$\pm$0.7	  & 18.45$\pm$0.42 & 6.96$\pm$0.51 &&  \nodata 	     &  6.37$\pm$0.48 &   25.8$\pm$0.51  &  9.82$\pm$1.28   &  8.14$\pm$1.41 \\ 
NGC0513 	           &  2.77$\pm$0.42 &  12.76$\pm$0.21 &   1.91$\pm$0.34  &  4.43$\pm$0.29 & 6.76$\pm$0.52  &&  6.27$\pm$0.16  &  1.09$\pm$0.09 &   6.54$\pm$0.49  & 14.50$\pm$0.37  & 27.49$\pm$0.56 \\   
MRK1034 NED01 &  \nodata	  &  18.68$\pm$0.32 &  $<$ 1.1   	  &  1.47$\pm$0.30 & 8.62$\pm$0.48  &&  \nodata 	      & $<$ 0.59           &  $<$ 0.77            & 10.20$\pm$0.81  & 24.41$\pm$2.00 \\ 
MRK1034 NED02 & 1.89$\pm$0.29 & 34.99$\pm$0.36 &  1.06$\pm$0.35  &  3.61$\pm$0.42 & 9.07$\pm$0.54 && 10.98$\pm$0.48  & $<$ 0.60      &   2.68$\pm$0.37  & 16.0 $\pm$1.14   & 51.71$\pm$2.00 \\ 
ESO545-G013	  &  5.32$\pm$0.22 &  10.05$\pm$0.16  &   4.80$\pm$0.25  & 10.50$\pm$0.30 & 3.59$\pm$0.41  &&  5.87$\pm$0.13  &  3.22$\pm$0.11 &  11.54$\pm$0.15  &  8.31$\pm$0.45  & 20.00$\pm$0.54 \\ 
NGC0931 	          & 10.70$\pm$0.48 &   5.47$\pm$0.43 &  14.30$\pm$0.78  & 15.41$\pm$0.43 & 4.86$\pm$0.61   &&  \nodata 	      & 13.67$\pm$0.43 &  42.60$\pm$0.38  & 11.97$\pm$1.28  & 13.72$\pm$1.70 \\   
NGC1125 	          &  6.07$\pm$0.22 &  16.37$\pm$0.20 &   5.09$\pm$0.26  & 15.55$\pm$0.23 &  10.99$\pm$0.27 &&  \nodata 	      &  9.69$\pm$0.31 &  40.36$\pm$0.53  & 23.94$\pm$0.97  & 31.32$\pm$0.78 \\ 
ESO033-G002	  &  5.54$\pm$0.27 &   2.13$\pm$0.19 &   6.34$\pm$0.16  &  9.22$\pm$0.18 & 3.99$\pm$0.29     &&  \nodata 	      &  5.27$\pm$0.17 &  14.39$\pm$0.18  &  2.79$\pm$0.65   &  4.53$\pm$0.44 \\ 
MRK0006 	         & 16.69$\pm$0.36 &  28.00$\pm$0.23 &   9.39$\pm$0.19  & 49.34$\pm$0.32 & 14.10$\pm$0.25  &&  \nodata 	      & 10.43$\pm$0.21 &  48.24$\pm$0.27  & 14.09$\pm$0.54  & 36.40$\pm$0.71 \\  
MRK0009 	         &  2.37$\pm$0.23 &   3.23$\pm$0.20  &   2.21$\pm$0.22  &  1.90$\pm$0.20 & 2.38$\pm$0.50     &&  3.44$\pm$0.15  &  2.24 $\pm$0.19 &   5.55$\pm$0.21  &  3.94$\pm$0.49  &  7.32$\pm$0.77 \\	 
NGC3516 	         & 13.33$\pm$0.38 &   8.07$\pm$0.25 &   7.88$\pm$0.50  & 17.72$\pm$0.33 & 5.86$\pm$0.35    &&  \nodata 	       & 10.39$\pm$0.33 &  46.92$\pm$0.35  &  9.52$\pm$0.96  & 22.14$\pm$0.54 \\ 
NGC3660 	         &  1.48$\pm$0.22 &   6.51$\pm$0.23 &   0.98$\pm$0.23  &  1.49$\pm$0.18 & 3.66$\pm$0.28     &&  \nodata 	      &  1.66$\pm$0.20 &   3.61$\pm$0.25    &  9.52$\pm$1.14  &  9.54$\pm$1.10 \\ 
NGC4501 	         &  $<$ 3.              &   7.02$\pm$0.27 &   $<$ 1.5  	      & 4.72$\pm$0.29 &   1.97$\pm$0.31     &&  \nodata 	      &  $<$ 3.6            &   4.22$\pm$0.28    &  7.9 $\pm$0.6    & 16.70$\pm$0.87 \\ 
NGC4602 	         &  $<$ 1.2             &   7.57$\pm$0.25 &  0.82$\pm$0.09	& 0.63$\pm$0.19 &  3.20$\pm$0.23    &&  \nodata 	      &  $<$ 1.2            &   $<$ 2.3              &  8.68$\pm$0.48  & 15.30$\pm$0.83 \\ 
TOLOLO1238-364 & 5.70$\pm$0.28 &  45.15$\pm$0.54 & 11.15$\pm$0.61 & 27.00$\pm$0.44 & 16.32$\pm$0.48 &&    \nodata	       &  5.35$\pm$0.67 &  21.21$\pm$0.58  & 32.80$\pm$1.62  & 44.99$\pm$1.62 \\ 
NGC4748         	&  9.87$\pm$0.18    &   7.37$\pm$0.19 &   6.75$\pm$0.19  & 15.93$\pm$0.20 & 7.61$\pm$0.22  && 16.33$\pm$0.47 & 20.00 $\pm$0.17 &  81.93$\pm$0.21  & 20.42$\pm$0.68  & 29.46$\pm$0.61 \\   
NGC4968 	        &  9.63$\pm$0.37   &  24.90$\pm$0.30 &  17.57$\pm$0.57  & 33.80$\pm$0.44 &  15.10$\pm$0.58 &&    \nodata	      & 10.60$\pm$0.55 &  33.70$\pm$0.45  & 14.90$\pm$1.14  & 22.36$\pm$1.15 \\  
MCG-03-34-063  &  $<$ 1.2           &   6.02$\pm$0.27 & $<$ 2.4         &  $<$ 3.        & 3.11$\pm$0.36       &&  \nodata 	      &  $<$ 3.              &   $<$ 3.                &  6.67$\pm$0.65  &  6.71$\pm$1.60 \\ 
MRK0817         & 1.53$\pm$0.38   &  3.83$\pm$0.22   &   1.86$\pm$0.19   &   4.58$\pm$0.41  &   2.76$\pm$0.60 &&  \nodata        &  3.60$\pm$0.44  &   6.53$\pm$0.29    &   $<$ 3.21          & $<$ 3.45      \\  
IRASF15091-2107 & 7.29$\pm$0.22 & 11.52$\pm$0.15 & 8.48$\pm$0.30   &  16.29$\pm$0.29   &  \nodata           &&  9.95$\pm$0.15 & 8.12$\pm$0.15 &  31.03$\pm$0.30  &  12.26$\pm$0.62 & 12.27$\pm$1.49 \\ 
ESO141-G055   & 3.45$\pm$0.22   & 2.24$\pm$0.21	    &2.25$\pm$0.24    &  5.62$\pm$0.27   & 1.75$\pm$0.52 &&    \nodata         &  1.62$\pm$0.12   &  7.26$\pm$0.12	 & 5.45$\pm$0.50   &  8.85$\pm$0.82  \\ 
NGC6890 	        &  2.92$\pm$0.28   &  11.32$\pm$0.32 &   5.77$\pm$0.29  &  6.57 $\pm$0.14 & 4.34$\pm$0.29   &&  \nodata 	       &  3.77$\pm$0.18 &  10.10 $\pm$0.17  & 16.97$\pm$0.72  & 26.54$\pm$0.77 \\ 
MRK0897      	&  $<$ 1.5              &  24.03$\pm$0.19 & 1.06$\pm$0.21	&  4.38 $\pm$0.13 & 14.91$\pm$0.29  && 15.52$\pm$0.24  &  $<$ 0.8           &    0.62$\pm$0.19  & 22.30$\pm$0.70  & 21.44$\pm$0.85 \\	
IRASF22017+0319 & 10.31$\pm$0.25 &   5.95$\pm$0.37 &   8.33$\pm$0.26  & 14.07 $\pm$0.29 & \nodata	          &&  6.10 $\pm$0.25  &  9.40$\pm$0.30 &  29.04$\pm$0.30  &  9.33$\pm$0.44  &   \nodata    \\ 
NGC7496 	       &  1.3 $\pm$0.4     &  48.08$\pm$0.37 & $<$ 1.8 	       & 6.67$\pm$0.35 & 23.48$\pm$0.50     &&  \nodata 	        &  $<$ 2.4           &   $<$ 2.4        & 39.47$\pm$1.17  & 44.58$\pm$2.11 \\ 
NGC7590 	        &  $<$ 3	             &   7.78$\pm$0.20 & $<$ 1.5 	       & 3.49$\pm$0.27 & 5.37$\pm$0.37       &&  \nodata 	        &  $<$ 1.2           &   5.60$\pm$0.25  & 15.69$\pm$0.83  & 26.50$\pm$0.95 \\ 
\enddata
\end{deluxetable}
\end{turnpage}

\clearpage

\begin{deluxetable}{lccccccccc}
\tabletypesize{\scriptsize} 
\tablecaption{Molecular Hydrogen lines and PAH emission feature at 11.25$\mu$m}\tablewidth{0pt} 
\tablehead{& \multicolumn{5}{c}{Line fluxes (10$^{-14}~erg~s^{-1} cm^{-2}$)} & \\
 \cline{2-6}\\
\colhead{NAME} &
\colhead{H$_2$ S(3)} & \colhead{ H$_2$ S(2)} & \colhead{H$_2$ S(1)} & \colhead{H$_2$ S(0)}&
\colhead{PAH} & \colhead{EQ.W.} \\
 &(9.67$\mu m$) &(12.28$\mu m$) &(17.03$\mu m$)&(28.22$\mu m$)&(11.25$\mu
 m$)&($\mu m$)\\
  & (1)       & (2)                & (3)             & (4)               & (5)                &(6)  }
\startdata
IRAS00198-7926 		& 2.31$\pm$0.51  &   4.22$\pm$0.36 &  3.92$\pm$0.29 &	 $<$ 1.47     &    24.7 & -0.117 \\
ESO012-G021		& 2.35$\pm$0.14  &   1.47$\pm$0.13 &  3.93$\pm$0.34 &	 $<$ 0.54     &    65.5 & -0.404 \\
IRAS00521-7054  	& 2.75$\pm$0.30  &   1.78$\pm$0.30 &  2.40$\pm$0.21 &	 $<$ 1.11     & \nodata & \nodata \\
ESO541-IG012    	& $<$ 0.65	 &   $<$ 0.72	   &  1.24$\pm$0.19 &	 $<$ 1.2      &   86.4 & -0.434 \\
NGC0424         	&   \nodata	 &   $<$ 1.14	   &  2.01$\pm$0.40  &	 $<$ 1.53     &   11.0 &  -0.021 \\ 
NGC0513         	&   \nodata	 &   $<$ 8.60	   &  2.67$\pm$0.29 & 1.54$\pm$0.10   &   63.0 &  -0.704 \\
MRK1034 NED01   	& $<$ 1.5	 &   $<$ 0.96	   &  3.50$\pm$0.36 &	 $<$ 2.4      &   68.8 &  -1.054 \\
MRK1034 NED02   	& 3.14$\pm$0.32  &   $<$ 0.72	   &  6.79$\pm$0.56 &	 $<$ 1.11     &  119.0 &  -0.754 \\  
ESO545-G013     	& 1.03$\pm$0.20  &   $<$ 0.48	   &  1.57$\pm$0.36 & 1.01$\pm$0.18   &   20.3 &  -0.494 \\
NGC0931         	&   \nodata	 &   $<$ 1.29	   &  $<$ 1.83      &	 $<$ 1.14     &   24.6 &  -0.064 \\
NGC1125         	&   \nodata	 &   $<$ 0.60	   &  1.28$\pm$0.32 &	 $<$ 1.53     &   49.9 &  -0.655 \\
ESO033-G002     	&   \nodata	 &   1.29$\pm$0.22 &  2.44$\pm$0.25 & 0.82$\pm$0.19   &   12.8 &  -0.080 \\
MRK0006         	&   \nodata	 &   1.10$\pm$0.24 &  4.26$\pm$0.26 & 1.49$\pm$0.29   &   18.1 &  -0.092 \\
MRK0009         	& 1.33$\pm$0.27  & $<$ 0.53   &  1.19$\pm$0.32 &	 $<$ 0.63     &   13.1 &  -0.078 \\
NGC3516         	&   \nodata	 &   1.00$\pm$0.39 &  3.93$\pm$0.41 &	 $<$ 1.05     &   22.1 &  -0.076 \\
NGC3660         	&   \nodata	 &   $<$ 0.69   &  2.13$\pm$0.24 &	 $<$ 0.75     &   21.7 &  -0.635 \\
NGC4501         	&   \nodata	 &   4.35$\pm$0.28 & 10.07$\pm$0.29 & 2.71$\pm$0.37   &   30.2 &  -0.997 \\
NGC4602         	&   \nodata	 &   1.84$\pm$0.19 &  2.44$\pm$0.23 & 2.94$\pm$0.26   &   24.1 &  -1.190 \\
TOLOLO1238-364  	&   \nodata	 &   $<$ 1.62	   &  3.44$\pm$0.46 &	 $<$ 1.74     &   94.3 &  -0.201 \\
NGC4748         	&   \nodata	 &   $<$ 0.57	   &  2.21$\pm$0.27 & 0.71$\pm$0.22   &   18.6 &  -0.615 \\
NGC4968         	&   \nodata	 &   $<$ 1	   &  3.34$\pm$0.39 &	 $<$ 1.35     &   69.3 &  -0.203 \\
MCG -03-34-063  	&   \nodata	 &   $<$ 0.5	   &  2.04$\pm$0.30 &	 $<$ 1.33     &   10.8 &  -1.043 \\
MRK0817                  &   \nodata      &   $<$ 0.6       &  1.55$\pm$0.45 &    $<$ 1.46     &	  15.8 &  -0.242 \\
IRASF15091-2107             & 2.06$\pm$0.24  &   1.40$\pm$0.21 &  4.04$\pm$0.24 &    $<$ 0.84     &	  28.4 &  -0.161 \\
ESO141-055              &   \nodata      &   $<$ 0.6       &  $<$ 1.09      &    $<$ 0.74     &	  16.2 &  -0.084 \\
NGC6890         	&   \nodata	 &   $<$ 0.96	   &  1.31$\pm$0.32 & 1.68$\pm$0.20   &   40.6 &  -0.382 \\
MRK0897         	& $<$ 0.57	 &   1.07$\pm$0.09 &  2.84$\pm$0.24 & 0.98$\pm$0.18   &   78.7 &  -1.279 \\
IRASF22017+0319 	& 1.84$\pm$0.24  &   0.88$\pm$0.25 &  1.90$\pm$0.27 &	 $<$ 0.76     &   24.0 &  -0.072 \\
NGC7496         	&   \nodata	&   $<$ 1.11	   &  4.95$\pm$0.33 &	 $<$ 1.70     &  105.0 &  -0.161 \\
NGC7590         	&   \nodata	&   1.63$\pm$0.25  &  2.64$\pm$0.28 & 2.36$\pm$0.33   &   38.8 &  -0.081 \\
\enddata
\tablenotetext{*}{Columns (1) to (4) give the line fluxes of the H$_2$ rotational lines in  units of 10$^{-14}~erg~s^{-1} cm^{-2}$.
Column (5) and (6) give the flux and the equivalent width of the PAH 11.25 $\mu$m emission feature, respectively.\\ 
}
\end{deluxetable}

\begin{deluxetable}{lcccccc}
\tabletypesize{\scriptsize} 
\tablecaption{Photometric fluxes and extendness factors} \label{tbl-4} \tablewidth{0pt}
\tablehead{\colhead{NAME} &  \colhead{F$_{19\mu m}$(SH)}  &  \colhead{F$_{19.5\mu m}$(LH)} & \colhead{R} & \colhead{Class} \\
                                          & (Jy)                       &       (Jy)                &      &      }
\startdata
IRAS00198-7926     & 0.574$\pm$0.022  & 0.729$\pm$0.015 &   1.27   &  II \\ 
ESO012-G021        & 0.146$\pm$0.016  & 0.174$\pm$0.009 &   1.19   &  II \\ 
IRAS00521-7054     & 0.507$\pm$0.021  & 0.531$\pm$0.010 &   1.05   &  I \\ 
ESO541-IG012       & 0.280$\pm$0.012  & 0.306$\pm$0.008 &   1.09   &  I \\ 
NGC0424            & 1.391$\pm$0.023  & 1.517$\pm$0.023 &   1.09   &  I \\ 
NGC0513            & 0.102$\pm$0.021  & 0.150$\pm$0.005 &   1.47   &  II \\ 
MRK1034 NED01      & 0.066$\pm$0.015  & 0.089$\pm$0.007 &   1.34   &  II \\ 
MRK1034 NED02      & 0.220$\pm$0.020  & 0.310$\pm$0.011 &   1.41   &  II \\ 
ESO545-G013        & 0.107$\pm$0.010  & 0.140$\pm$0.006 &   1.31   &  II \\ 
NGC0931            & 0.664$\pm$0.024  & 0.839$\pm$0.011 &   1.26   &  II \\ 
NGC1125            & 0.235$\pm$0.014  & 0.448$\pm$0.011 &   1.91   &  III \\
ESO033-G002        & 0.356$\pm$0.012  & 0.387$\pm$0.007 &   1.09   &  I \\ 
MRK0006            & 0.576$\pm$0.011  & 0.577$\pm$0.012 &   1.00   &  I \\ 
MRK0009            & 0.322$\pm$0.014  & 0.350$\pm$0.007 &   1.09   &  I \\ 
NGC3516            & 0.588$\pm$0.014  & 0.672$\pm$0.019 &   1.14   &  I \\ 
NGC3660            & 0.067$\pm$0.012  & 0.087$\pm$0.012 &   1.30   &  II \\ 
NGC4501            & 0.032$\pm$0.013  & 0.054$\pm$0.012 &   1.69   &  III \\ 
NGC4602            & 0.042$\pm$0.008  & 0.101$\pm$0.012 &   2.40   &  III \\ 
TOLOLO1238-364     & 1.290$\pm$0.043  & 1.522$\pm$0.021 &   1.18   &  II \\ 
NGC4748            & 0.089$\pm$0.010  & 0.223$\pm$0.008 &   2.50   &  III \\ 
NGC4968            & 0.812$\pm$0.029  & 0.817$\pm$0.014 &   1.00   &  I \\
MCG-03-34-063      & 0.020$\pm$0.010  & 0.042$\pm$0.009 &   2.09   &  III \\ 
MRK817     	   & 0.670$\pm$0.016  & 0.778$\pm$0.014 &   1.16   &  II \\ 
IRASF15091-2107	   & 0.337$\pm$0.012  & 0.388$\pm$0.008 &   1.15   &  I \\ 
ESO141-055         & 0.249$\pm$0.013  & 0.288$\pm$0.008 &   1.16   &  II \\ 
NGC6890            & 0.328$\pm$0.019  & 0.381$\pm$0.007 &   1.16   &  II \\ 
MRK0897            & 0.120$\pm$0.010  & 0.164$\pm$0.006 &   1.37   &  II \\ 
IRASF22017+0319    & 0.533$\pm$0.020  & 0.583$\pm$0.010 &   1.09   &  I \\ 
NGC7496            & 0.658$\pm$0.031  & 0.806$\pm$0.013 &   1.22   &  II \\ 
NGC7590        	   & 0.032$\pm$0.010  & 0.096$\pm$0.011 &   3.00   &  III \\ 
\enddata
\end{deluxetable}

\begin{deluxetable}{lcccrrr}
\tabletypesize{\scriptsize} 
\tablecaption{H$_2$ Estimated Temperatures and Masses} \label{tbl-1} \tablewidth{0pt}
\tablehead{\colhead{NAME} &  \colhead{T(0-1)}  &  \colhead{T(1-2)} & \colhead{T(2-3)} & \colhead{H$_2$ Mass} \\
                                          & (K)                       &       (K)                &  (K)                  &  (10$^{6}$ M$_\sun$) }
\startdata
IRAS00198-7926     &	  $>$ 229  &    573       &   241        &   $<$ 42.3  \\ 
ESO012-G021         &	  $>$ 422  &    300       &   351        &   $<$ 4.19  \\ 
IRAS00521-7054     &	  $>$ 209  &    434       &   346        &   $<$ 26.7 \\ 
ESO541-IG012       &	  $>$ 160  &   $<$ 374  &  \nodata   &   $<$ 4.37 \\ 
NGC0424               &	  $>$ 173  &   $<$ 370  &	 \nodata   &   $<$ 0.9  \\ 
NGC0513               &	   191        &  \nodata    &\nodata      &    0.88     \\ 
MRK1034 NED01   &	   182        &   $<$ 264  &   \nodata   &    11.8    \\ 
MRK1034 NED02   &	  $>$ 173   &   $<$ 192  &  $>$ 185  &   $<$ 25.6 \\ 
ESO545-G013        &	   184        &   $<$ 275  &  $>$ 402  &    5.18    \\ 
NGC1125               &	  $>$ 150  &   $<$ 334  &	 \nodata    &   $<$ 0.73 \\
ESO033-G002        &	   241       &    356       & \nodata      &    1.4	\\ 
MRK0006               &	      236    &       257    &	 \nodata    &   $<$ 2.8 \\ 
MRK0009               &	  $>$ 198  &   $<$ 325  & $>$  435   &   $<$ 4.8 \\ 
NGC3516               &	  $>$ 271  &       256    &	 \nodata    &   $<$ 0.48 \\ 
NGC3660               &	  $>$ 137  &   $<$ 282 & \nodata      &    2.0     \\ 
NGC4501               &	   270       &    321       & \nodata       &    0.91    \\ 
NGC4602               &	   150       &    439       &	 \nodata     &    0.84    \\ 
TOLOLO1238-364  &	  $>$ 202  &   $<$ 335  &	 \nodata     &   $<$ 0.97 \\ 
NGC4748               &	   246       &   $<$ 257  &	 \nodata     &    0.82    \\ 
NGC4968               &	  $>$ 222  &   $<$ 273  &	  \nodata    &   $<$ 0.65 \\
MCG-03-34-063      &	  $>$ 183  &   $<$ 252  &	 \nodata     &   $<$ 2.7  \\ 
MRK817     	      &	  $>$ 162  &   $<$ 305  &	  \nodata    &   $<$ 0.28 \\ 
IRASF15091-2107  &	  $>$ 198  &    290       &   339          &   $<$ 3.8  \\ 
NGC6890               &	   147       &   $<$ 431  &	 \nodata     &    0.43  \\ 
MRK0897               &	   238       &    301       &  $<$ 239     &    3.6	\\ 
IRASF22017+0319  &	  $>$ 223  &    332       &   396          &   $<$ 4.0 \\ 
NGC7496               &	  $>$ 239  &   $<$ 244  &	  \nodata    &   $<$ 13.5 \\ 
NGC7590        	      &	   164       &    388       & \nodata       &    1.0     \\ 
\enddata
\end{deluxetable}

\begin{deluxetable}{lcccccccc}
\tabletypesize{\scriptsize} 
\tablecaption{Table A1 - Rare ionic fine structure lines} 
\tablewidth{0pt}
\tablehead{& \multicolumn{8}{c}{Line fluxes (10$^{-14}~erg~s^{-1} cm^{-2}$)} \\
 \cline{2-9}\\
\colhead{NAME} & \colhead{[ArV]} & \colhead{[MgV]} & \colhead{[FeII] } & \colhead{[FeIII]}& \colhead{[FeII]} & \colhead{[FeII]} & \colhead{[FeII]} & \colhead{[FeII]}  \\
                                          & (13.10$\mu m$)  & (13.52$\mu m$)   &(17.94$\mu m$)   & (22.93$\mu m$)  & (24.52$\mu m$)  & (25.99$\mu m$) & (35.35$\mu m$)  & (35.77$\mu m$)  }
\startdata
NGC0513       & \nodata   &\nodata & \nodata  &\nodata   & \nodata  &1.41$\pm$0.07& \nodata  & \nodata \\
MRK1034 NED01   & \nodata   &\nodata & \nodata  &\nodata   & \nodata  &1.04$\pm$0.27& \nodata  & \nodata\\
MRK1034 NED02   & \nodata   &\nodata & \nodata  &\nodata   & \nodata  &1.95$\pm$0.35& \nodata  & \nodata\\
ESO545-G013     & \nodata   &\nodata & \nodata  &\nodata   &0.35$\pm$0.10 &\nodata&\nodata& \nodata\\
NGC1125         & \nodata   &\nodata & \nodata	&2.39$\pm$0.28 &\nodata	  &\nodata& \nodata & \nodata\\
MRK0006         & \nodata   & \nodata &2.07$\pm$0.15&\nodata  &  \nodata  &\nodata  &	\nodata	  & \\
MRK0009         & \nodata   & \nodata & 	 &   \nodata  &  \nodata  &\nodata  &8.09$\pm$0.77& \\
NGC3660         & \nodata   &\nodata& \nodata	 & \nodata & 1.40$\pm$0.20& \nodata  & \nodata 	   & \\
NGC4501         & \nodata   &\nodata& \nodata	 & \nodata &\nodata	  & 2.62$\pm$0.28&	   & \\
TOLOLO1238-364  & \nodata   &2.59$\pm$0.24& \nodata& \nodata &\nodata	 &\nodata &  \nodata	   & \\
NGC4968         &  \nodata  &2.44$\pm$0.22& \nodata &\nodata&\nodata&\nodata     &\nodata & \nodata \\
MCG-03-34-063   & \nodata   &	\nodata   & \nodata &\nodata&	    &\nodata 	&9.10$\pm$1.48 & \nodata\\
MRK0817         & \nodata   &	\nodata   & \nodata &\nodata&  $<$ 1.0 &\nodata &14.10$\pm$1.25& \nodata\\
IRASF15091-2107&1.37$\pm$0.19&$<$ 0.96&  \nodata  & \nodata &\nodata&\nodata	&\nodata & \nodata\\
MRK0897         & \nodata   &	\nodata  &\nodata &1.27$\pm$0.12&\nodata  & 1.49$\pm$0.19&\nodata & \nodata\\
NGC7496         & \nodata   &	\nodata  &\nodata &\nodata  &	\nodata   &\nodata&\nodata   &5.72$\pm$1.88 \\
\enddata
\end{deluxetable}

\begin{deluxetable}{lccccc}
\tabletypesize{\scriptsize} 
\tablecaption{Table A2 - Equivalent widths of key lines} 
\tablewidth{0pt}
\tablehead{\colhead{NAME} & \colhead{[NeII]12.81$\mu$m} & \colhead{[NeV]14.32$\mu m$} & \colhead{[NeIII]15.56$\mu m$ } & \colhead{H$_2$ 17.02$\mu$m}& \colhead{[OIV]25.89$\mu m$}  \\
             & ($\mu$m)   & ($\mu$m)   & ($\mu$m)   & ($\mu$m)  &  ($\mu$m)   }
\startdata
IRAS00198-7926   &  -0.009 & -0.023  &  -0.029    &  -0.009   &  -0.065  \\
ESO012-G021	 &  -0.055 & -0.023  &  -0.048    &  -0.032   &  -0.159  \\
IRAS00521-7054   &  -0.008 & -0.010  &  -0.016    &  -0.006   &  -0.027  \\
ESO541-IG012	 &  -0.006 & -0.008  &  -0.008    &  -0.004   &  -0.043  \\
NGC0424 	 &  -0.005 & -0.011  &  -0.013    &  -0.003   &  -0.044  \\
NGC0513 	 &  -0.072 & -0.017  &  -0.049    &  -0.023   &  -0.072  \\
MRK1034 NED01	 &  -0.129 &  \nodata&  -0.025    &  -0.041   & \nodata  \\
MRK1034 NED02	 &  -0.107 & -0.012  &  -0.018    &  -0.025   &  -0.014  \\
ESO545-G013	 &  -0.108 & -0.064  &  -0.119    &  -0.017   &  -0.178  \\
NGC0931 	 &  -0.007 & -0.017  &  -0.024    & \nodata   &  -0.112  \\
NGC1125 	 &  -0.077 & -0.027  &  -0.090    &  -0.004   &  -0.104  \\
ESO033-G002	 &  -0.003 & -0.018  &  -0.028    &  -0.008   &  -0.098  \\
MRK0006 	 &  -0.064 & -0.020  &  -0.100    &  -0.009   &  -0.190  \\
MRK0009 	 &  -0.010 & -0.007  &  -0.006    &  -0.004   &  -0.035  \\
NGC3516 	 &  -0.013 & -0.017  &  -0.033    &  -0.008   &  -0.129  \\
NGC3660 	 &  -0.095 & -0.015  &  -0.029    &  -0.033   &  -0.065  \\
NGC4501 	 &  -0.107 &  \nodata&  -0.215    &  -0.248   &  -0.164  \\
NGC4602 	 &  -0.122 & -0.027  &  -0.043    &  -0.056   & \nodata  \\
TOLOLO1238-364   &  -0.047 & -0.013  &  -0.029    &  -0.004   &  -0.025  \\
NGC4748 	 &  -0.096 & -0.113  &  -0.248    &  -0.034   &  -0.658  \\
NGC4968 	 &  -0.032 & -0.027  &  -0.048    &  -0.005   &  -0.084  \\
MCG-03-34-063	 &  -0.162 &  \nodata& \nodata    &  -0.126   & \nodata  \\
MRK0817 	 &  -0.006 & -0.003  &  -0.007    &  -0.002   &  -0.016  \\
IRASF15091-2107  &  -0.036 & -0.034  &  -0.057    &  -0.013   &  -0.159  \\
ESO141-G055	 &  -0.007 & -0.010  &  -0.023    & \nodata   &  -0.064  \\
NGC6890 	 &  -0.038 & -0.021  &  -0.024    &  -0.005   &  -0.043  \\
MRK0897 	 &  -0.128 & -0.012  &  -0.057    &  -0.026   &  -0.004  \\
IRASF22017+0319  &  -0.015 & -0.020  &  -0.037    &  -0.005   &  -0.121  \\
NGC7496 	 &  -0.089 &  \nodata&  -0.016    &  -0.009   & \nodata  \\
NGC7590 	 &  -0.098 &  \nodata&  -0.113    &  -0.061   &  -0.090  \\
\enddata
\end{deluxetable}


\clearpage

\begin{figure}
\centerline{\includegraphics[width=8cm]{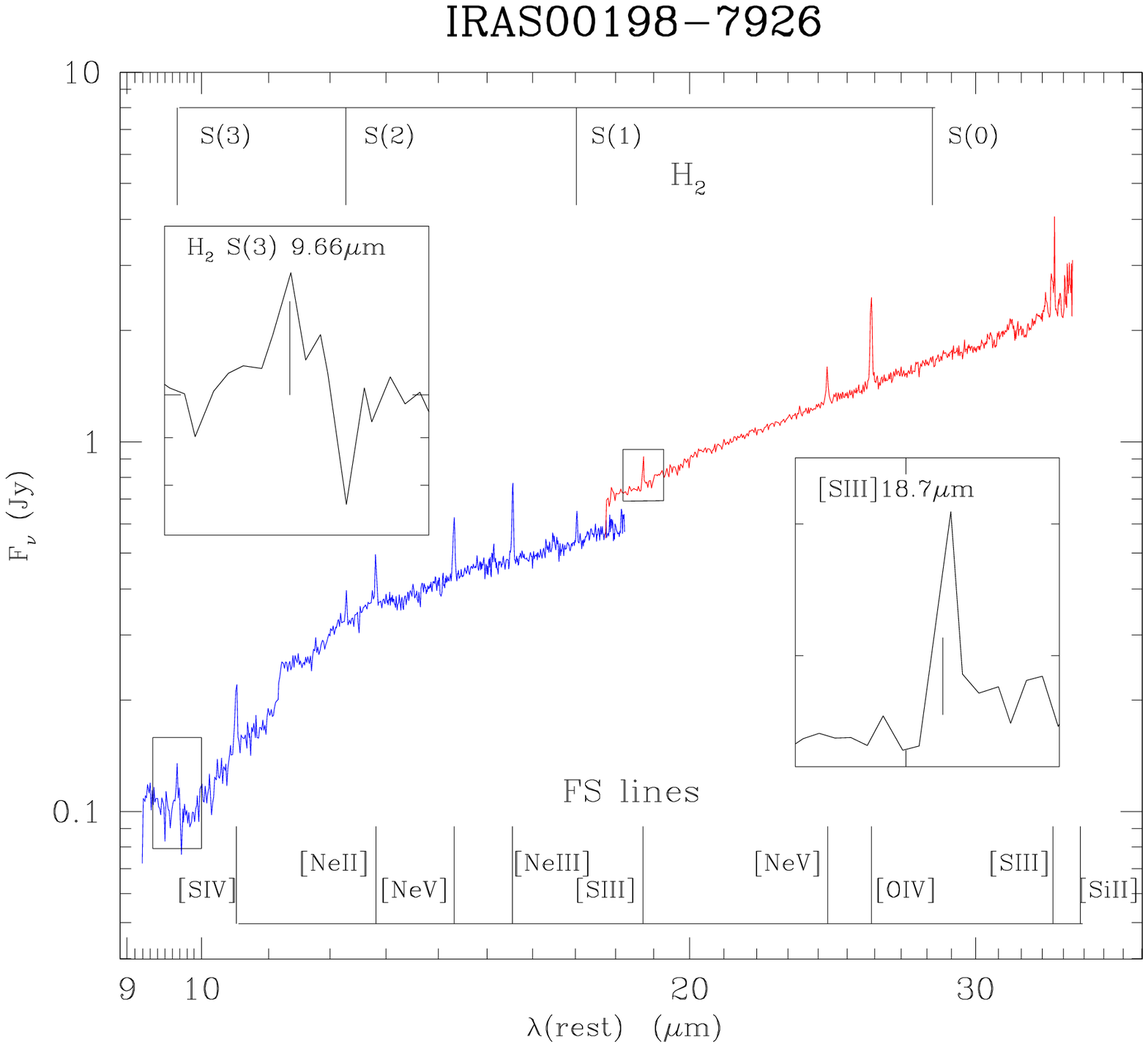}\includegraphics[width=8cm]{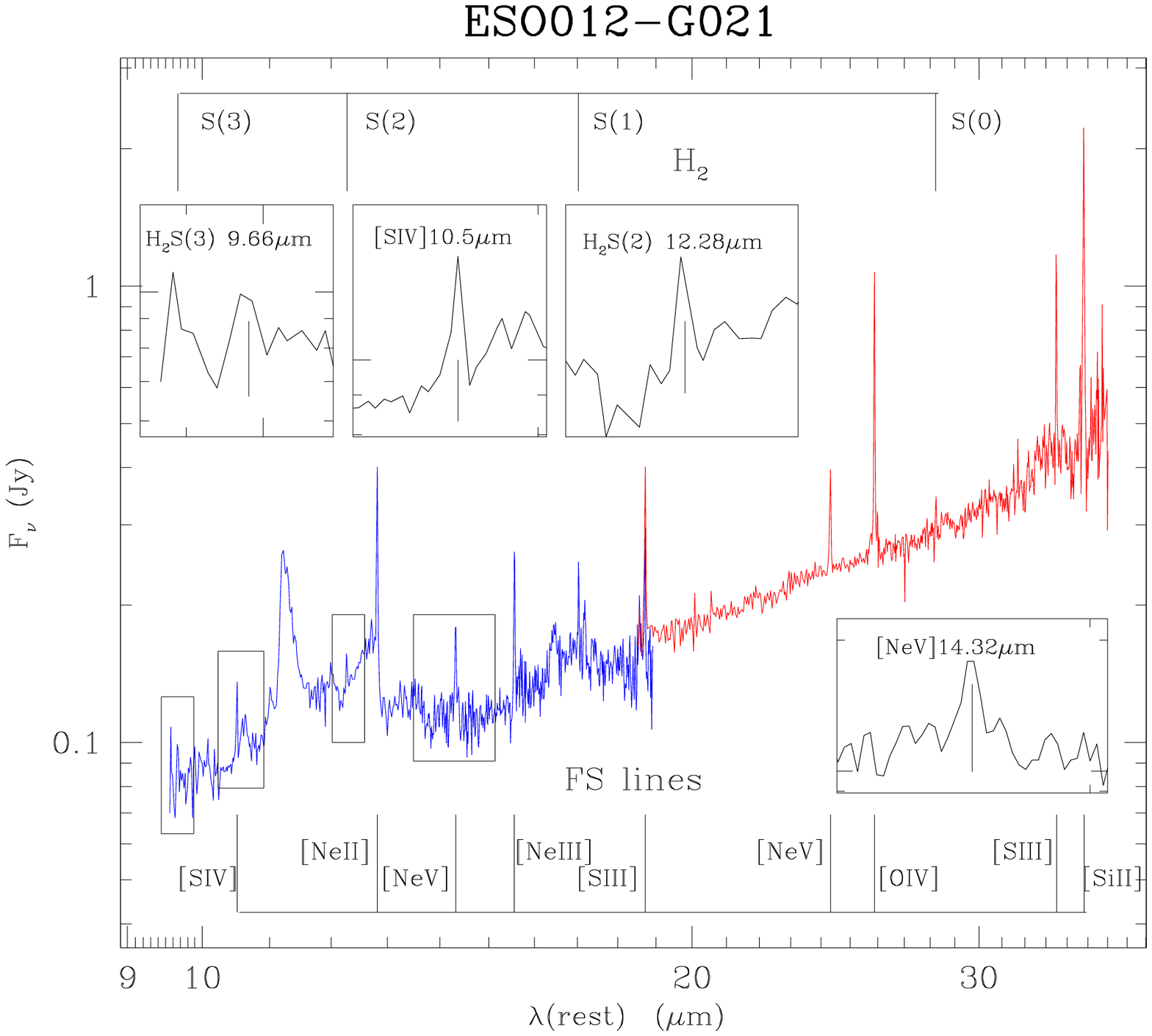}}

\centerline{\includegraphics[width=8cm]{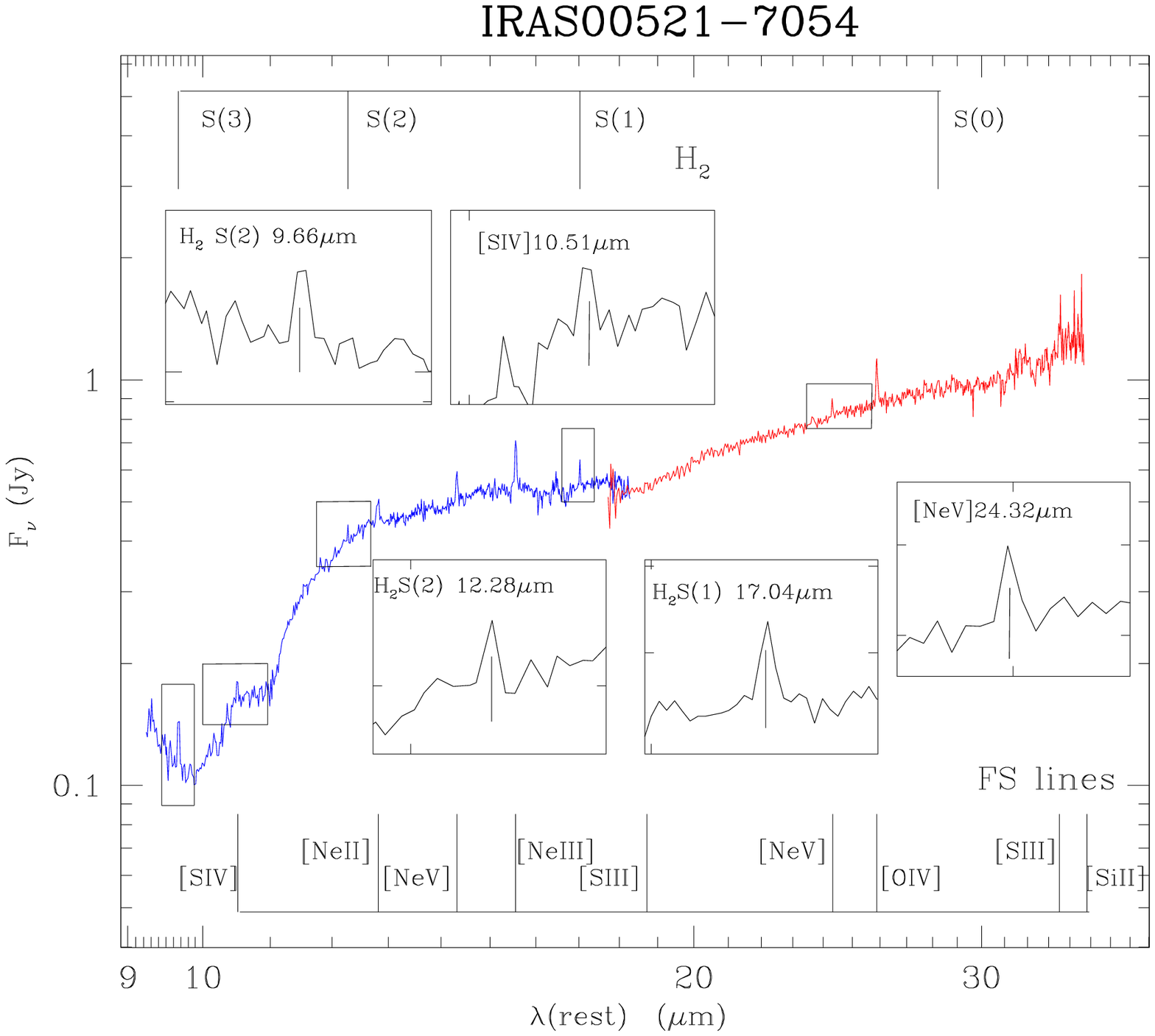}\includegraphics[width=8cm]{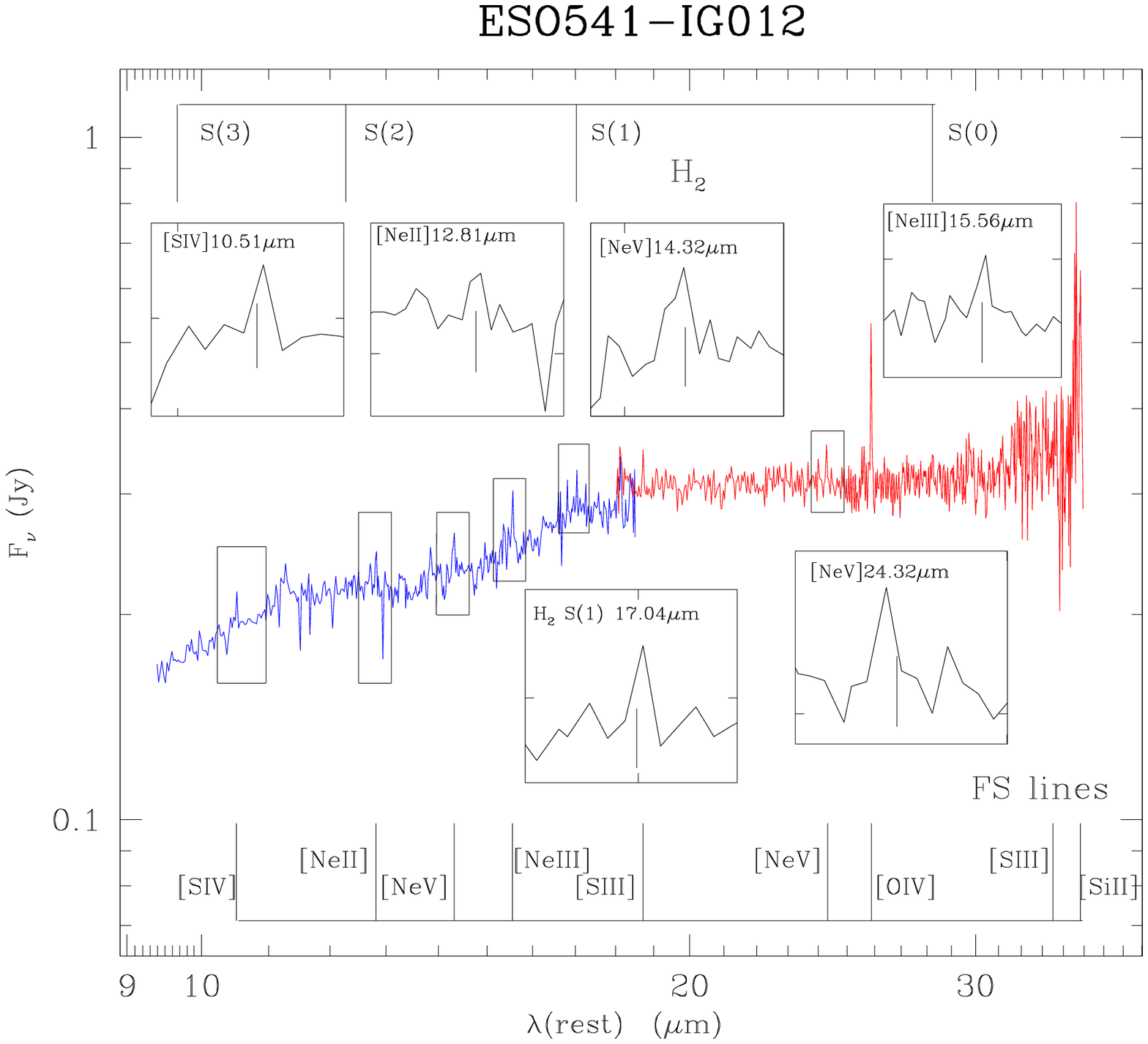}}

\centerline{\includegraphics[width=8cm]{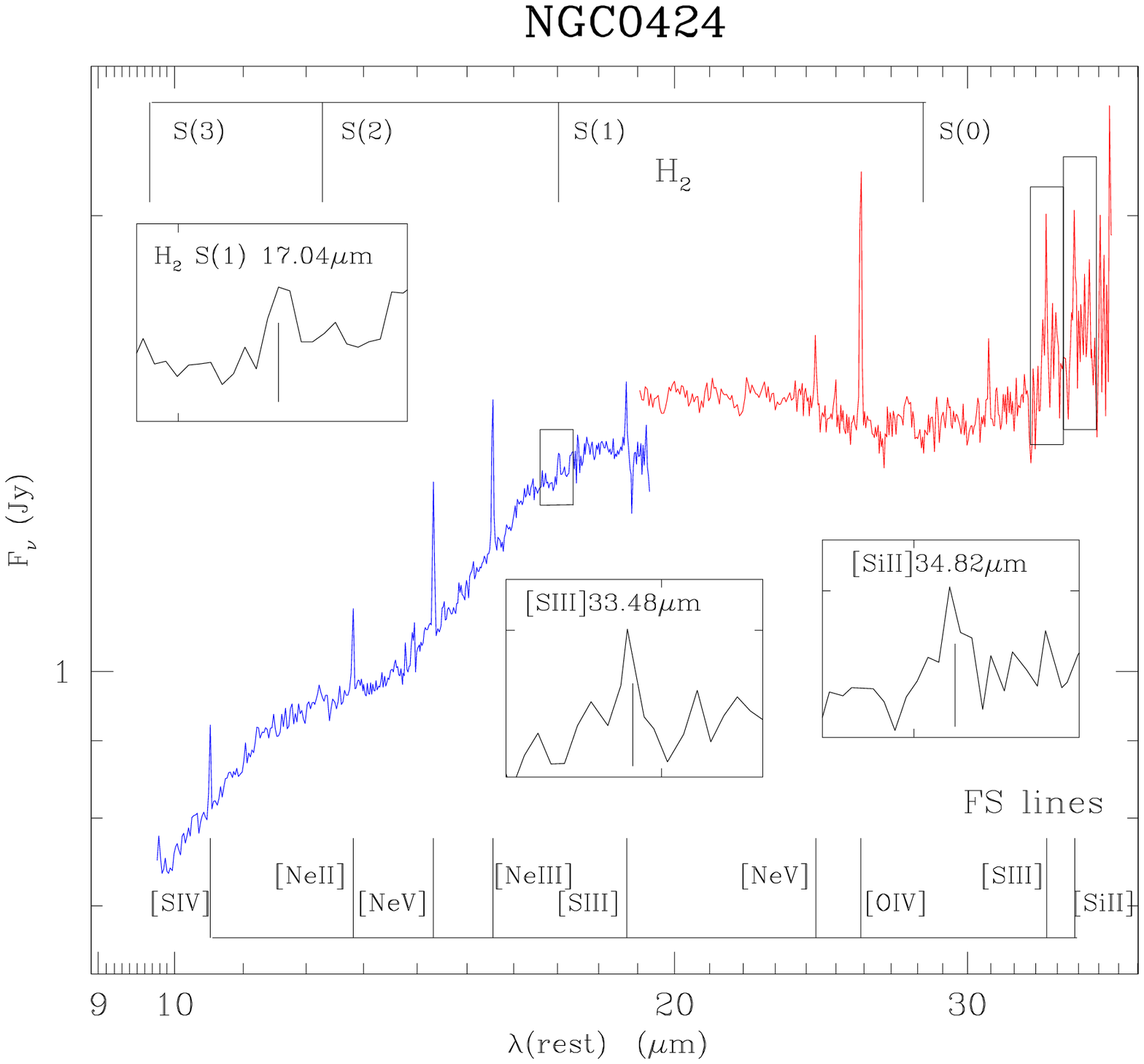}\includegraphics[width=8cm]{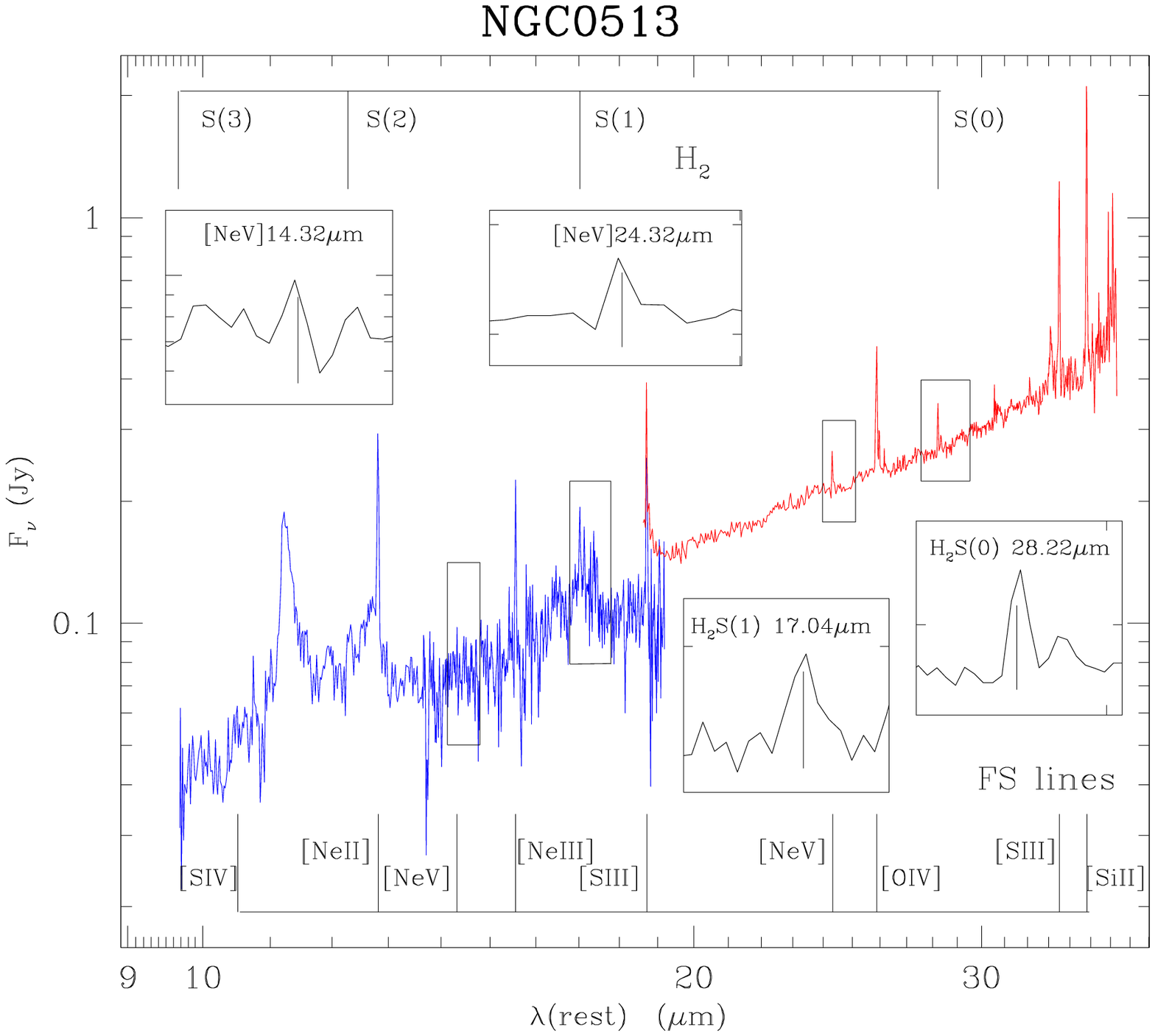}}
\caption{Spitzer IRS SH and LH specra of the observed Seyfert galaxies. Wavelengths have 
been shifted to the galaxies rest frames. For the fainter lines (those with a S/N $\lesssim$ 10.) 
the blow-ups with their theoretical wavelengths are also given. }
\end{figure}

\clearpage

\begin{figure}
\centerline{\includegraphics[width=8cm]{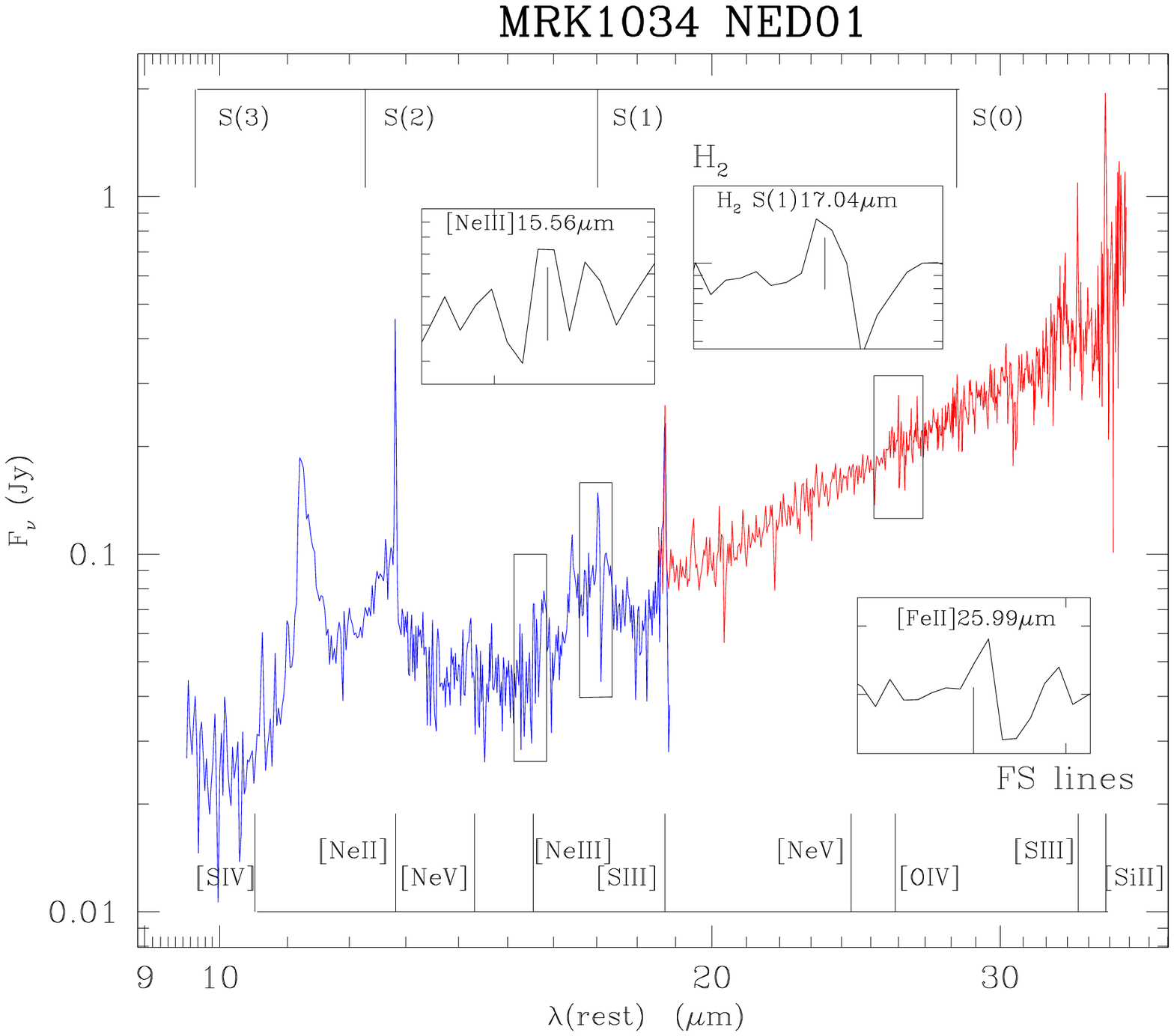}\includegraphics[width=8cm]{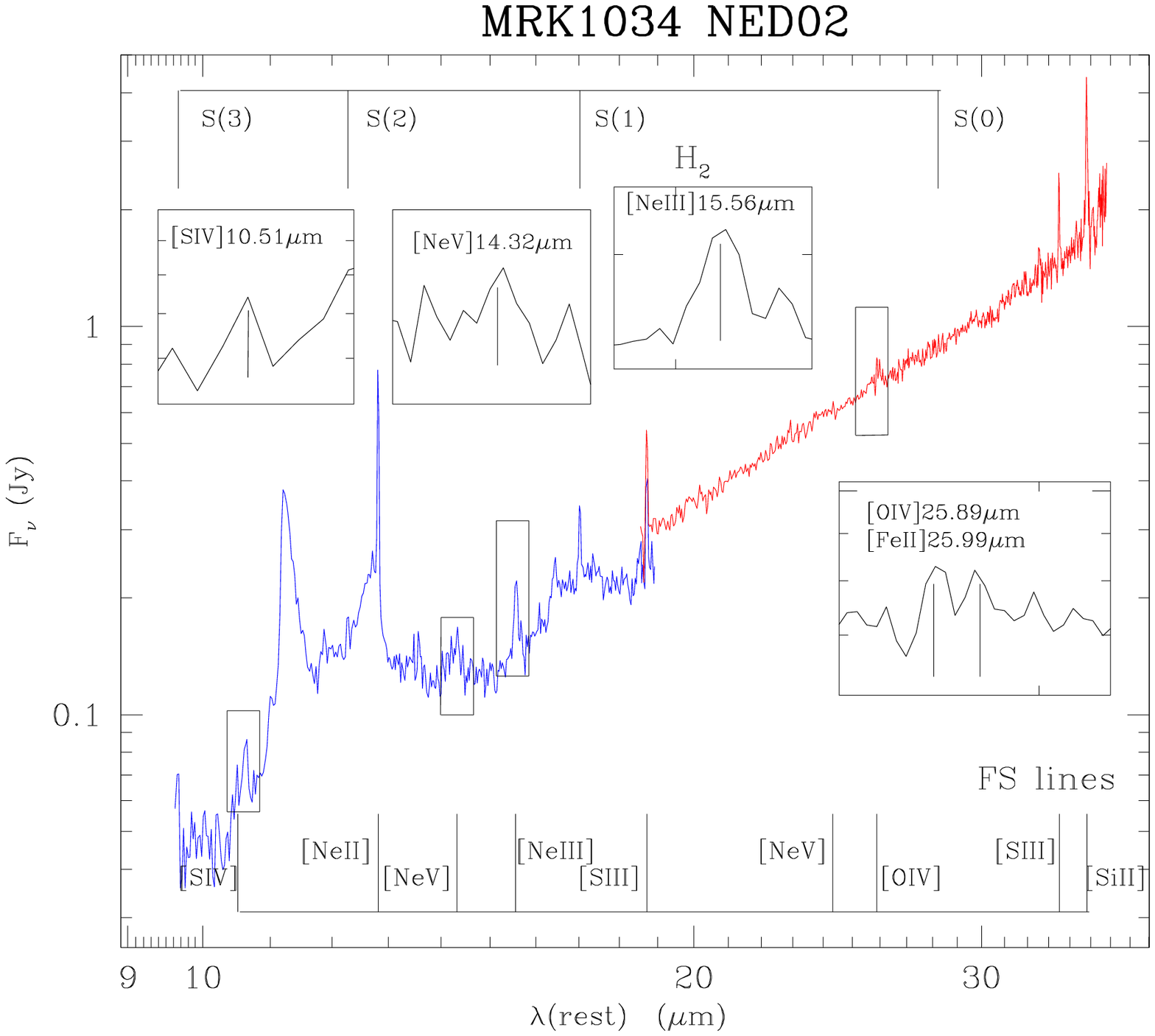}}

\centerline{\includegraphics[width=8cm]{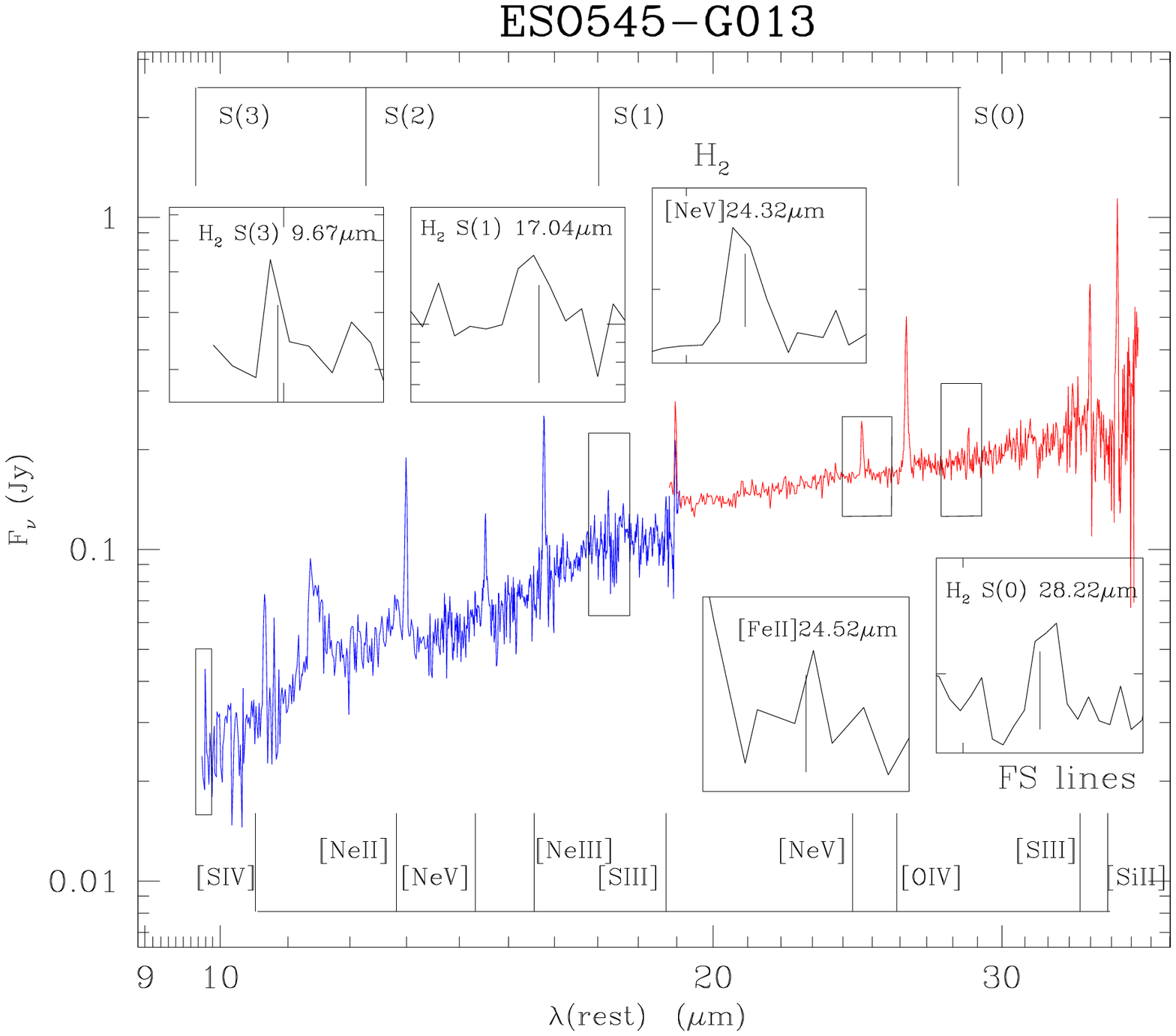}\includegraphics[width=8cm]{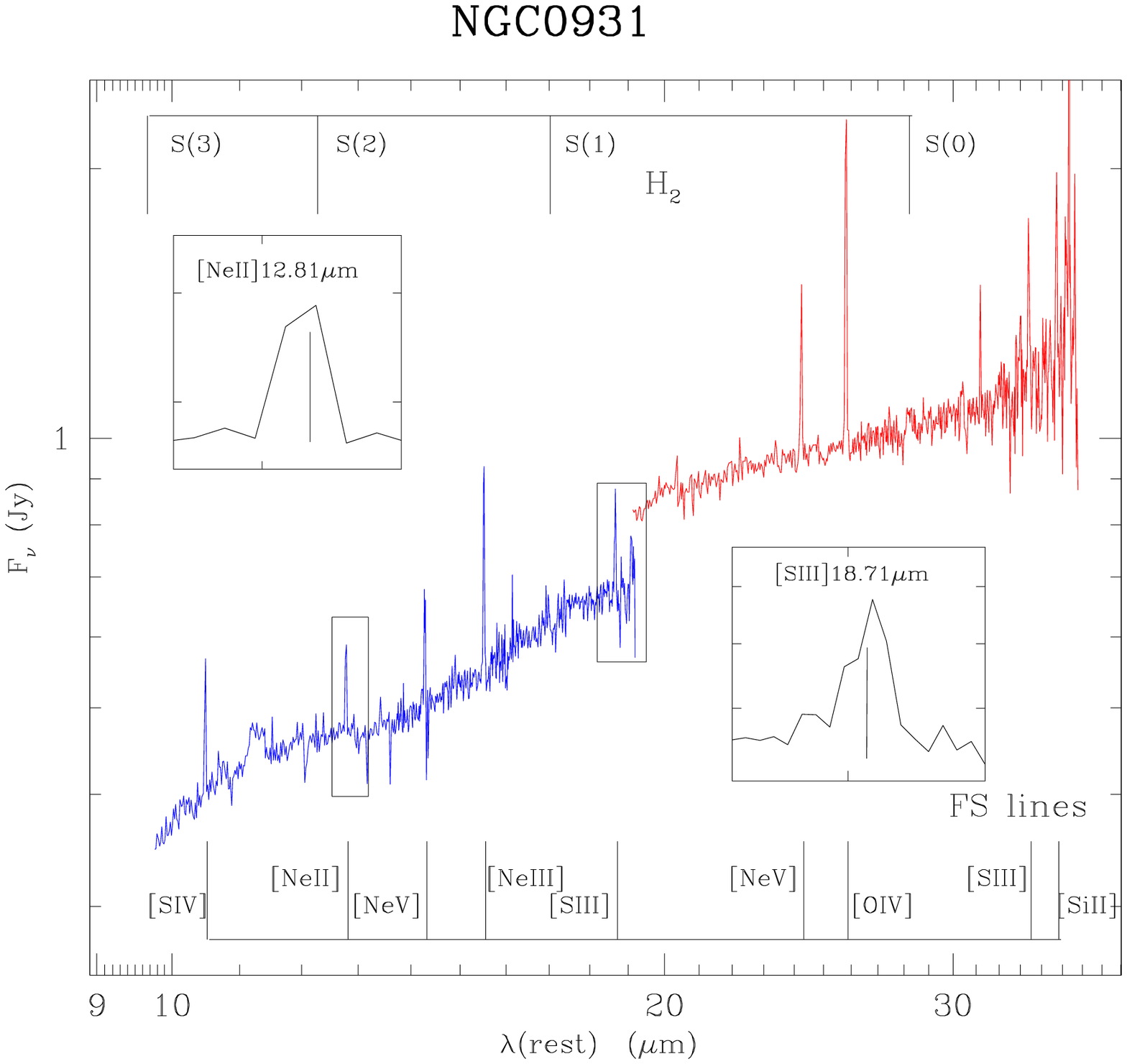}}

\centerline{\includegraphics[width=8cm]{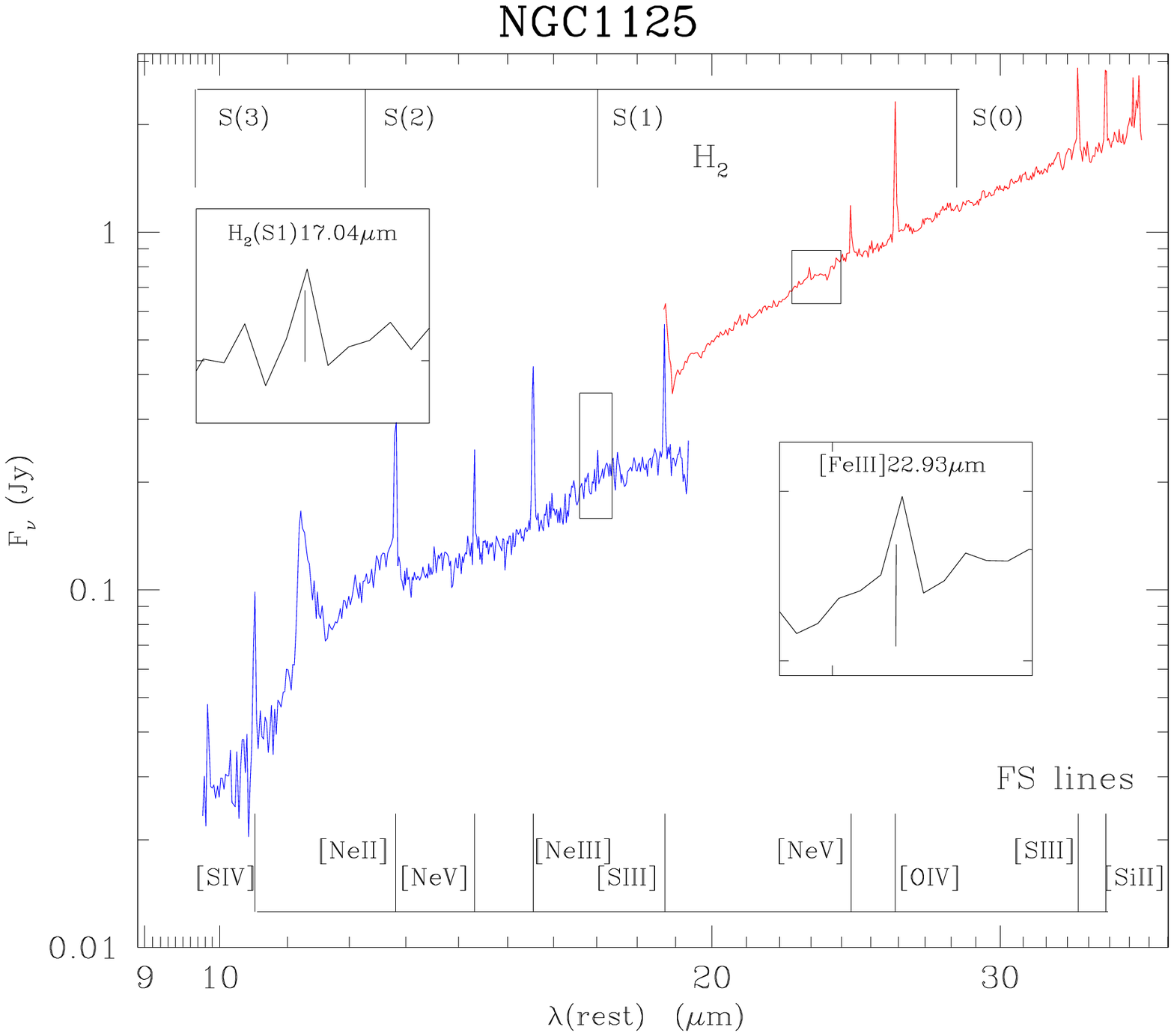}\includegraphics[width=8cm]{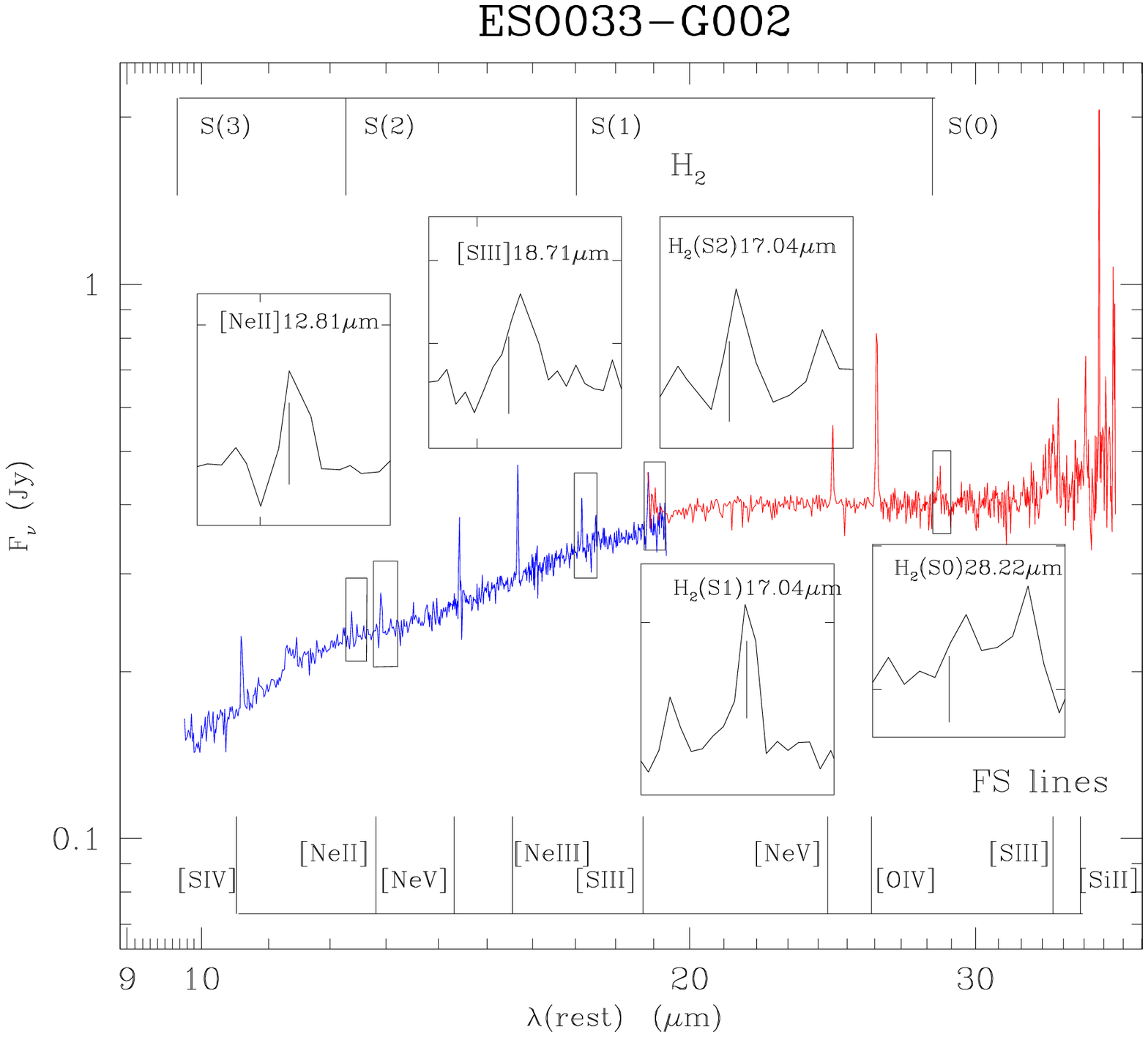}}
\end{figure}
\clearpage

\begin{figure}
\centerline{\includegraphics[width=8cm]{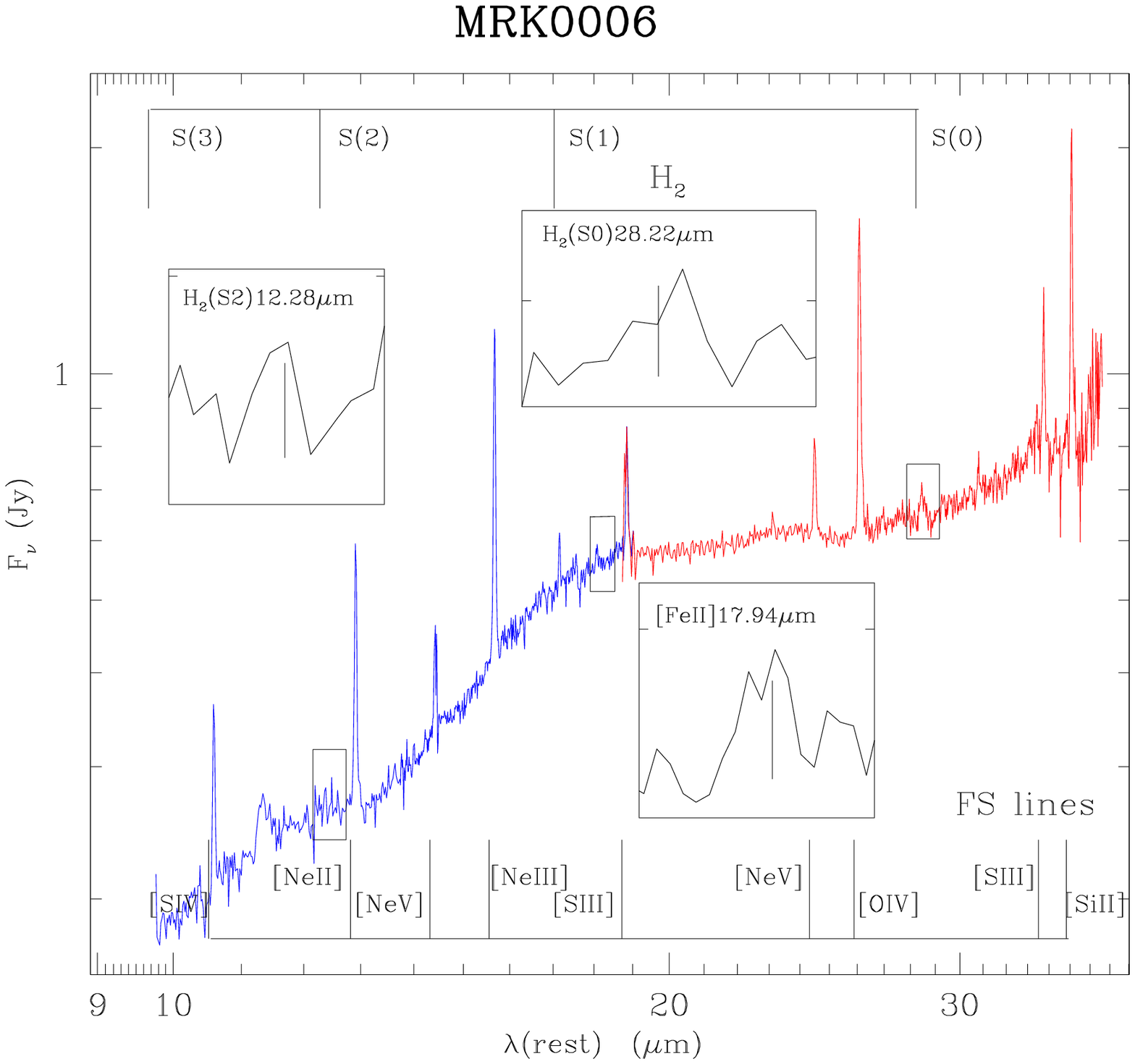}\includegraphics[width=8cm]{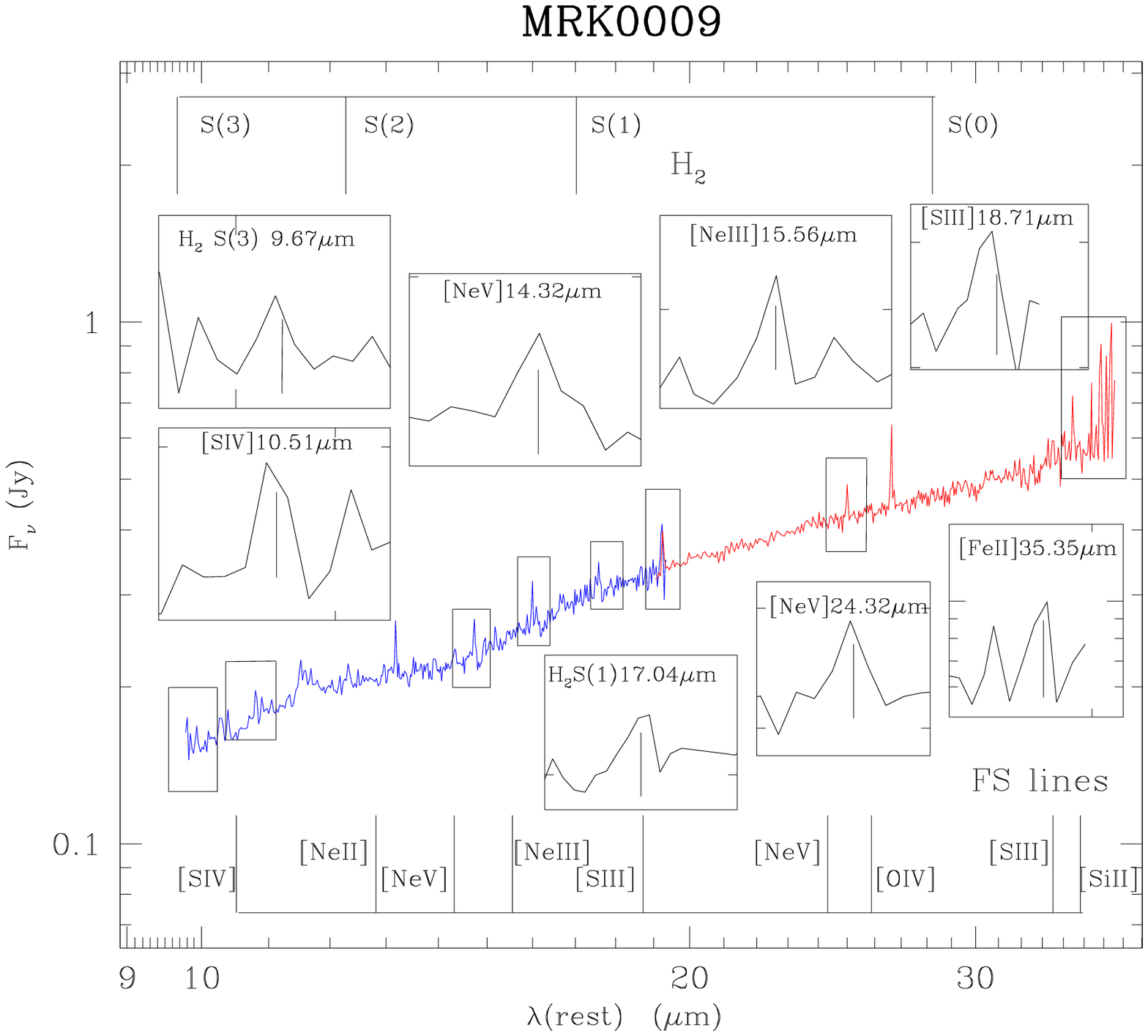}}

\centerline{\includegraphics[width=8cm]{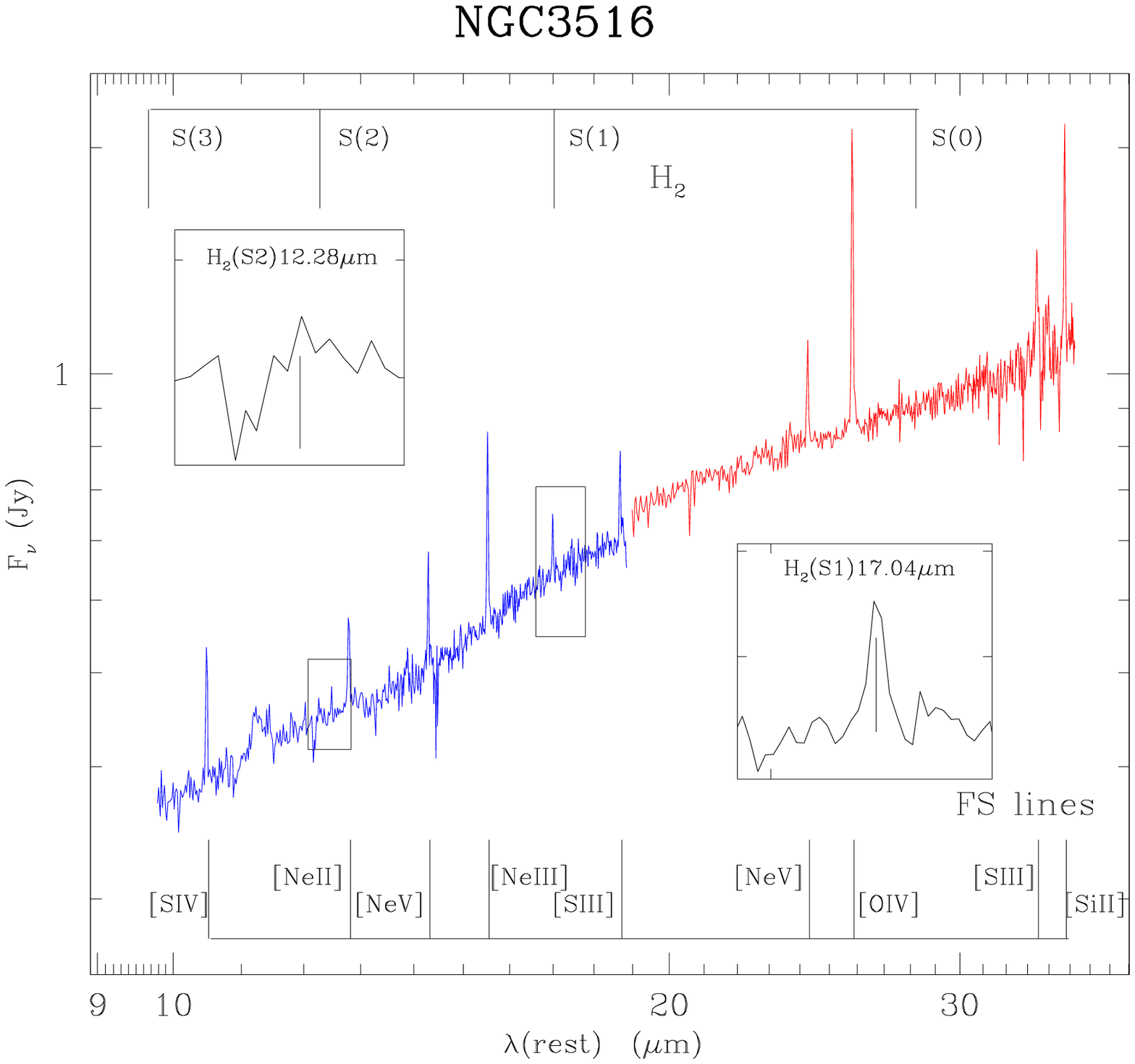}\includegraphics[width=8cm]{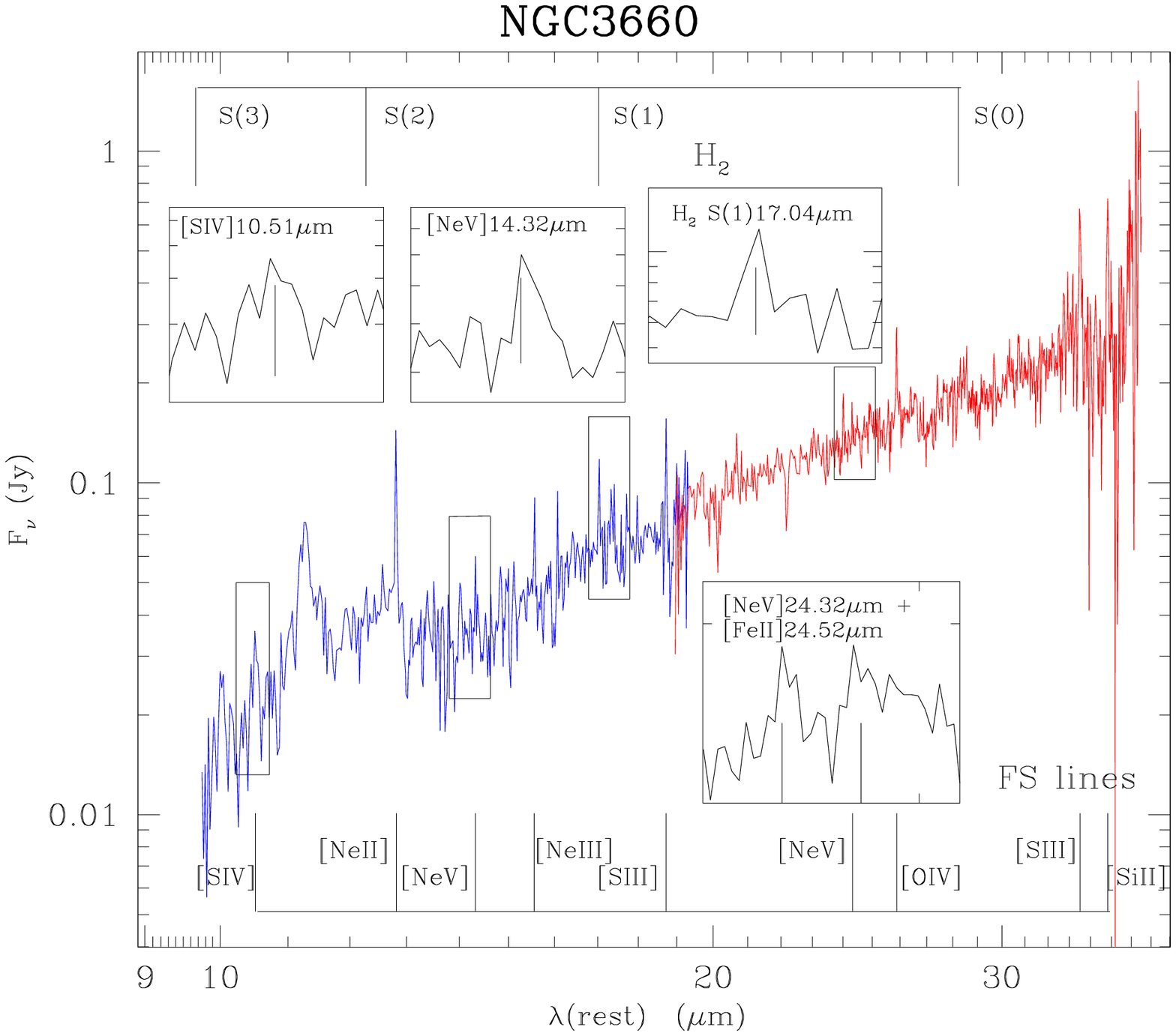}}

\centerline{\includegraphics[width=8cm]{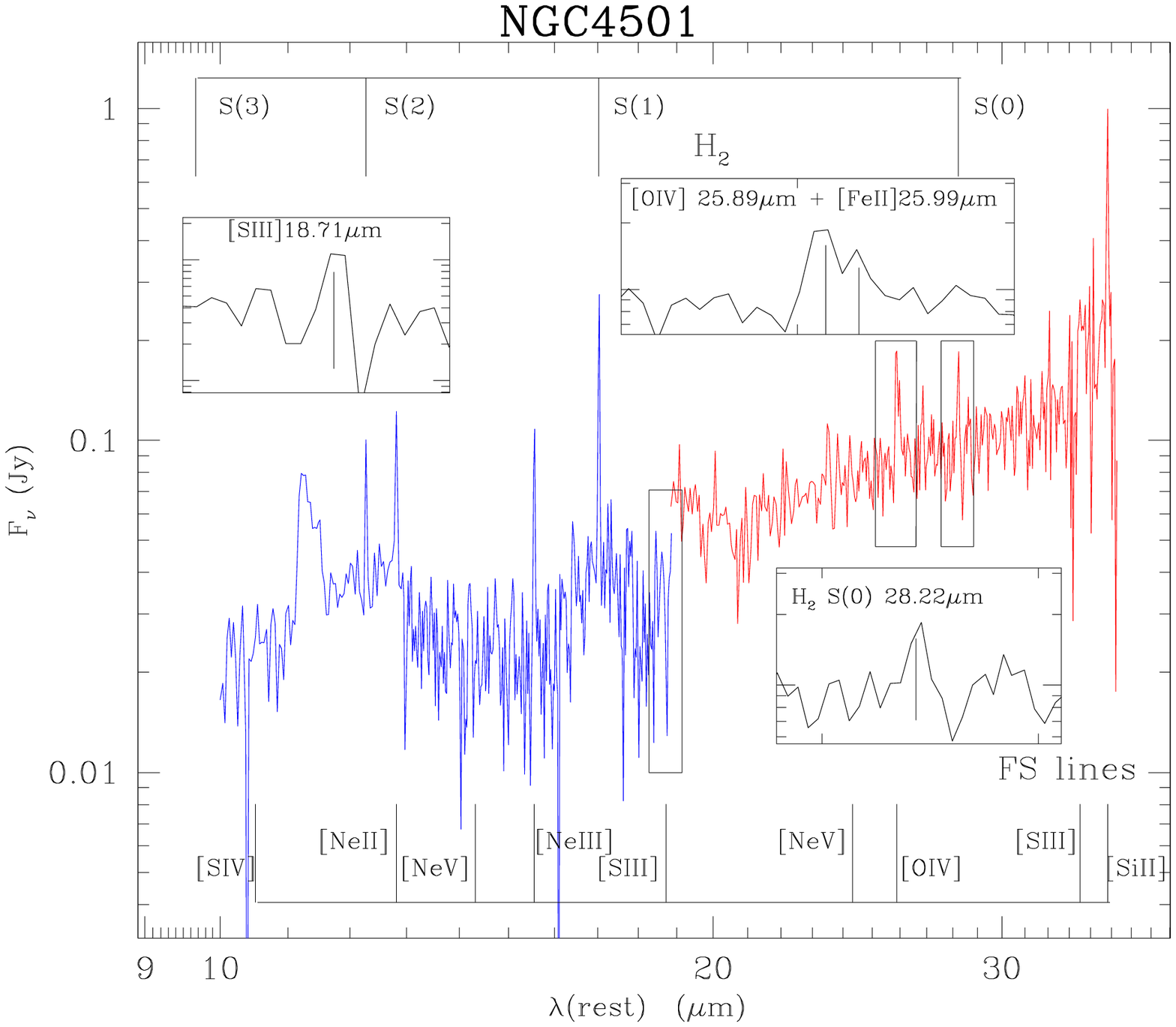}\includegraphics[width=8cm]{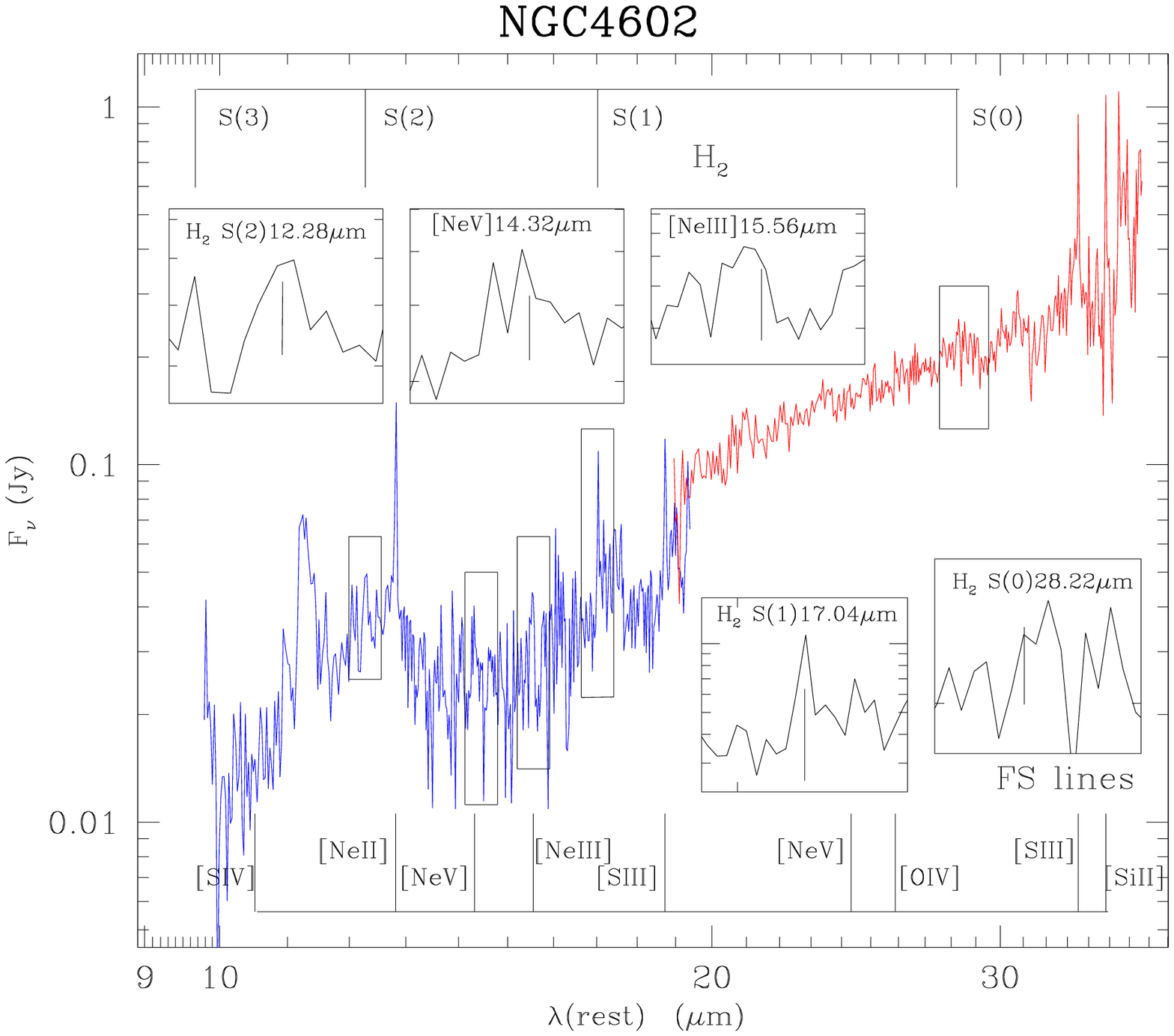}}
\end{figure}
\clearpage

\begin{figure}
\centerline{\includegraphics[width=8cm]{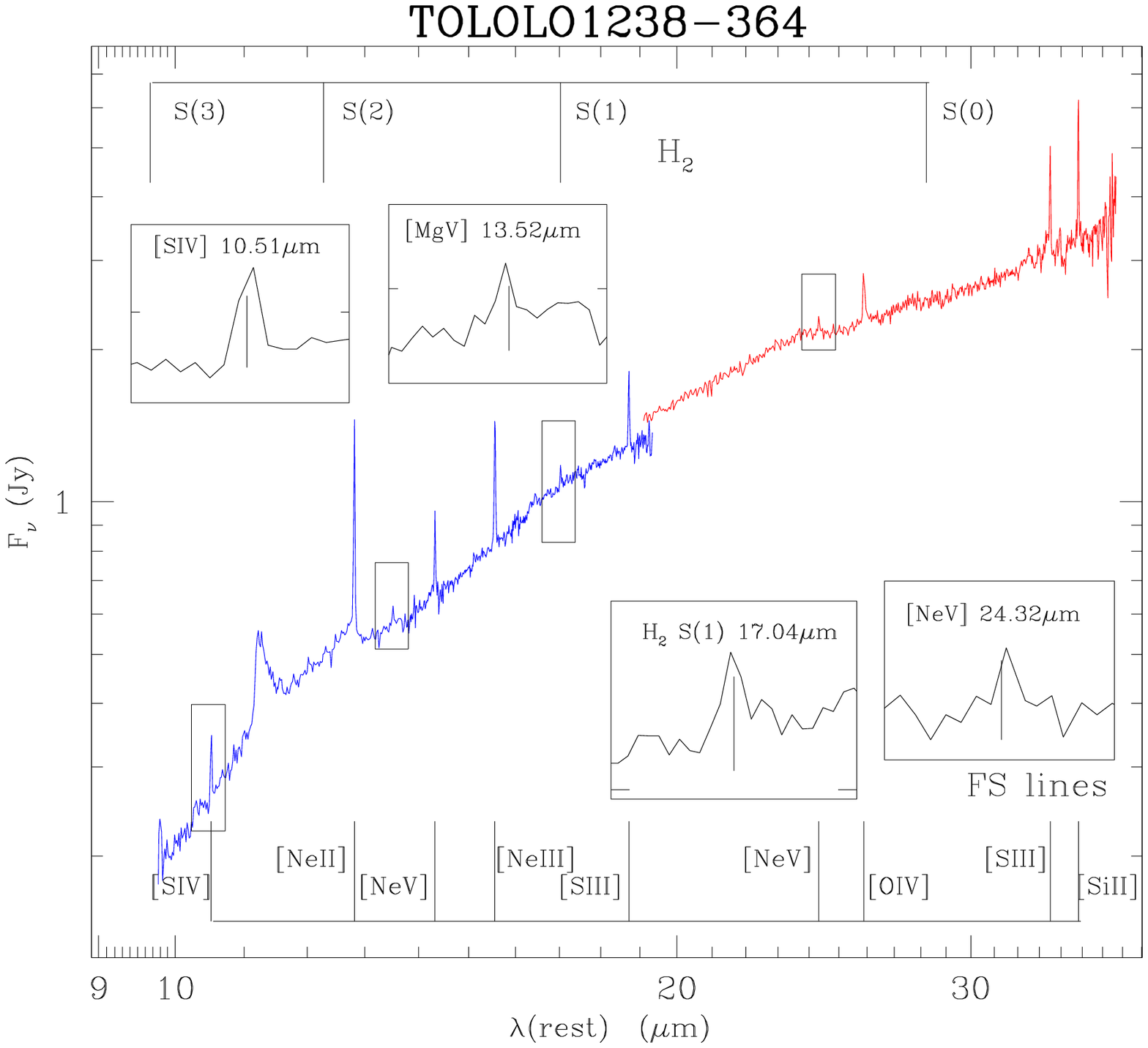}\includegraphics[width=8cm]{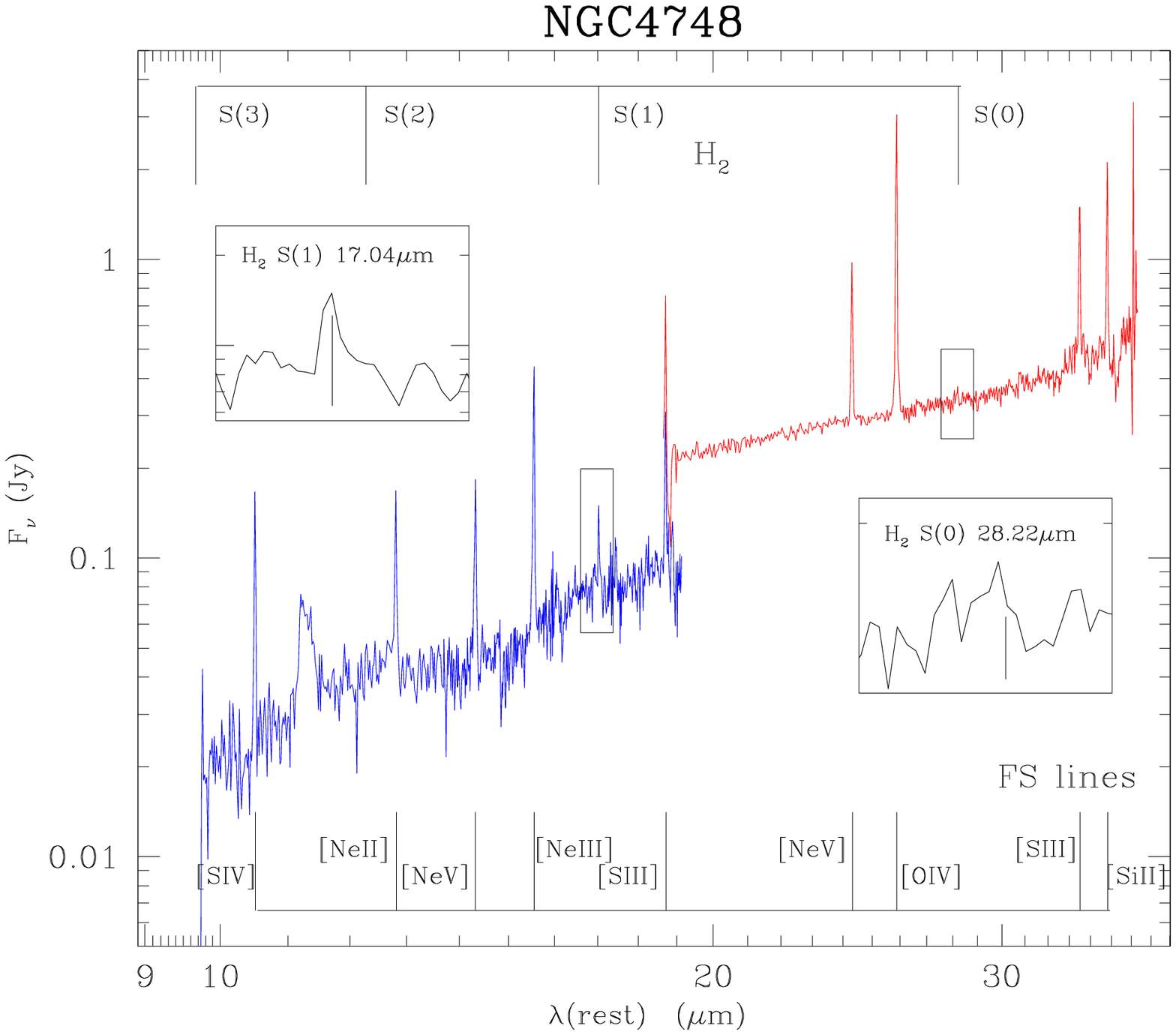}}

\centerline{\includegraphics[width=8cm]{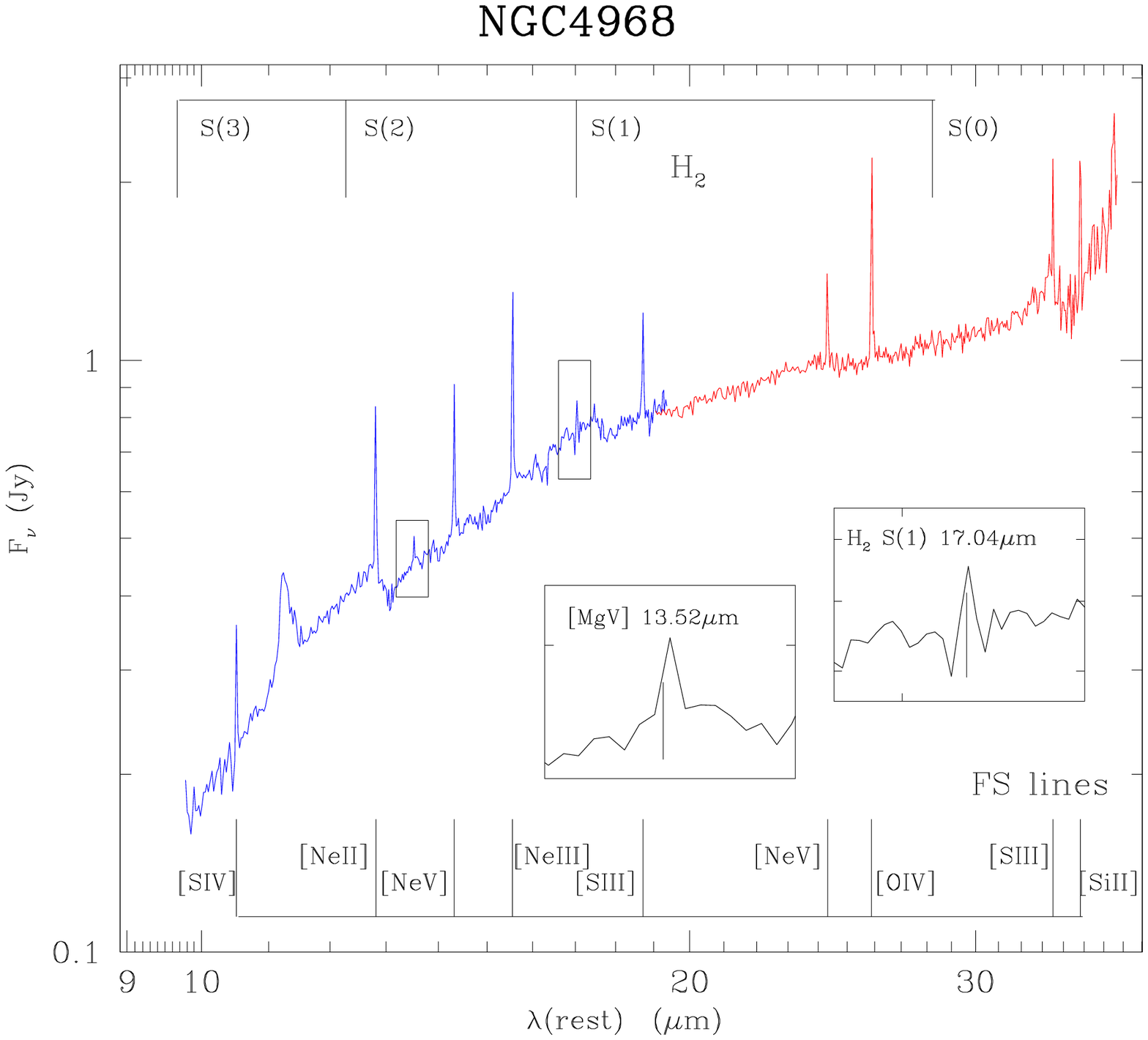}\includegraphics[width=8cm]{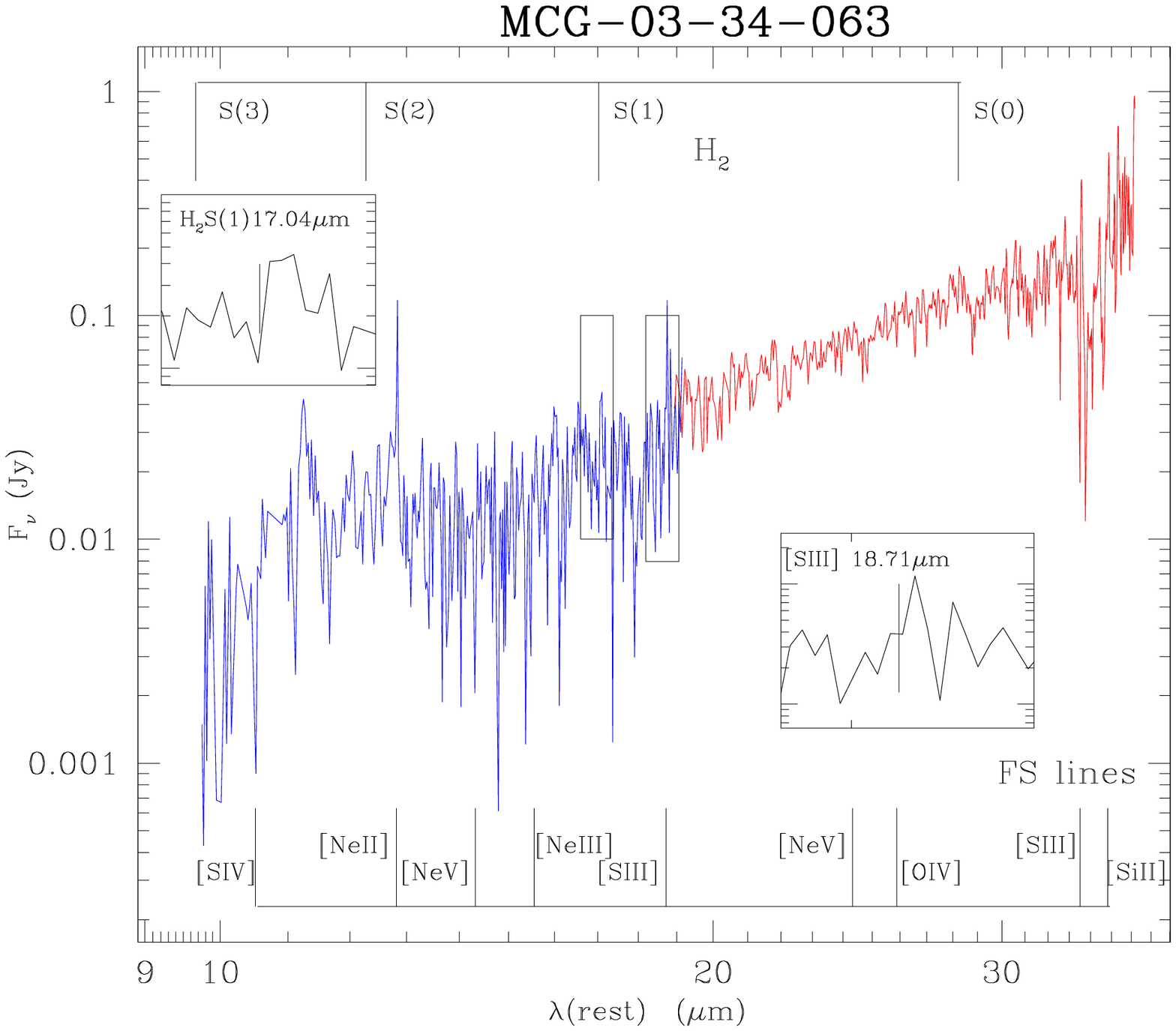}}

\centerline{\includegraphics[width=8cm]{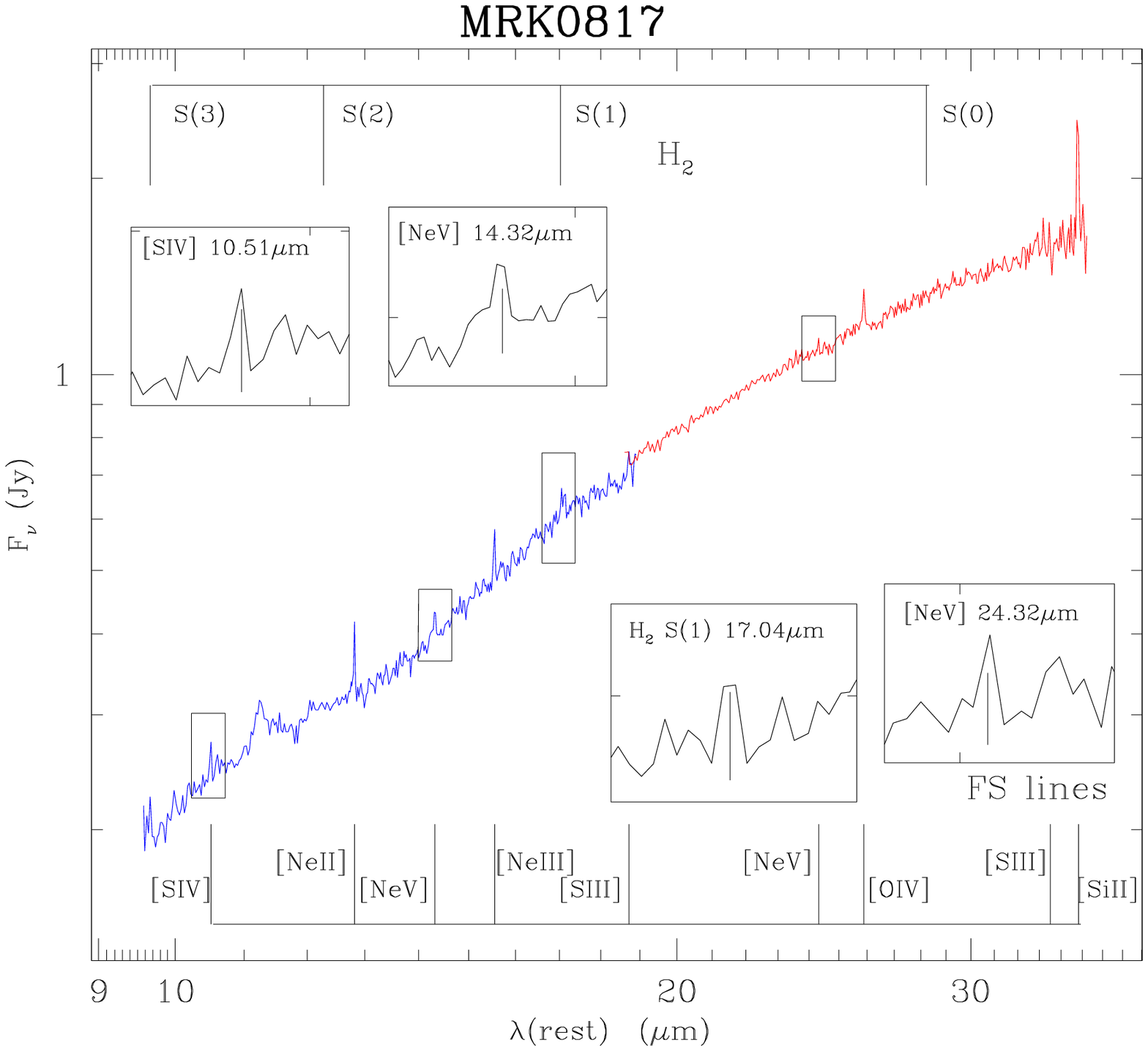}\includegraphics[width=8cm]{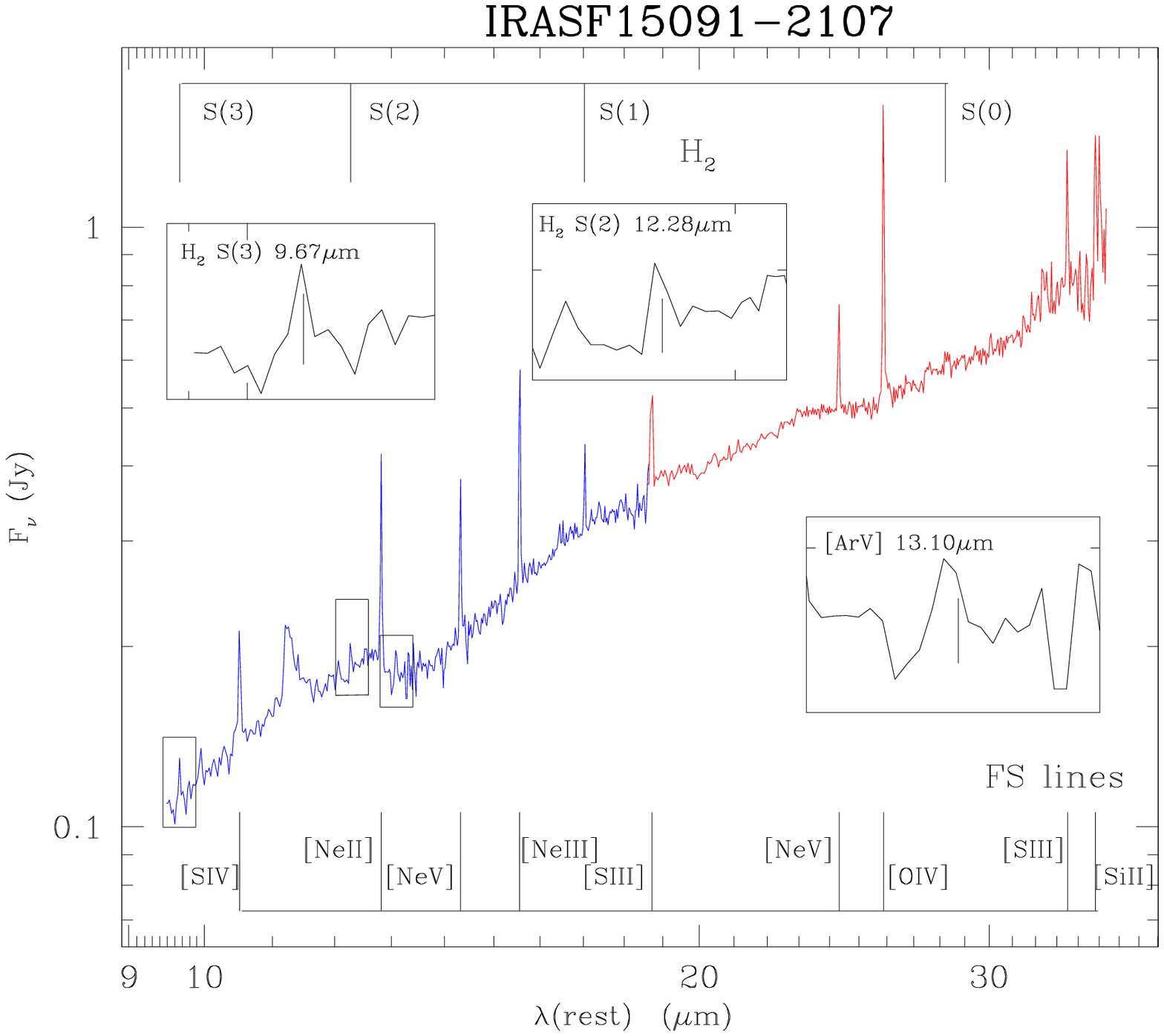}}
\end{figure}
\clearpage

\begin{figure}
\centerline{\includegraphics[width=8cm]{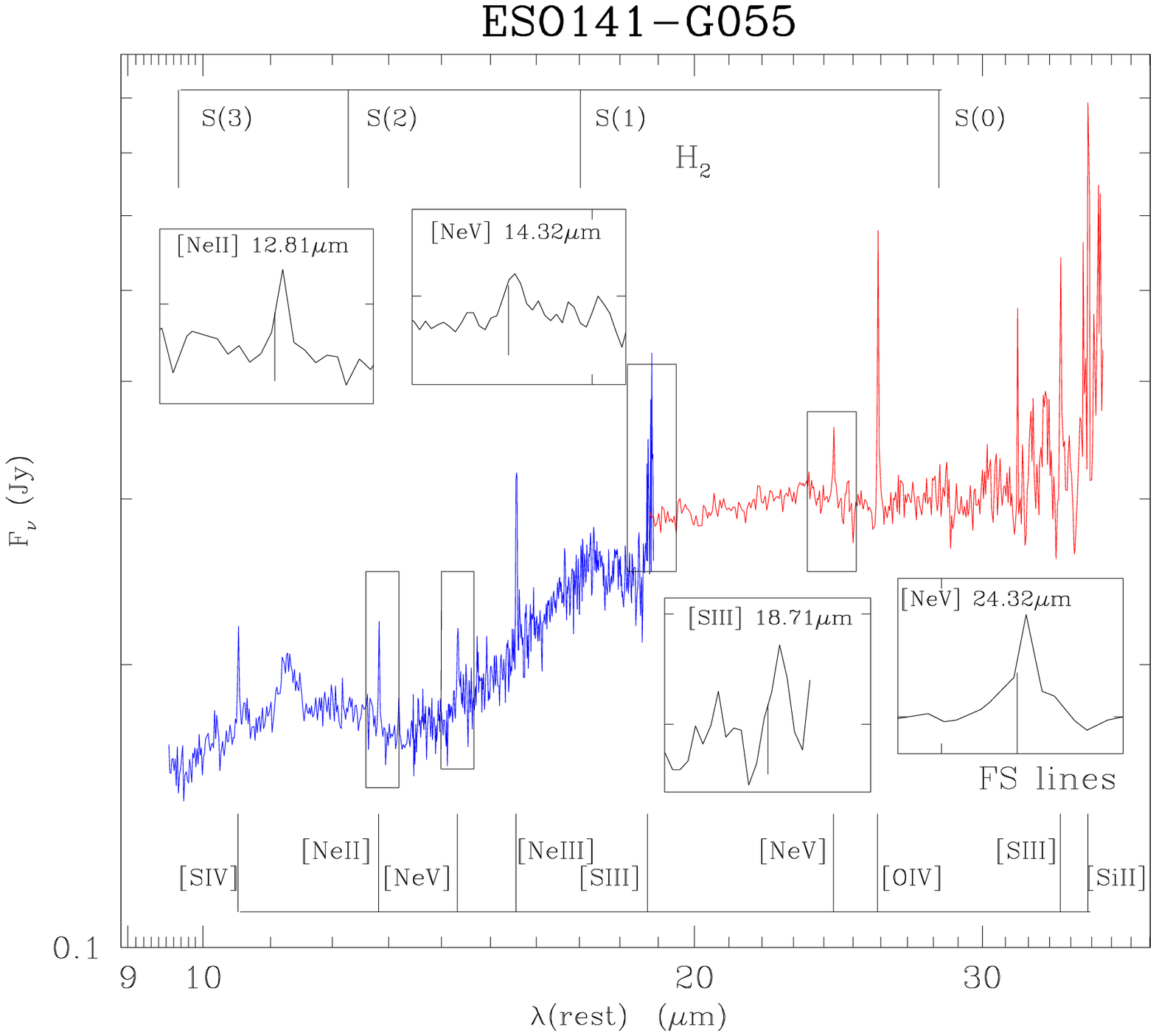}\includegraphics[width=8cm]{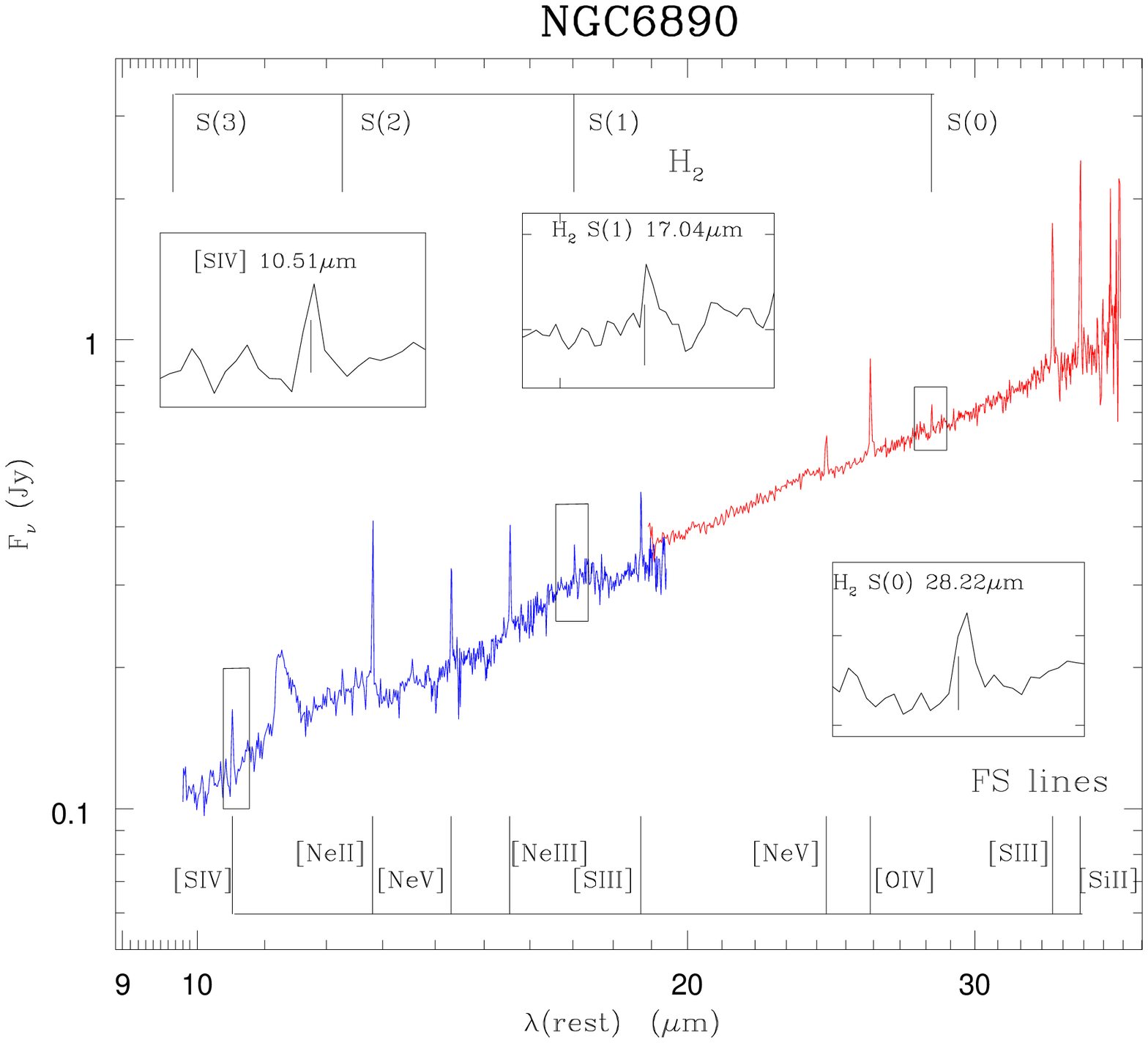}}

\centerline{\includegraphics[width=8cm]{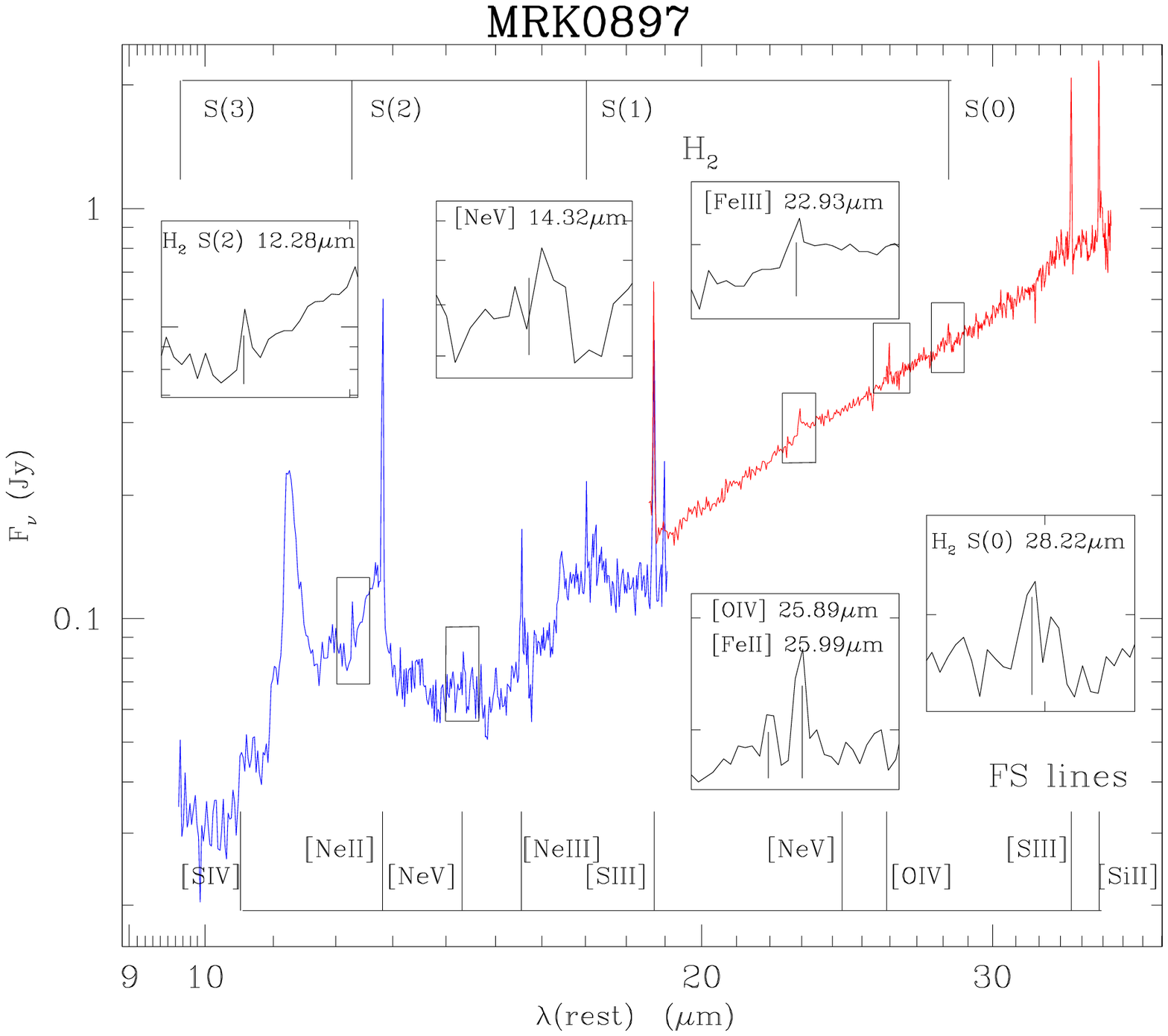}\includegraphics[width=8cm]{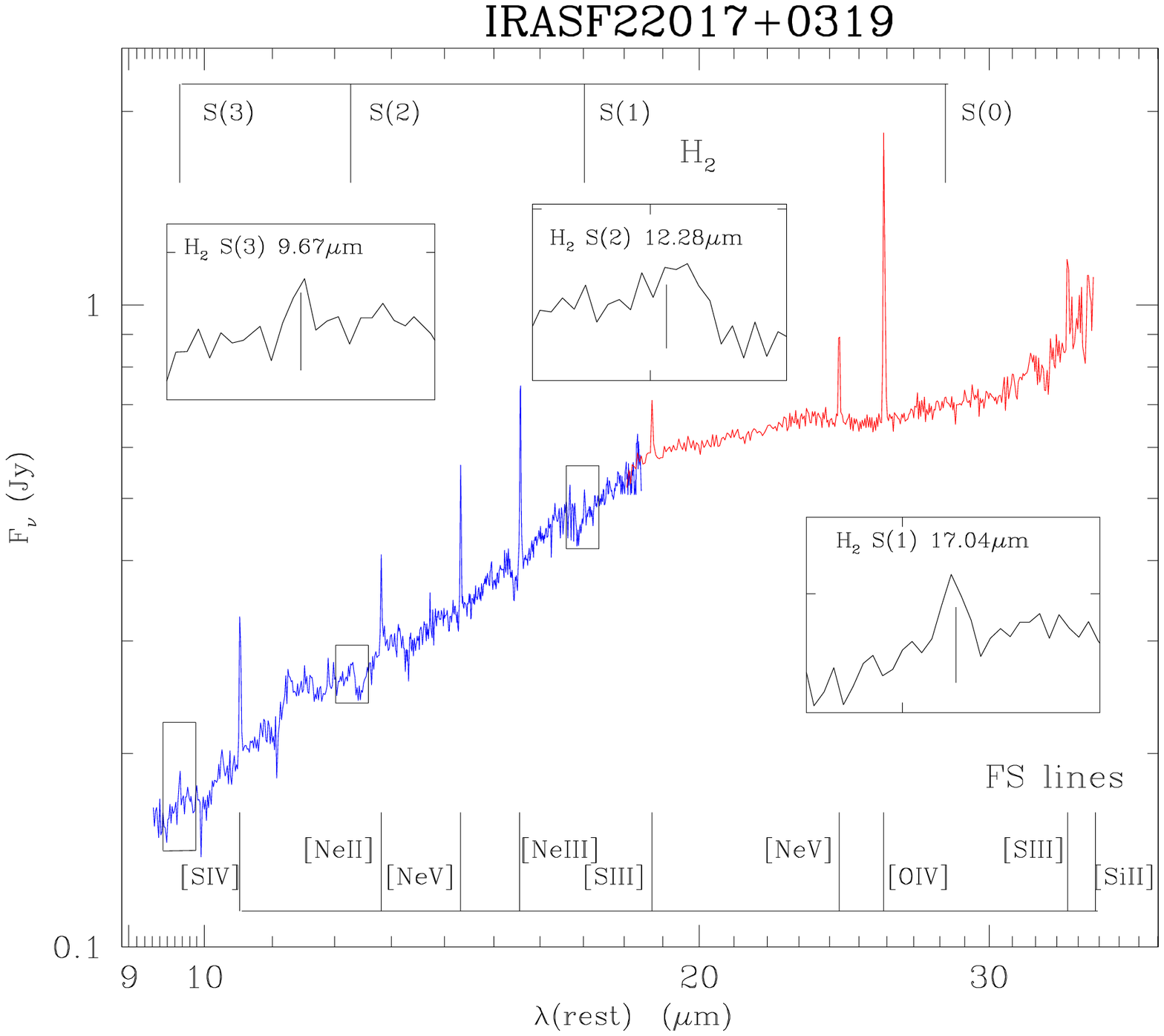}}

\centerline{\includegraphics[width=8cm]{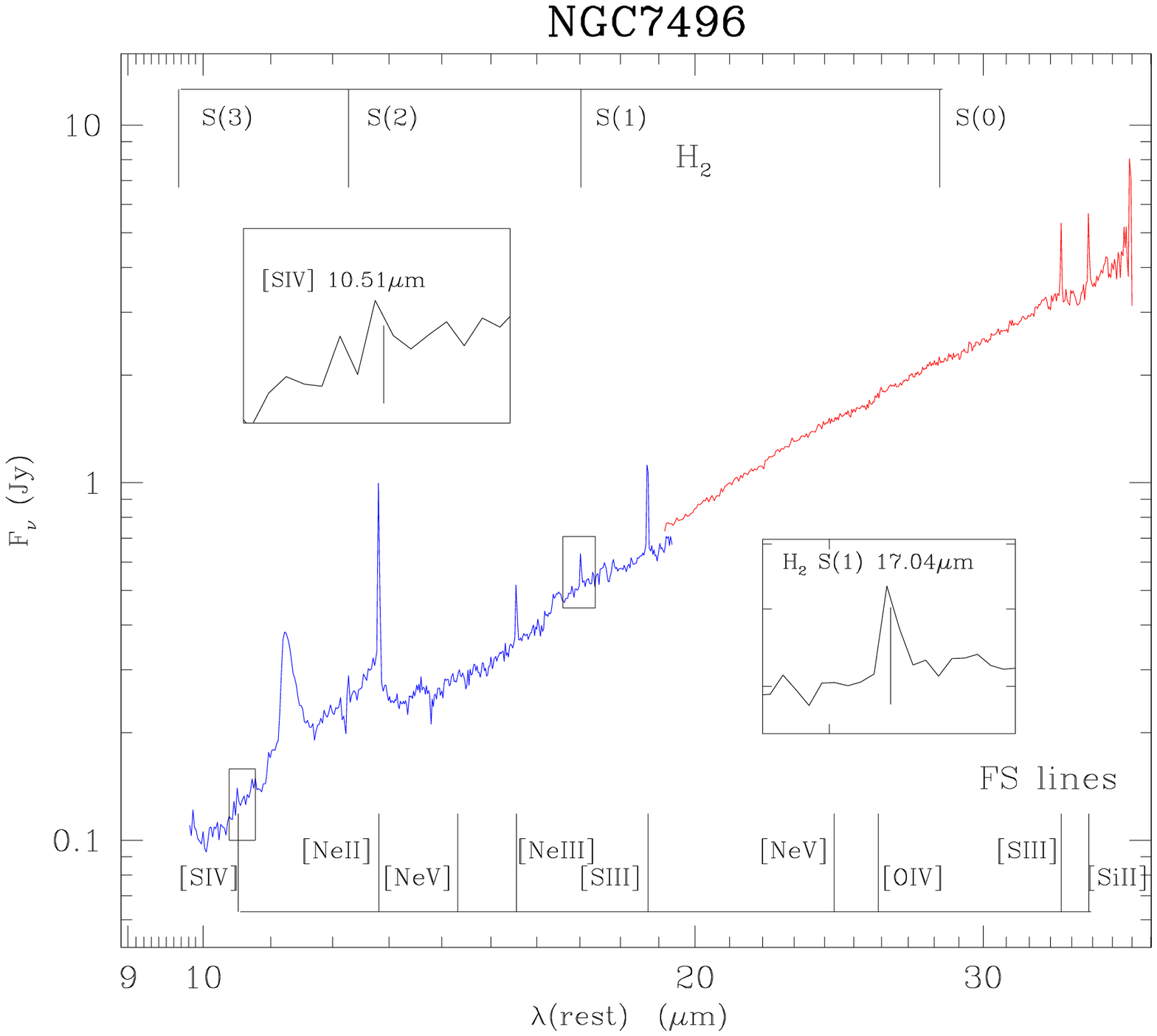}\includegraphics[width=8cm]{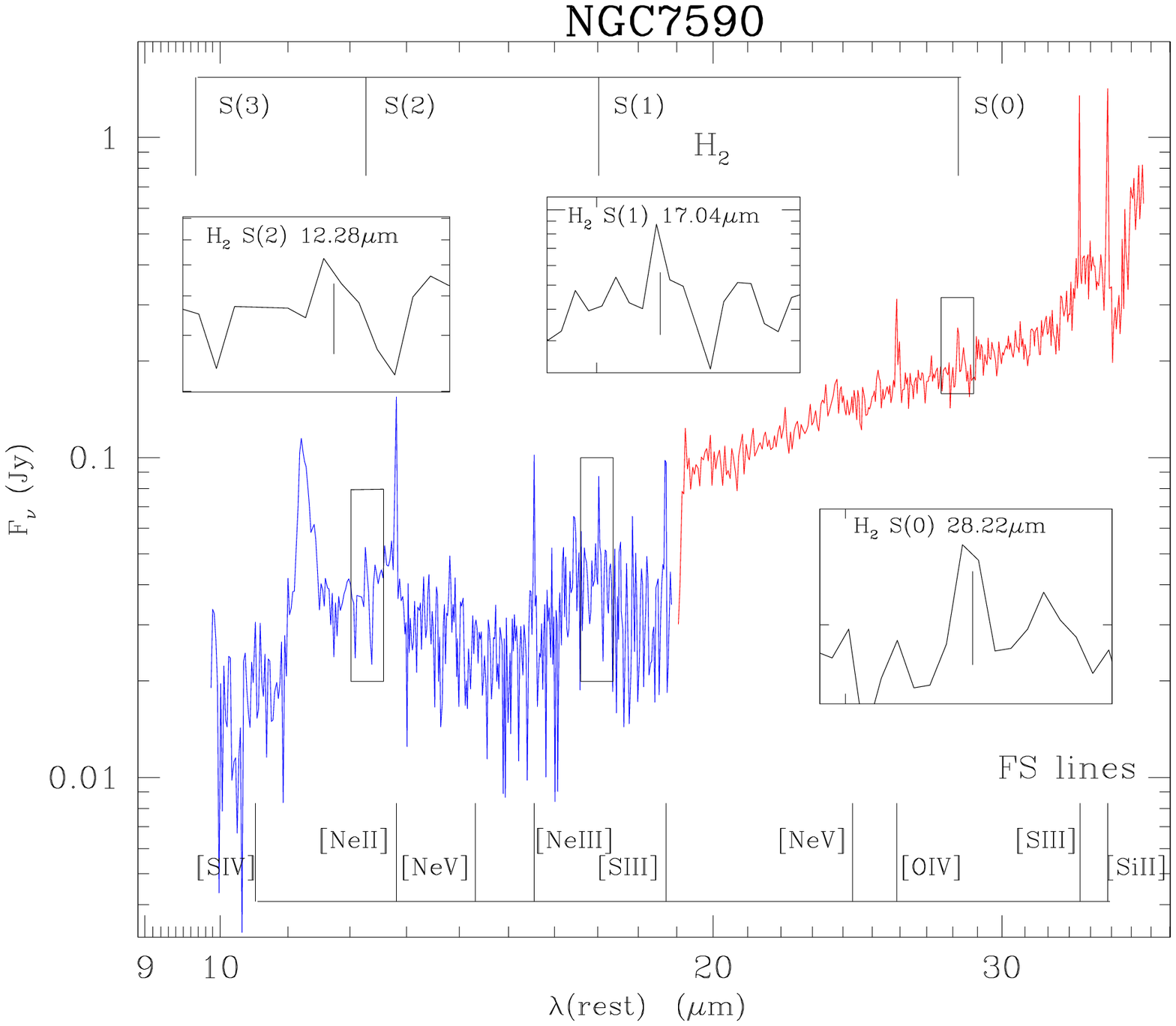}}
\end{figure}


\begin{figure}
\includegraphics[angle=0,scale=.80]{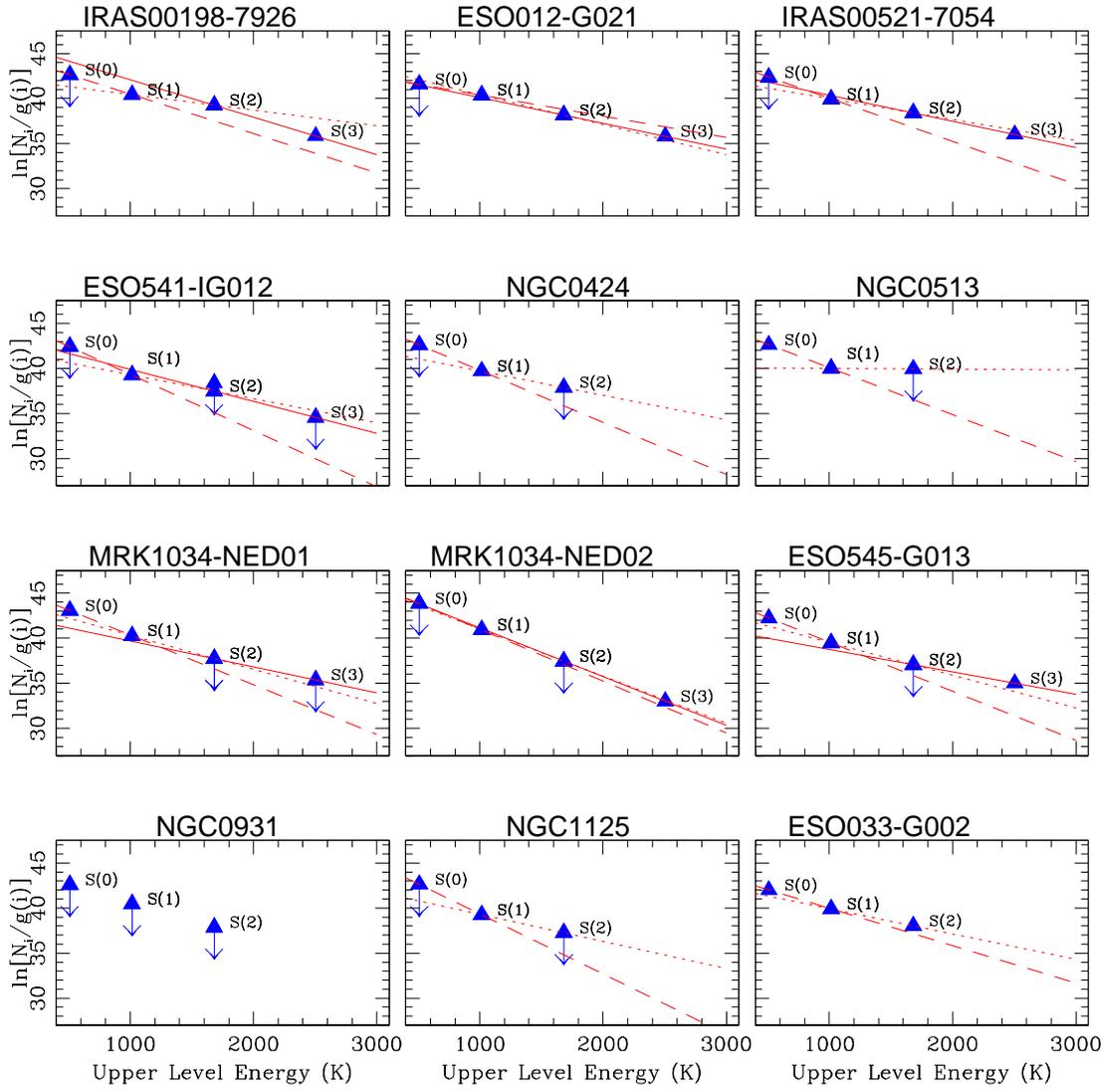}
\caption{H$_2$ excitation diagrams. For each measured line, the natural logarithm of the level population normalized to its statistical weight is plotted against the upper level  energy (in temperature units). For each pair of adjacent transitions the connecting line is shown, whose inverse value represents the gas temperature: the dashed line connects the S(0) and S(1) detections, the dotted line the S(1) and S(2), the solid line the S(2) and S(3). Upper limits have been used to obtain limiting slopes and hence limiting temperatures and masses (see text).}
\end{figure}

\begin{figure}
\epsscale{.80}
\includegraphics[angle=0,scale=.80]{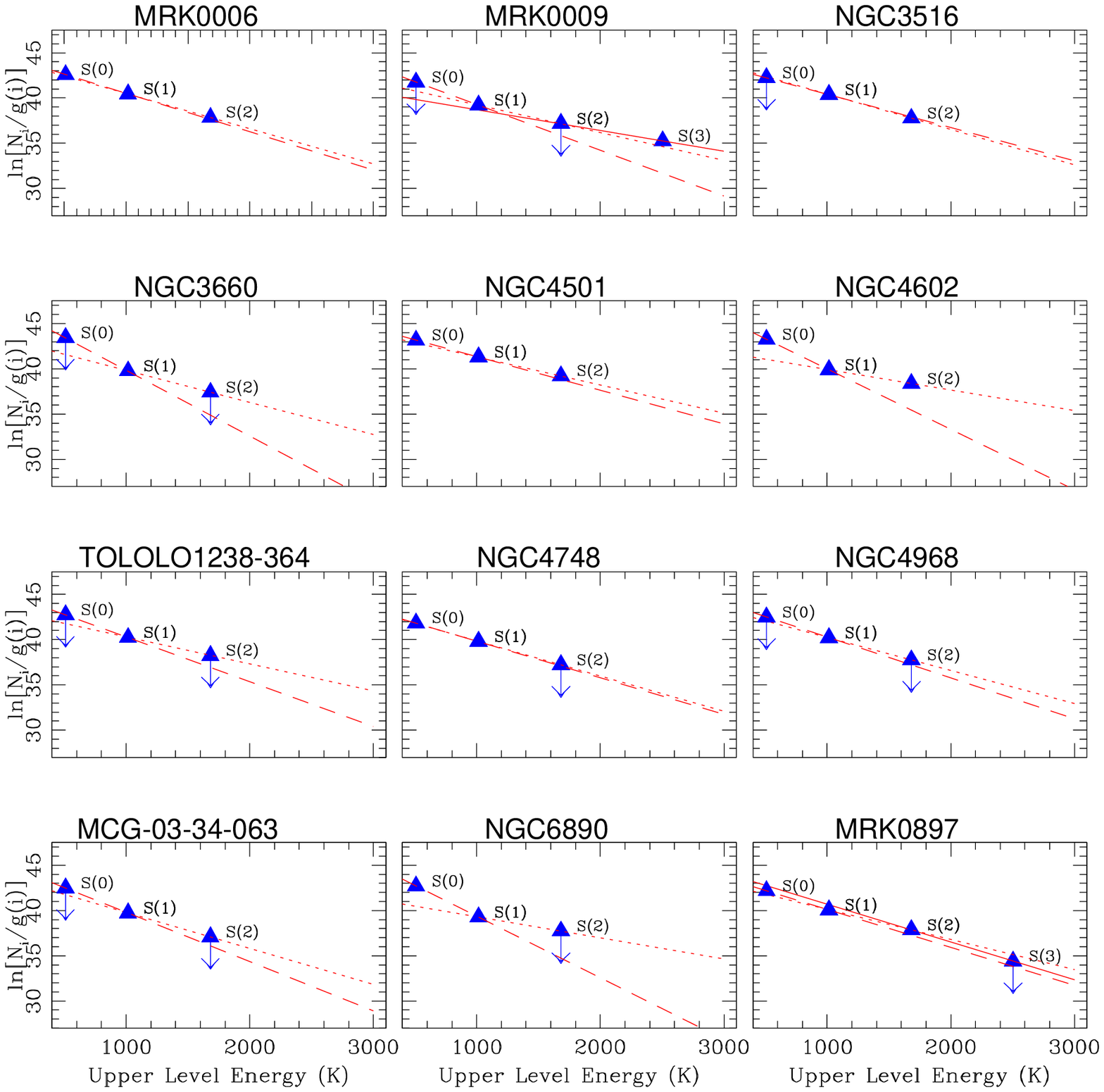}
\end{figure}

\begin{figure}
\includegraphics[angle=0,scale=.80]{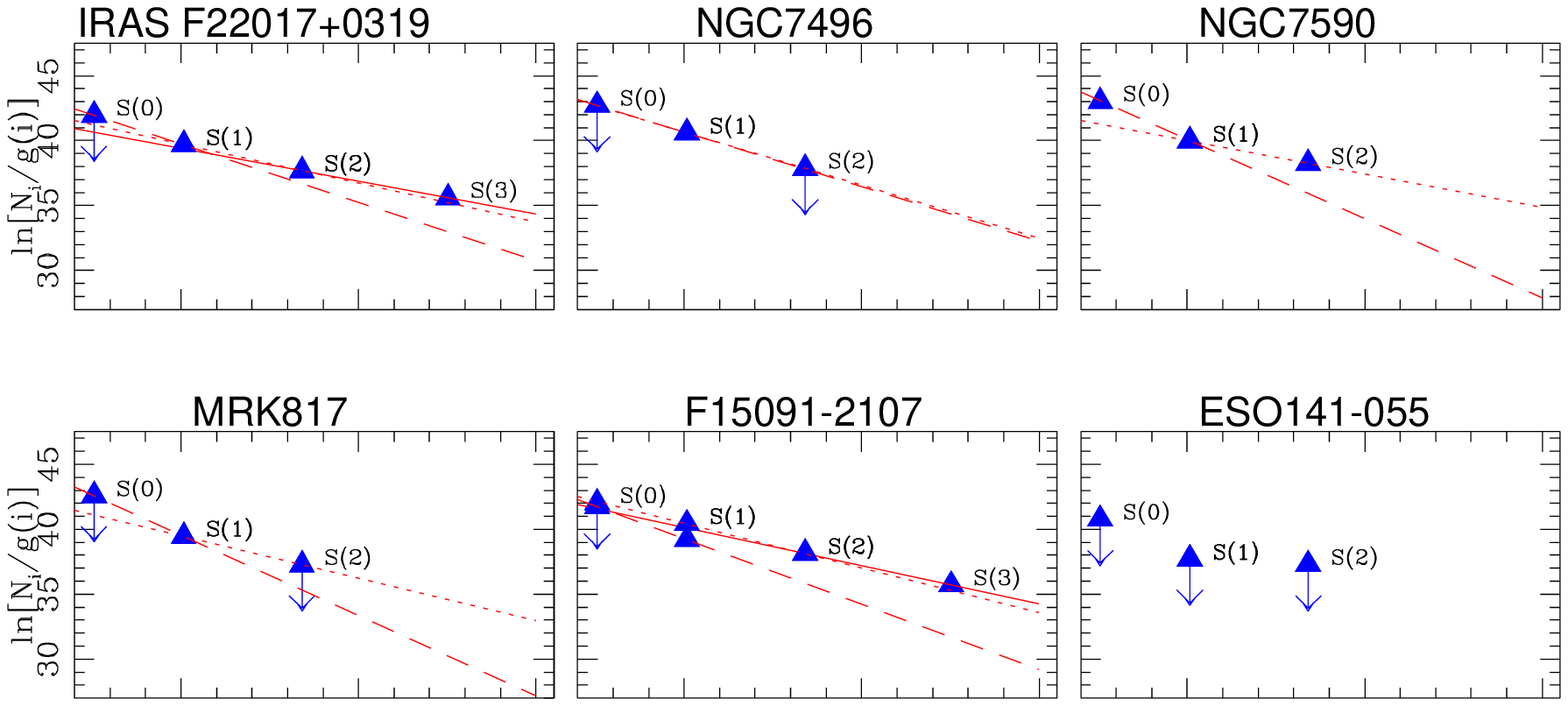}
\end{figure}



\begin{figure}
\centerline{\includegraphics[width=9cm]{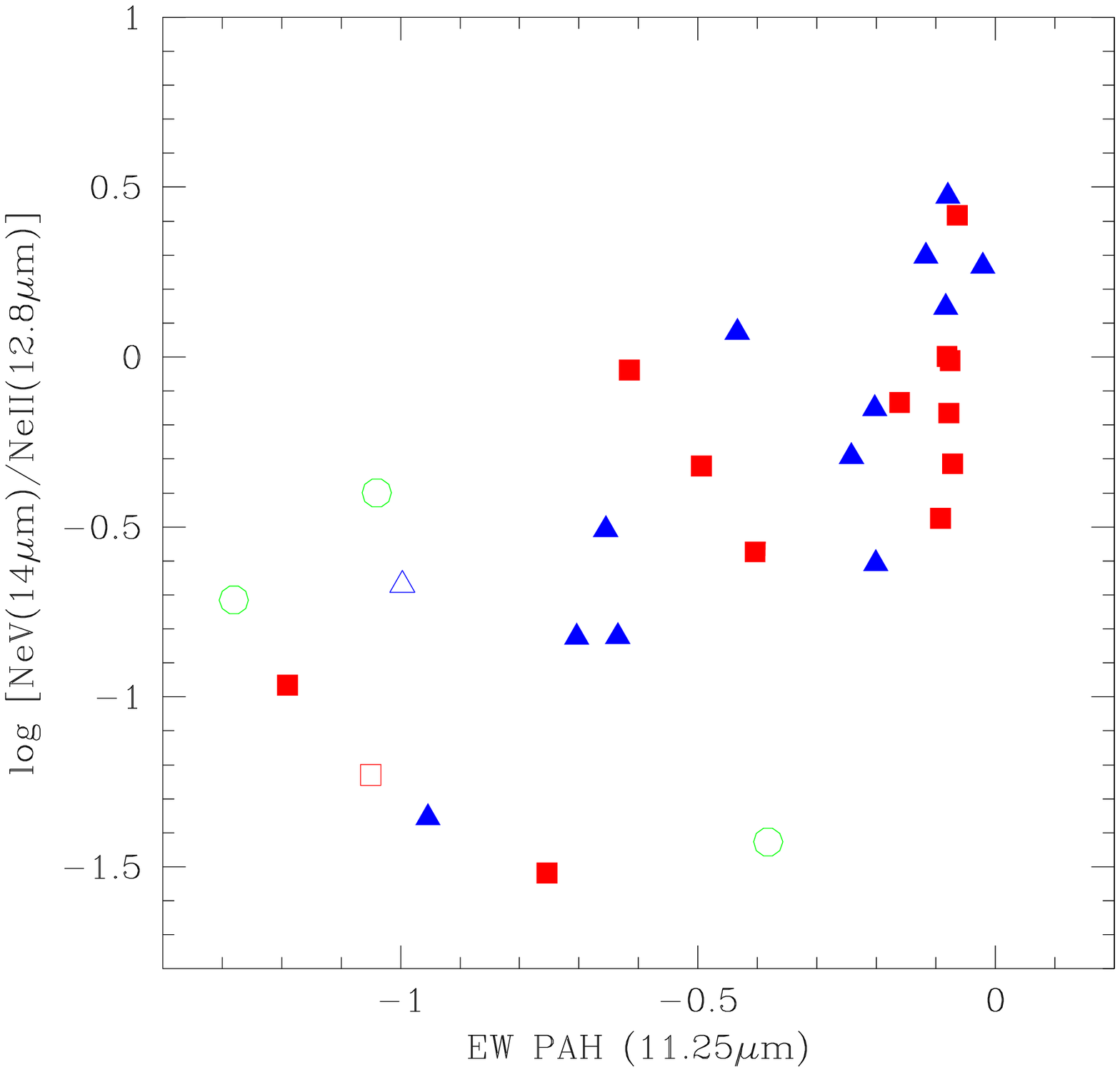}\includegraphics[width=9cm]{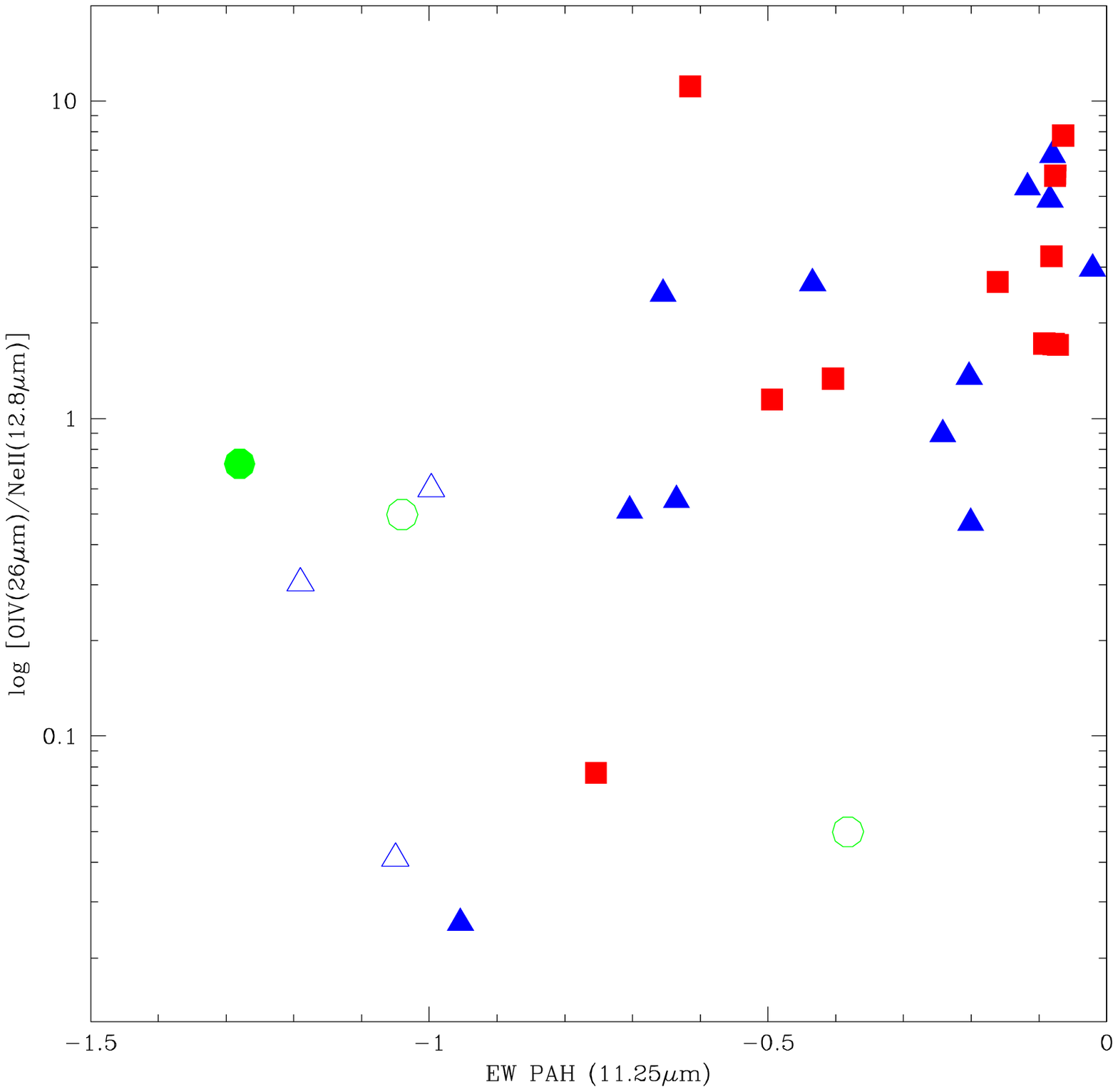}}
\caption{\textbf{a:}[NeV]14.3$\mu$m/[NeII]12.8$\mu$m line ratio versus the PAH 11.25$\mu$m 
equivalent width. Solid triangles show the good detections of Seyfert type 2 galaxies and solid squares 
of Seyfert 1's, open triangles and open squares show upper limits of Seyfert 2's and 1's, respectively. 
The circles indicate MCG-3-34-63, NGC7496 and NGC7590, which are not classified as a Seyfert 
galaxies (see text).   
\textbf{b:}
[OIV]14.3$\mu$m/[NeII]12.8$\mu$m line ratio versus the PAH 11.25$\mu$m equivalent width. 
Symbols and lines as in the previous figure.}  
\end{figure}


\begin{figure}
\centerline{\includegraphics[width=9cm]{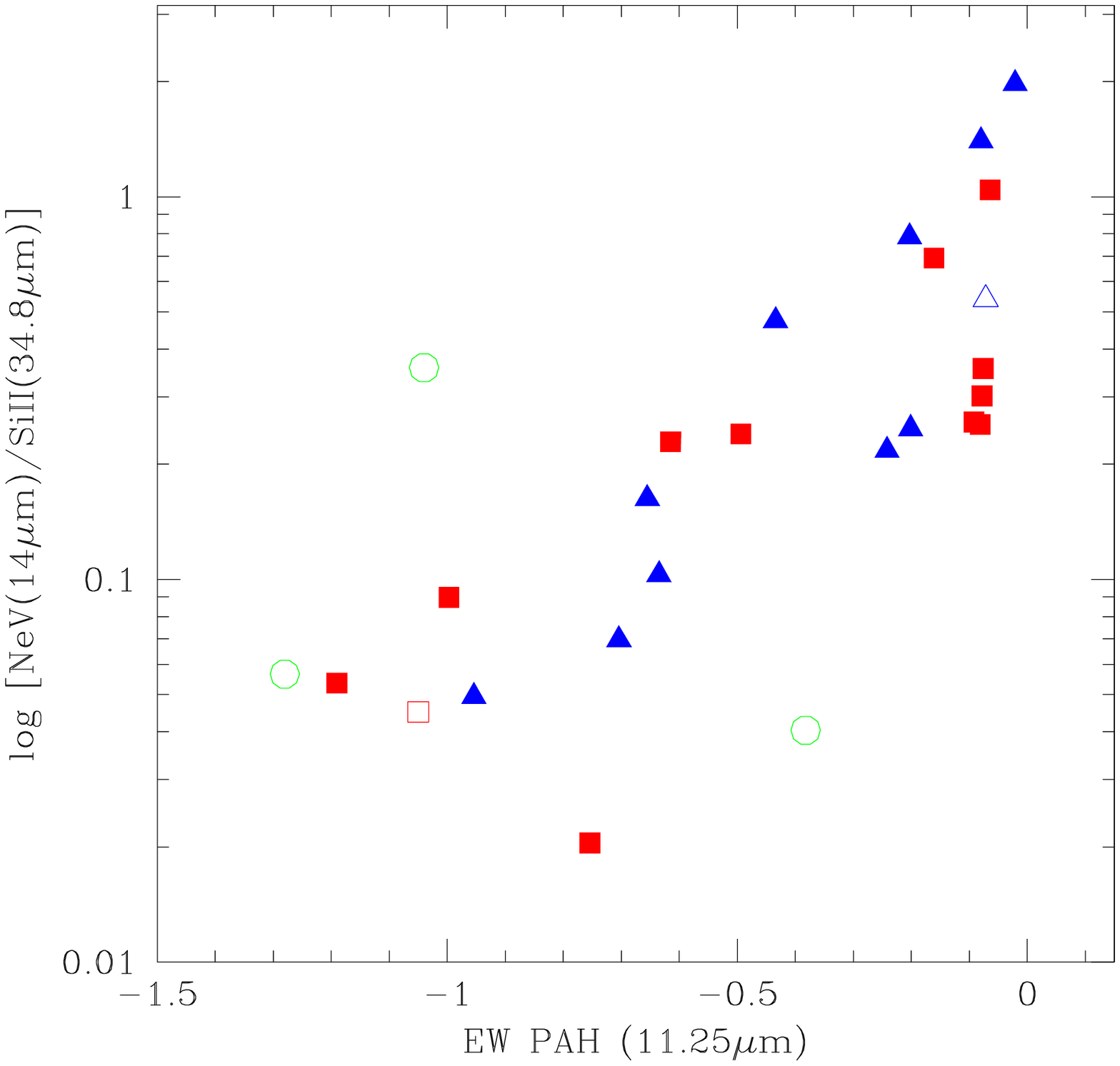}\includegraphics[width=9cm]{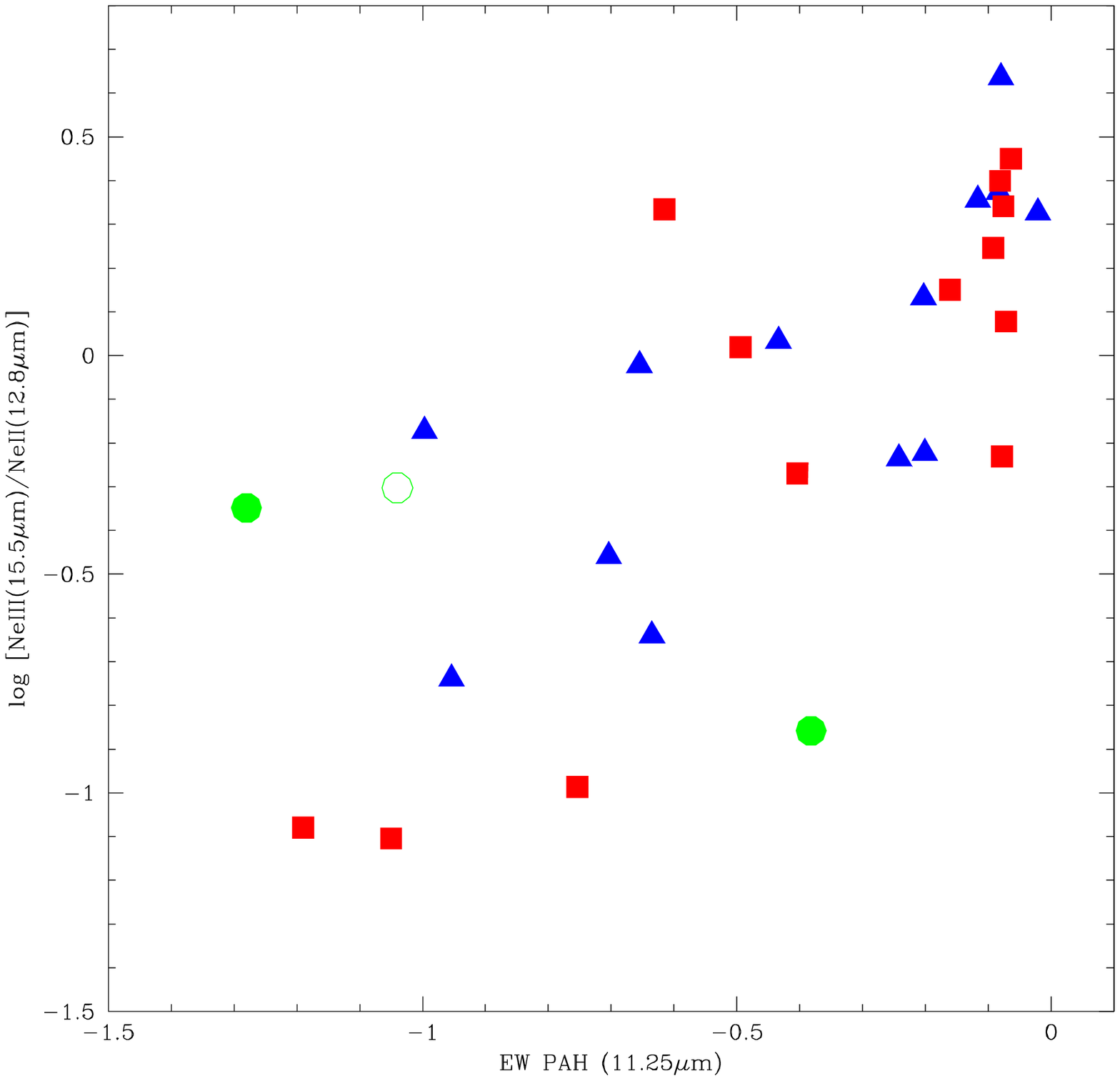}}
\caption{\textbf{a:}[NeV]14.3$\mu$m/[SiII]34.8$\mu$m line ratio versus the PAH 11.25$\mu$m equivalent width. 
 Symbols and lines as in the previous figure. 
 \textbf{b:} 
 [NeIII]15.5$\mu$m/[NeII]12.8$\mu$m line ratio versus the PAH 11.25$\mu$m 
 equivalent width. Symbols and lines as in the previous figure.}
\end{figure}


\begin{figure}
\centerline{\includegraphics[width=9cm]{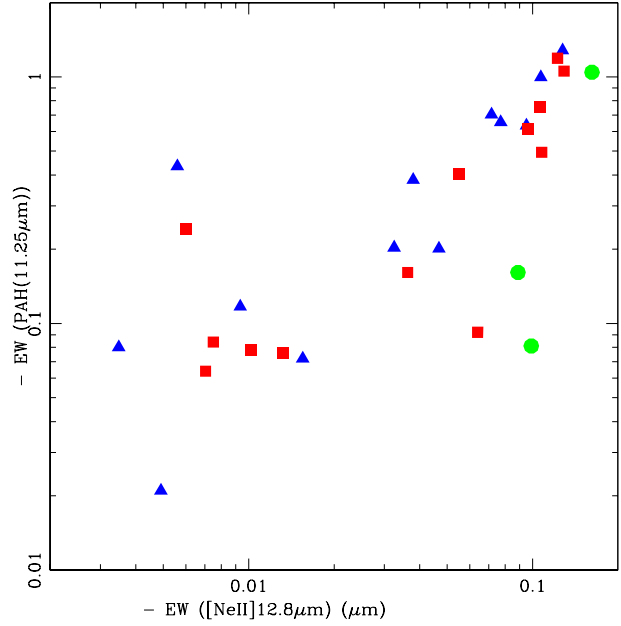}}
\caption{[NeII]12.8$\mu$m  equivalent width versus the PAH equivalent width. 
Symbols in the previous figure. We note that, 
for graphical reasons, in this diagram and in the following ones in which the equivalent widths, 
covering a large range, are shown, the logarithm of the inverse of the actual 
equivalent width is plotted.}
\end{figure}


\begin{figure}
\centerline{\includegraphics[width=9cm]{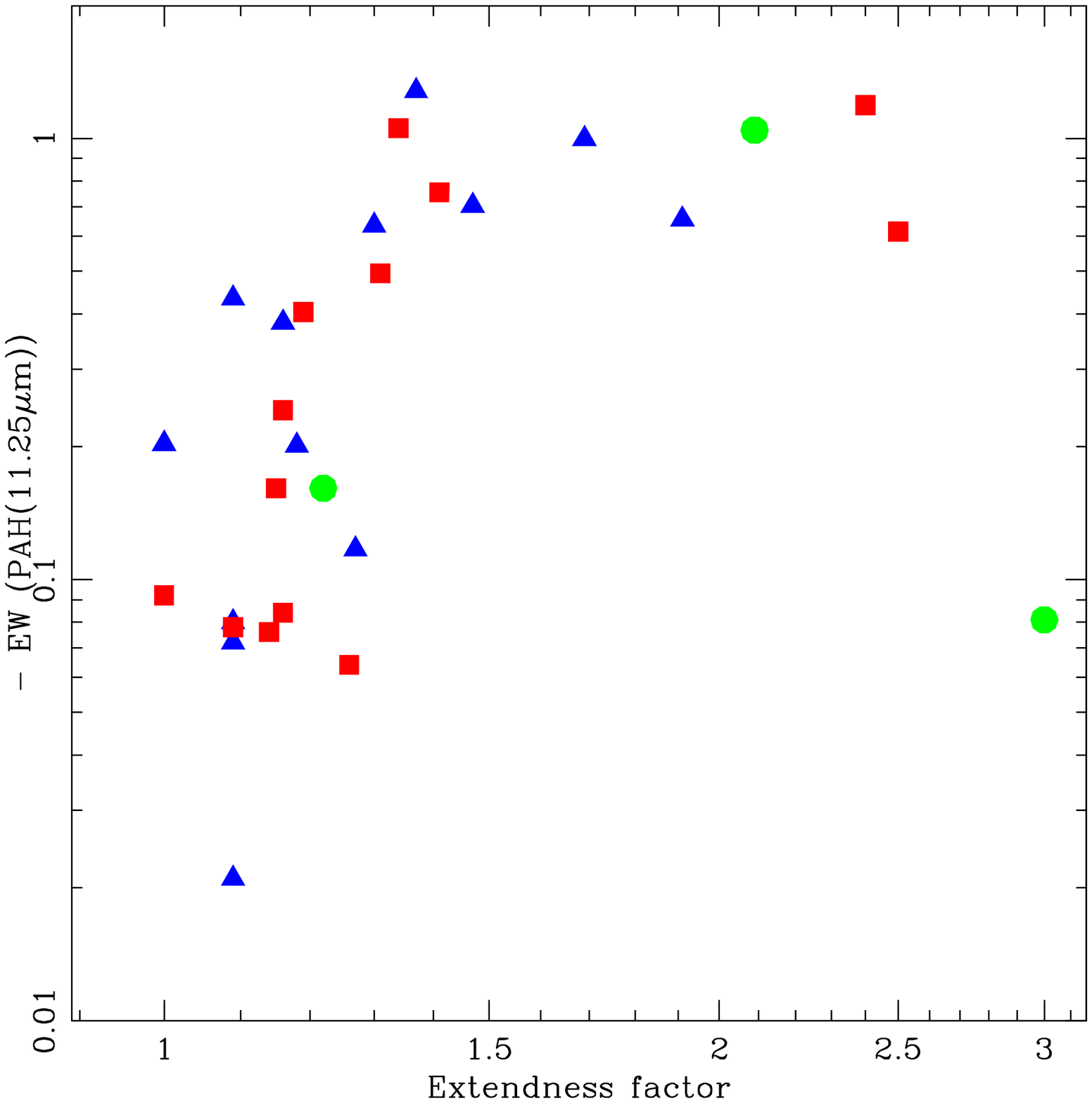}\includegraphics[width=9cm]{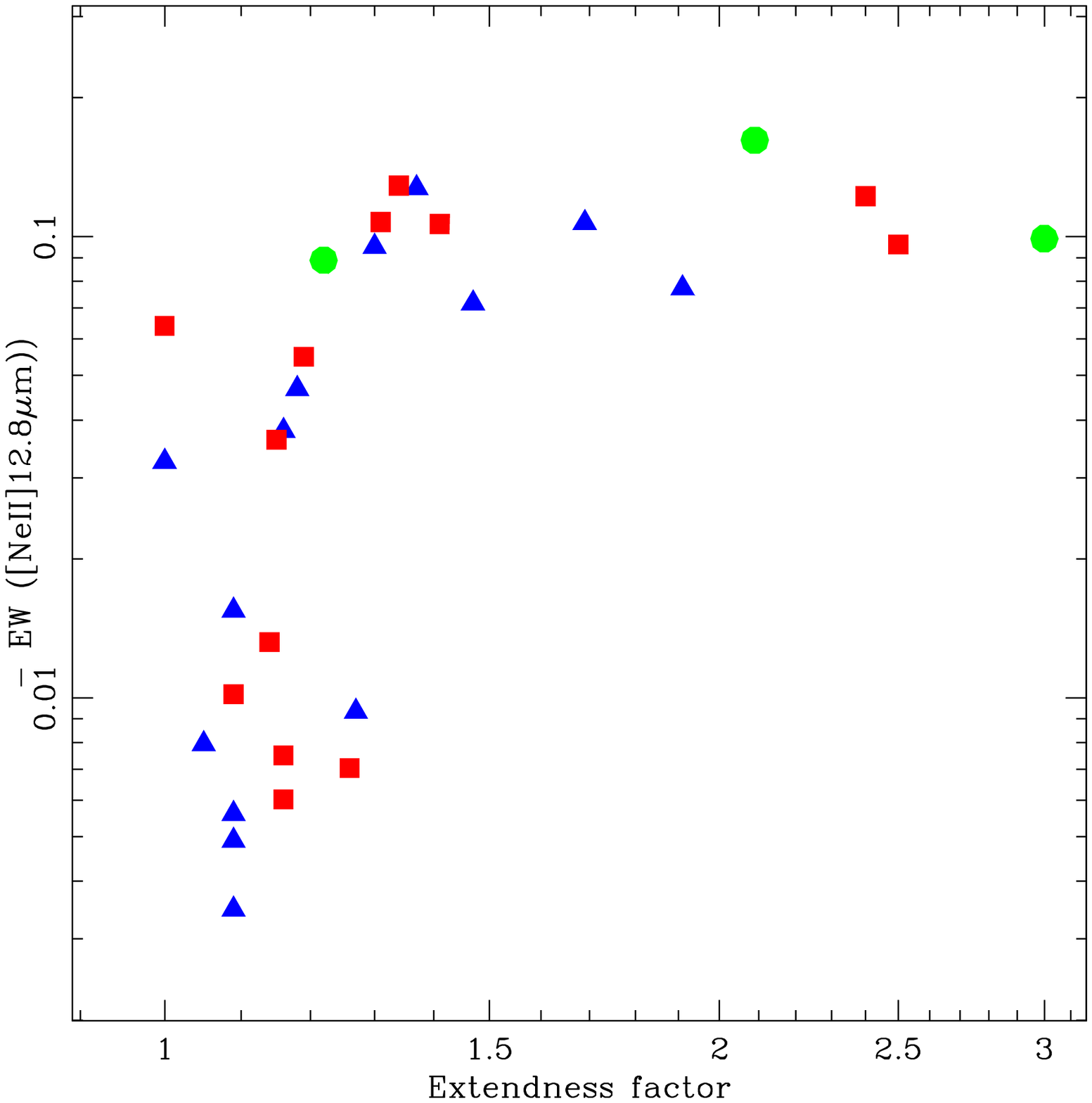}}
\caption{PAH11.25$\mu$m and [NeII]12.8$\mu$m equivalent widths versus extendness parameter.
Extendness of 1.0 corresponds to completely unresolved (point-like) 19$\mu$m continuum,
which is therefore completely AGN-dominated. In both figures the non-linear effects of
extended emission from the host galaxy become important for extendness parameters above 1.3. 
Symbols as in the previous figure. }
\end{figure}


\begin{figure}
\centerline{\includegraphics[width=9cm]{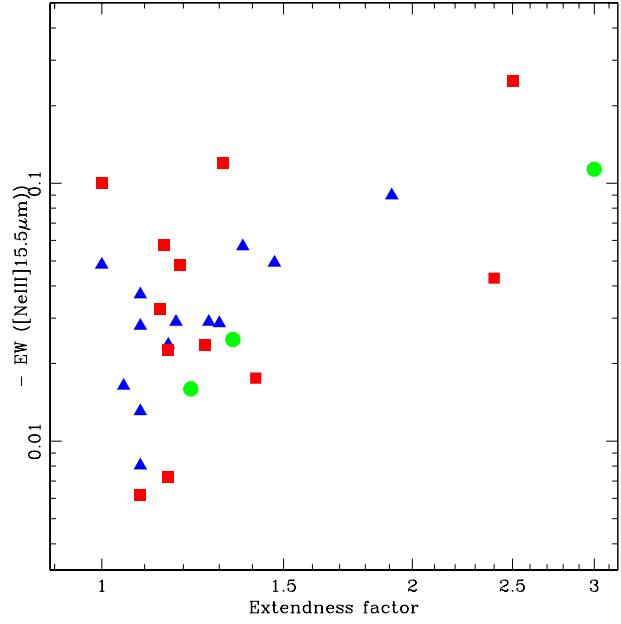}}
\caption{[NeIII]15.5$\mu$m line equivalent width versus extendness parameter. 
Symbols as in the previous figure. }
\end{figure}



\begin{figure}
\centerline{\includegraphics[width=9cm]{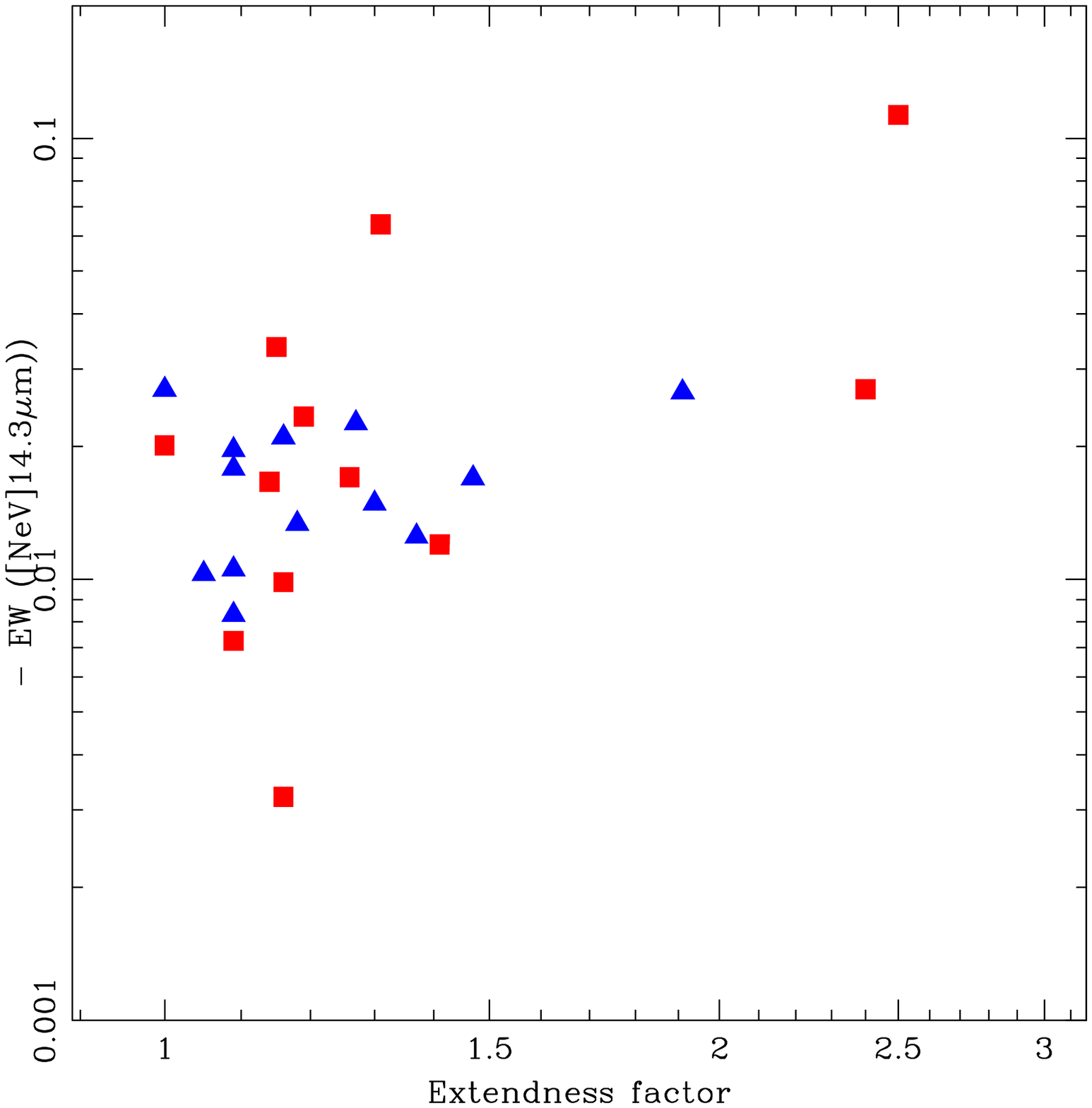}\includegraphics[width=9cm]{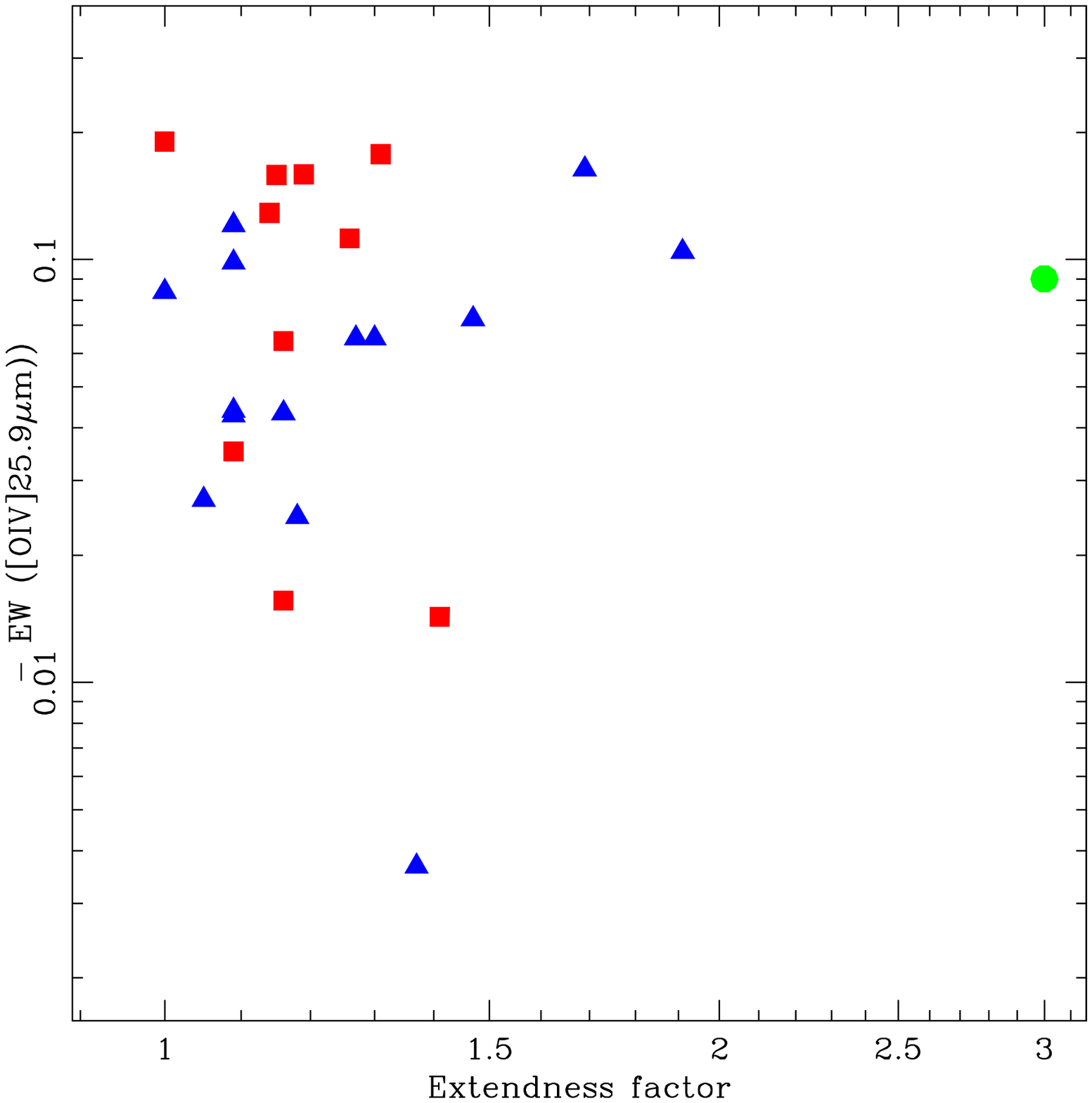}}
\caption{[NeV]14.3$\mu$m  and [OIV]25.9$\mu$m line equivalent widths versus extendness parameter.
As expected, there are no significant correlations, because the equivalent width
is the ratio of high-ionization emission to the hot dust continuum, and both of
these arise mostly from the AGN. Symbols as in the previous figure. }
\end{figure}


\begin{figure}
\centerline{\includegraphics[width=9cm]{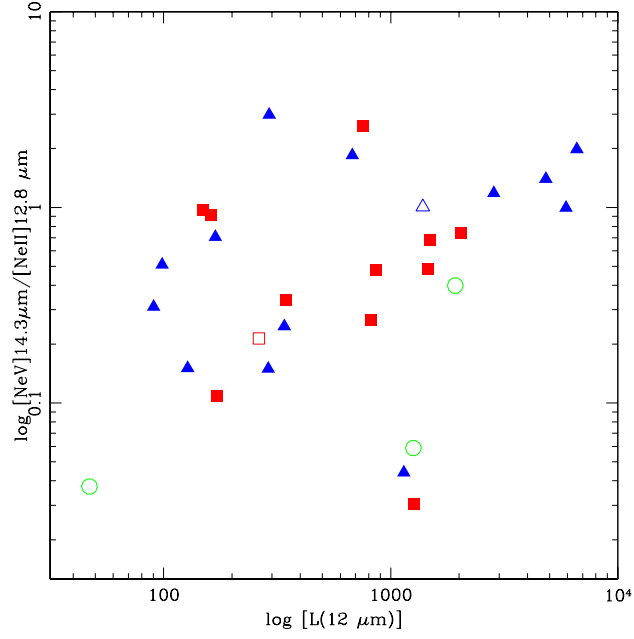}}
\caption{[NeV]14.3$\mu$m/[NeII]12.8$\mu$m  line ratio versus 12$\mu$m luminosity. }
\end{figure}


\begin{figure}
\centerline{\includegraphics[width=9cm]{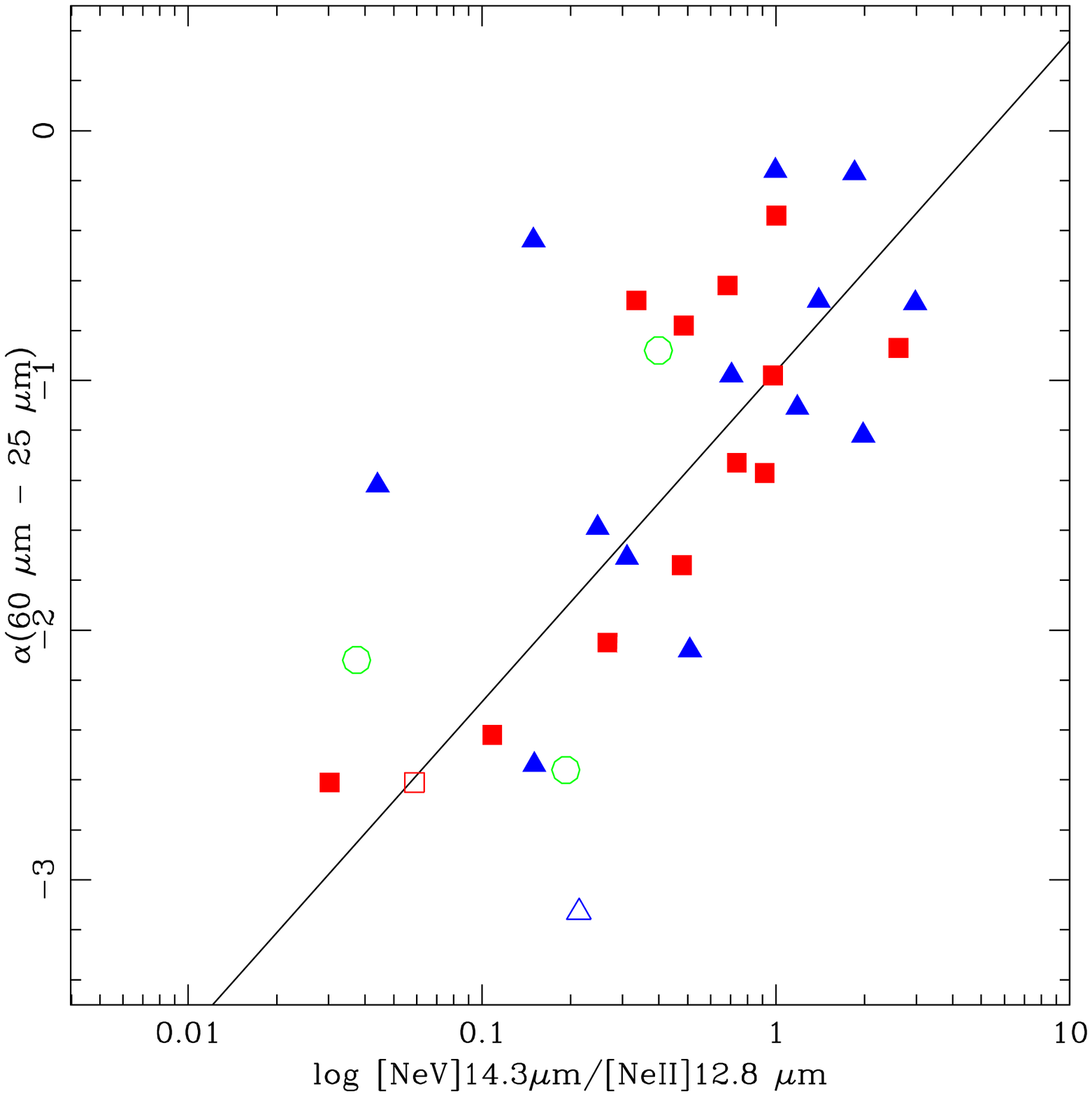}\includegraphics[width=9cm]{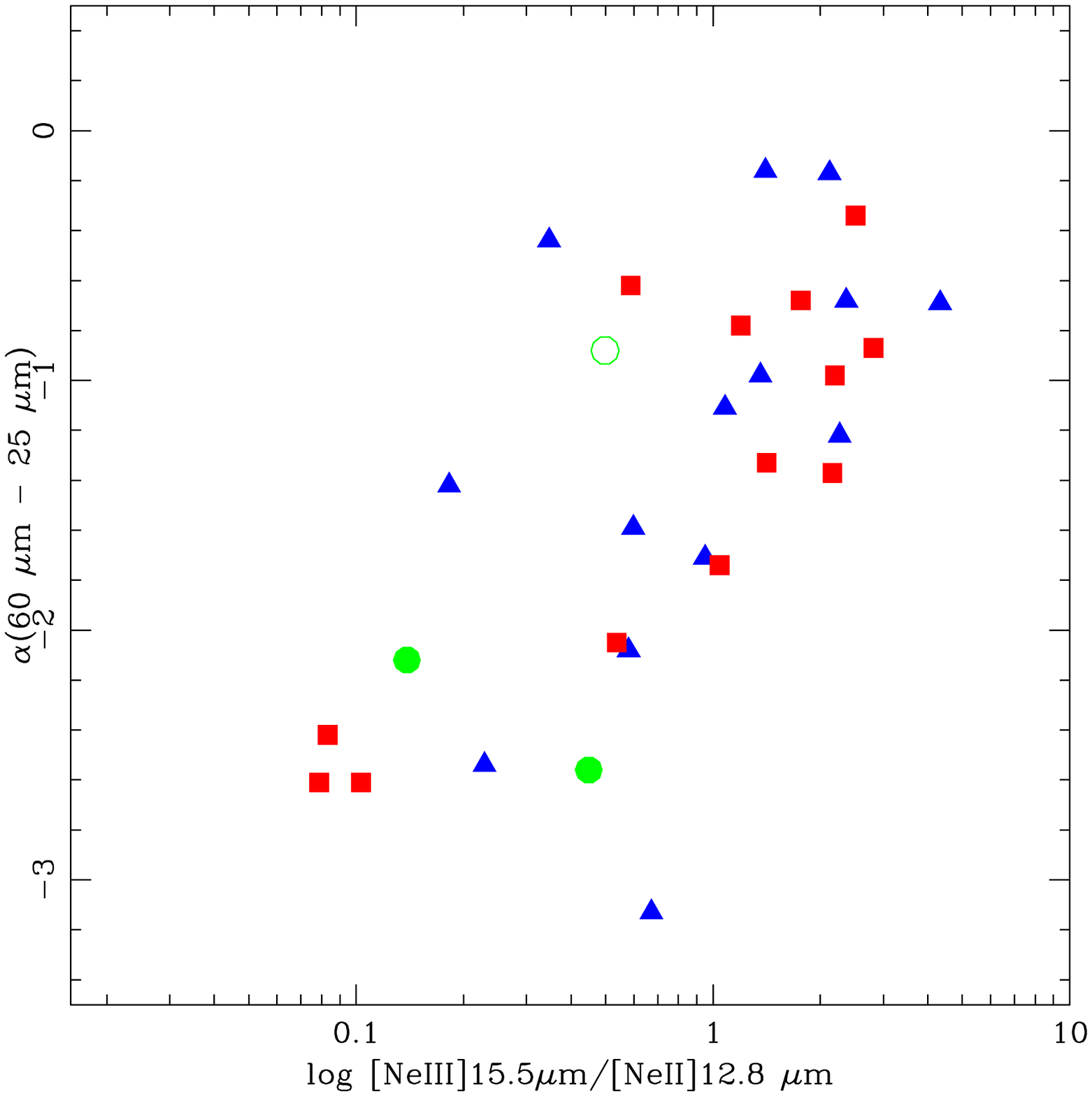}}
\caption{\textbf{a:} [NeV]14.3$\mu$m/[NeII]12.8$\mu$m  line ratio versus the 25-60 $\mu$m spectral index. A correlation is present for all Seyfert's data, with a slope of 1.33 using the weighted least squares fit and 1.32 $\pm$ 0.77 (1 $\sigma$) using the bootstrap method. \textbf{b:} [NeIII]15.5$\mu$m/[NeII]12.8$\mu$m line ratio versus the 25-60 $\mu$m spectral index.
 }
\end{figure}


\begin{figure}
\centerline{\includegraphics[width=9cm]{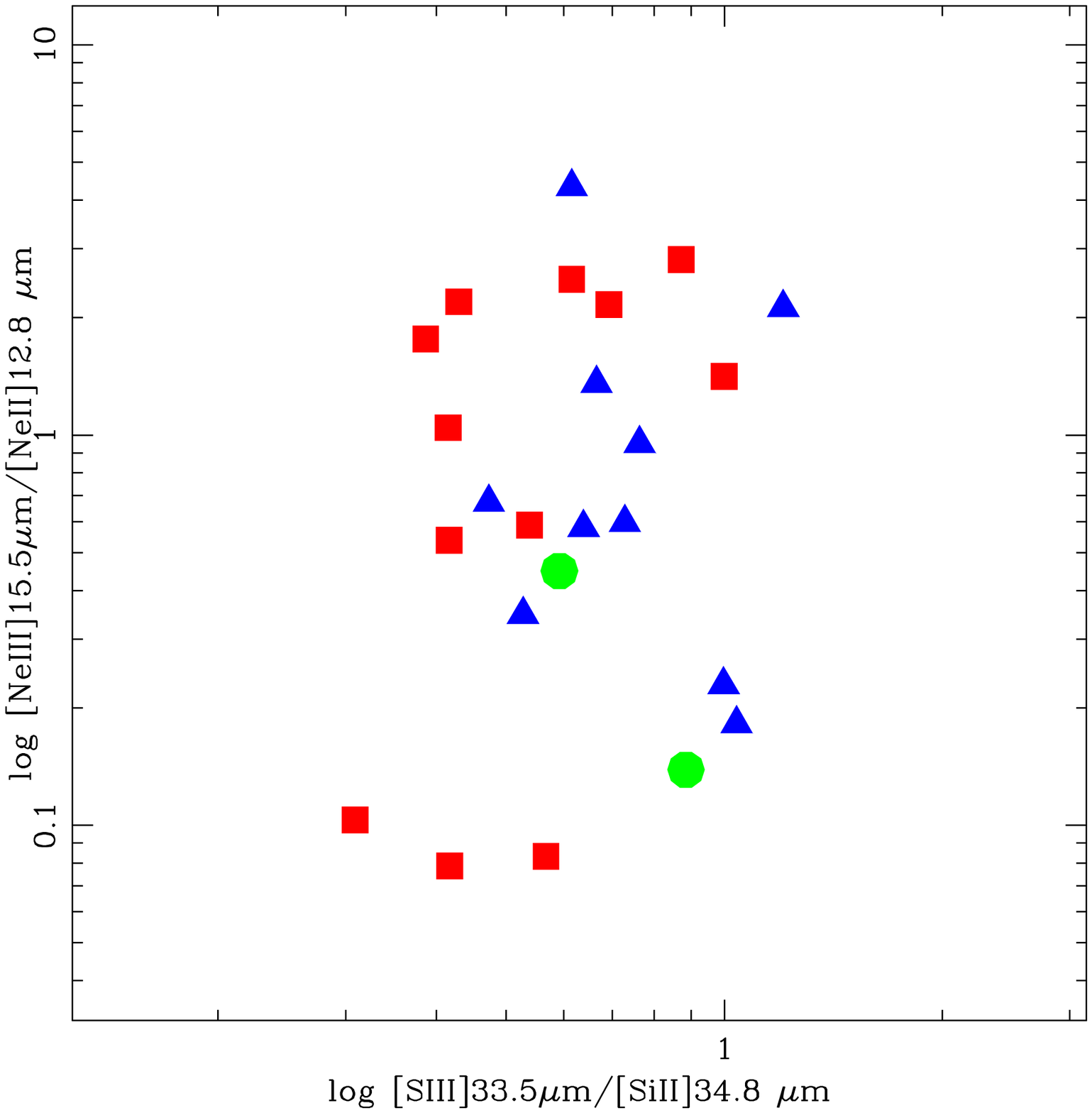}\includegraphics[width=9cm]{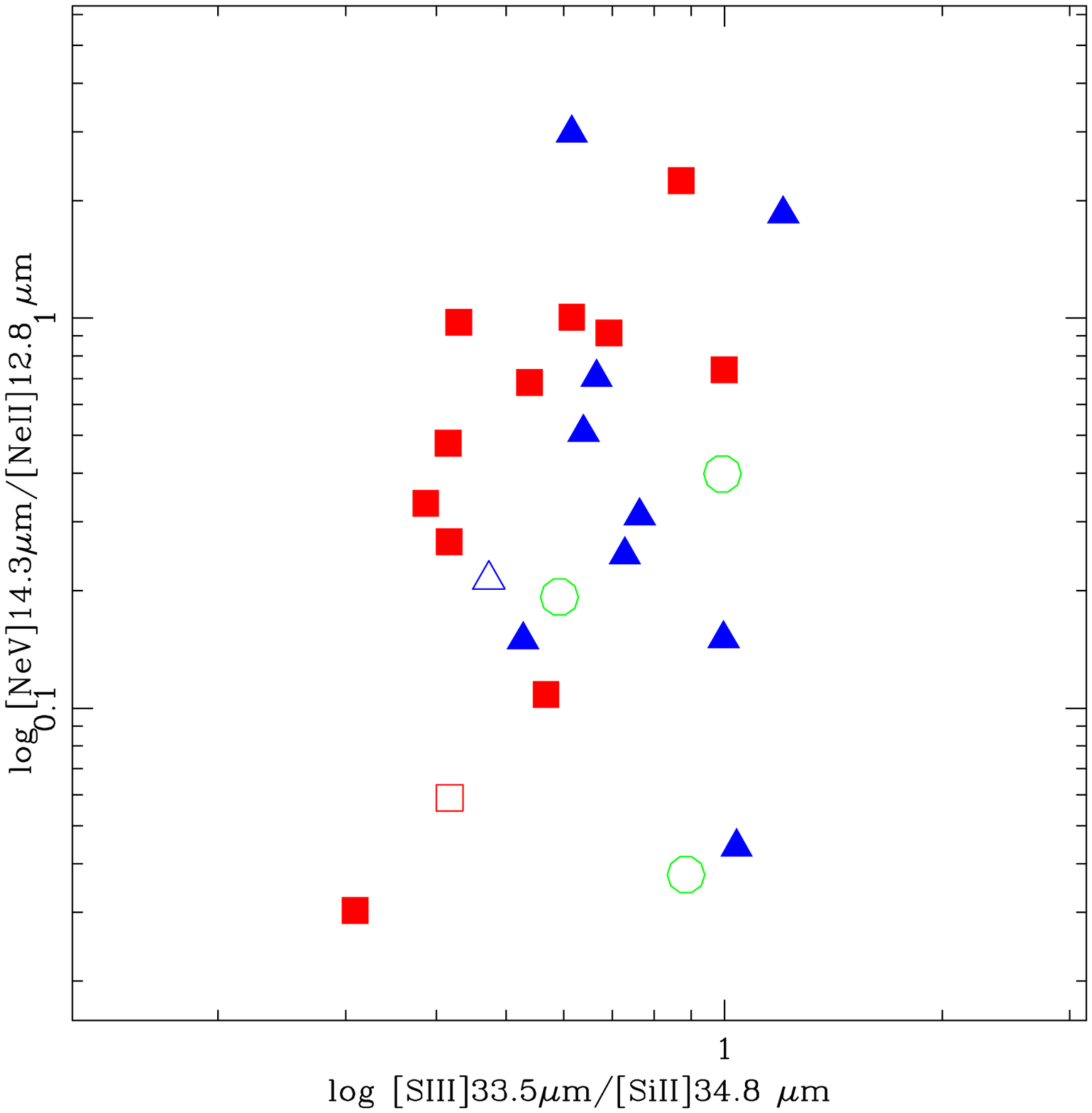}}
\caption{\textbf{a:}[NeV]14.3$\mu$m/[NeII]12.8$\mu$m line ratio versus the 
[SIII]33.5$\mu$m/[SiII]34.8$\mu$m line ratio. Symbols  as in the previous figures. 
\textbf{b:} [NeIII]15.5$\mu$m/[NeII]12.8$\mu$m line ratio versus the 
[SIII]33.5$\mu$m/[SiII]34.8$\mu$m line ratio. Symbols  as in the previous figures. 
Although uncorrelated with ionization level,
the [SIII]/[SiII] ratio does roughly separate the Seyfert 1's and 2's, although with substantial
overlap. }
\end{figure}



\begin{figure}
\centerline{\includegraphics[width=9cm]{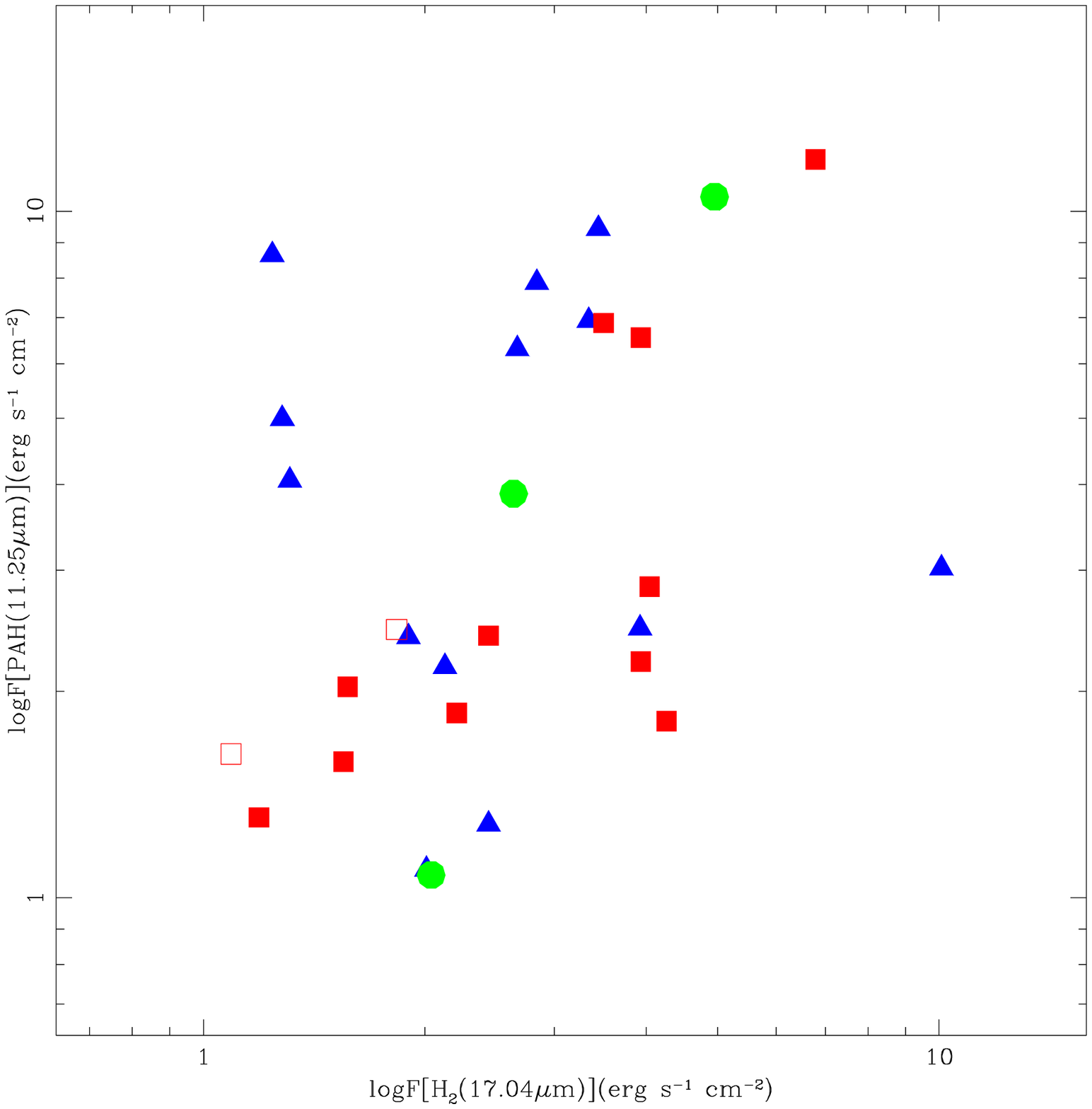}\includegraphics[width=9cm]{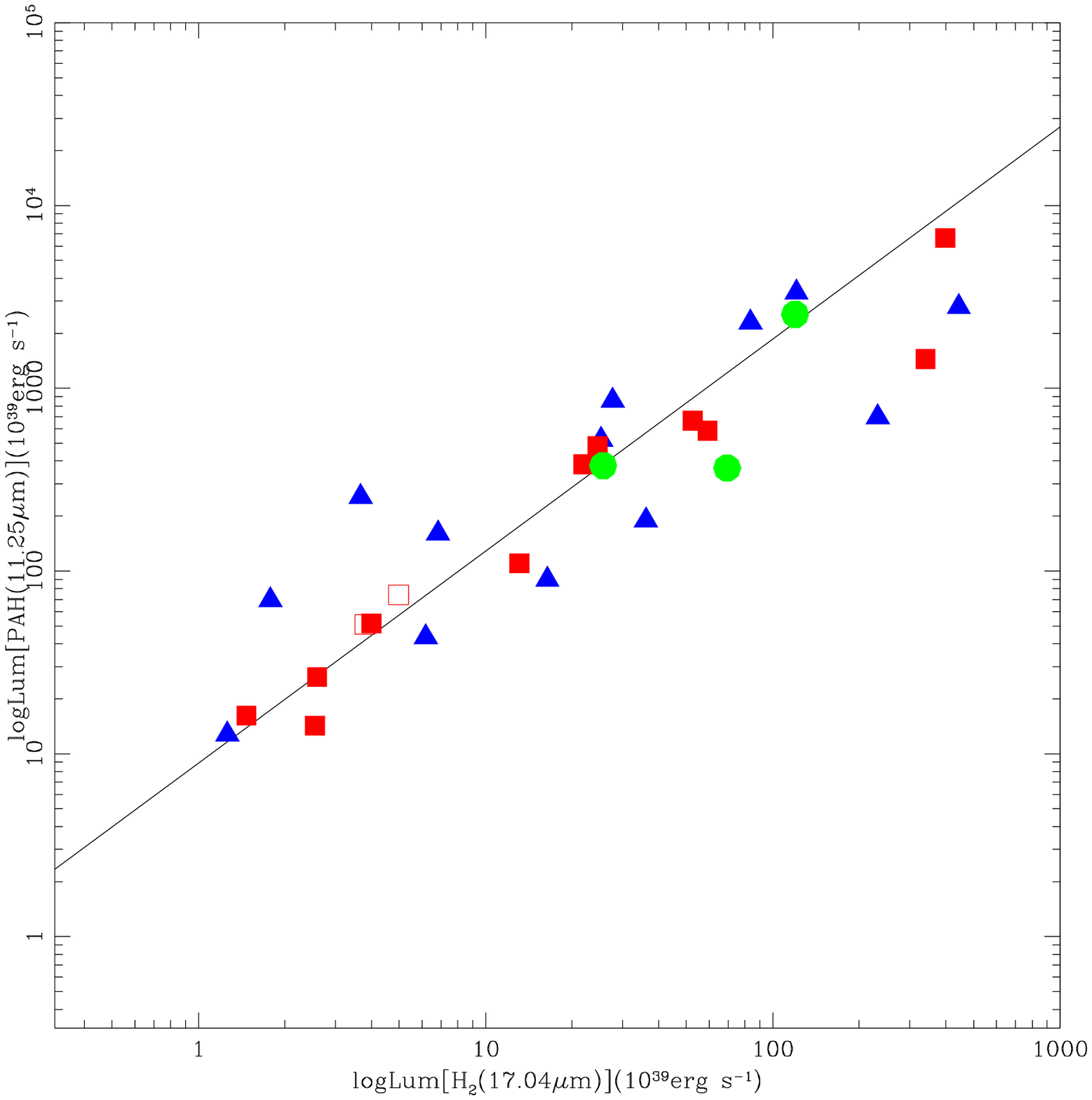}}
\caption{\textbf{a:} H$_2$ 17.04 $\mu$m line flux versus PAH 11.25$\mu$m integrated flux. 
 No correlation is present. \textbf{b:} H$_2$ 17.04 $\mu$m line luminosity 
versus PAH 11.25$\mu$m luminosity.The solid line is the least squares fit to all the Seyfert's data 
with a slope of 1.16, using a weighted least squares fit and 1.14 $\pm$ 0.13, using the 
bootstrap method. }
\end{figure}



\begin{figure}
\includegraphics[angle=0,scale=.80] {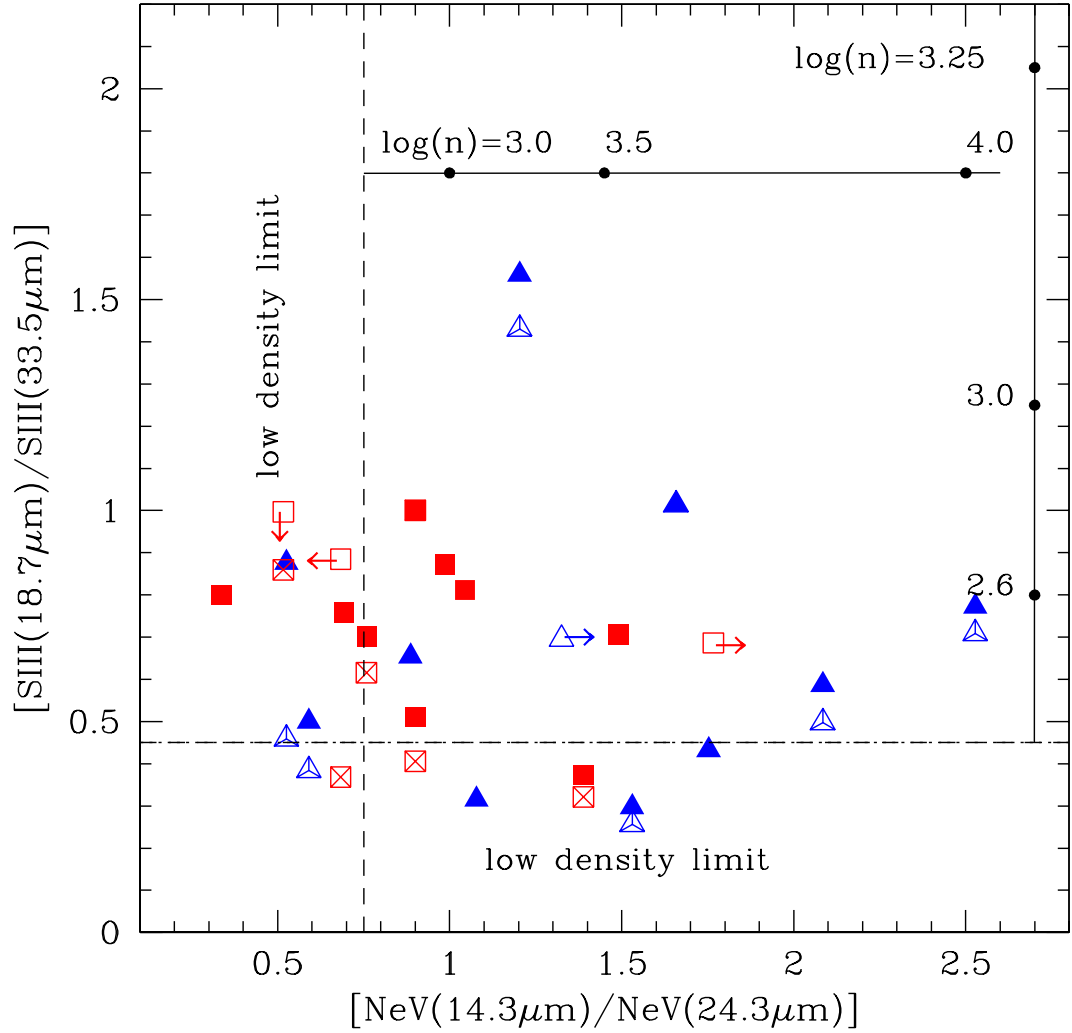}
\caption{[NeV]14.3$\mu$m/[NeV]24.3$\mu$m line ratio versus the 
[SIII]18.7$\mu$m/[SIII]33.5$\mu$m line ratio. Symbols  as in
the previous figures, except for the fact that we indicate here as open crossed symbols 
those for which the [SIII]18.71$\mu$m line was measured only with the small (SH) 
aperture and the above filled symbols directly above them show the corrected [SIII] line ratio for that object. 
The dashed lines show the low density limits (see text). 
The solid lines at the top and at the right give the corresponding electron densities. }
\end{figure}

\end{document}